\documentclass[aps,twocolumn,prb,superscriptaddress,amsmath,amssymb,longbibliography]{revtex4-2} 
\usepackage{dcolumn}
\usepackage{amsmath}
\usepackage{mathrsfs}
\usepackage{txfonts}
\usepackage{bm}
\usepackage[T1]{fontenc}
\usepackage{xspace}
\usepackage{ulem}
\usepackage{comment}
\usepackage{braket}
\usepackage{lipsum}
\usepackage{bbold}
\usepackage{mathtools}
\usepackage{booktabs}
\ifx\pdfoutput\undefined
\usepackage[dvipdfmx]{graphicx}
\usepackage[dvipdfmx]{hyperref}
\usepackage[dvipdfmx]{color}
\usepackage[dvipdfmx]{xcolor}
\else
\usepackage{graphicx}
\usepackage{hyperref}
\usepackage{color}
\usepackage{xcolor}
\usepackage[version=3]{mhchem}
\fi
\hypersetup{
        colorlinks=true,
        citecolor=blue,
        urlcolor=blue,
        linkcolor=blue
}

\begin{document}
\let\emph\textit

\title{
    Dynamical spin correlations in kagome antiferromagnets: comparison of Abrikosov fermion and Schwinger boson approaches beyond mean field
}
\author{Daiki Sasamoto}
\email[sasamoto.daiki.r6@dc.tohoku.ac.jp]{}
\author{Joji Nasu}
\affiliation{
  Department of Physics, Graduate School of Science, Tohoku University, Sendai, Miyagi 980-8578, Japan
}

\date{\today}
\begin{abstract}
Quantum spin liquids exhibit fractionalized spin excitations as a consequence of strong quantum many-body effects and have therefore attracted considerable attention.
The kagome antiferromagnetic Heisenberg model is a promising candidate for realizing a quantum spin-liquid ground state; however, the nature of its excitation spectrum remains controversial, particularly for the presence of a spin gap and the gauge structure coupled to fractional quasiparticles.
To address these issues, parton approaches have been extensively employed in theoretical studies, where spin operators are represented in terms of either fermionic or bosonic quasiparticles; the former approach is known as the Abrikosov fermion method, while the latter is referred to as the Schwinger boson method.
Thus far, these approaches have been pursued independently, and it has remained unclear how the results obtained from these two different frameworks compare with each other, particularly with respect to the spin dynamics and gauge structure of the kagome antiferromagnet.
Here, we investigate the dynamical spin structure factor of the antiferromagnetic Heisenberg model with a Dzyaloshinskii-Moriya interaction on a kagome lattice, which is relevant to the material herbertsmithite, by employing both the Abrikosov fermion and Schwinger boson approaches.
We find that the dynamical spin structure factor obtained from the Abrikosov fermion mean-field theory exhibits dome-shaped features, and that its continuum structure significantly depends on the gauge structure of the spin-liquid ansatz.
On the other hand, the Schwinger boson mean-field theory yields a concave-down structure in the low-energy region of the dynamical spin structure factor, which is distinct from that obtained using the Abrikosov fermion approach.
Moreover, the low-energy gap is substantially reduced, and the low-energy spectral weight is enhanced, by incorporating many-body effects beyond the mean-field approximation, which is consistent with experimental observations.
Our results suggest the importance of many-body effects in the Schwinger boson theory for capturing the low-energy spin dynamics of kagome antiferromagnets.
\end{abstract}
\maketitle


\section{Introduction}
\label{introduction}
In strongly correlated electron systems, the essential many-body effects are often revealed most clearly not in static quantities but in the structure and dynamics of elementary excitations. A canonical example is spin-charge separation in one dimension, where an injected electron fractionalizes into independent spin and charge modes~\cite{Voit-1995,Jompol-2009}. Another celebrated example is the fractional quantum Hall effect, in which strong interactions produce quasiparticles with fractional charge~\cite{Laughlin-1983,dePicciotto-1997,Stormer-1999}. These examples highlight that identifying the excitation spectrum and its fractionalized nature is crucial for understanding strongly correlated quantum matter.
Quantum spin liquids provide another important setting in which strong correlations give rise to fractionalized excitations~\cite{
Balents-2010,
Savary-Balents-2016,
Zhou-Kanoda-2017,
Knolle-Moessner-2019,
Wen-Yu-2019,
Broholm-Cava-2020,
Clark-Abdeldaim-2021}. They arise in Mott insulators, where the charge degrees of freedom are frozen by strong electron-electron interactions and the remaining spin degrees of freedom evade conventional long-range magnetic order even at zero temperature because of strong quantum fluctuations and geometrical frustration~\cite{Imada-1998,Ramirez-1994}. The idea of a quantum spin liquid goes back to Anderson's proposal of a resonating-valence-bond ground state for the triangular-lattice antiferromagnetic Heisenberg model~\cite{Anderson-1973}. Although it is now established that the nearest-neighbor triangular-lattice Heisenberg antiferromagnet develops $120^{\circ}$ magnetic order, Anderson's proposal initiated the modern search for microscopic models realizing quantum spin liquids. Among them, the antiferromagnetic Heisenberg model on the kagome lattice has long been regarded as one of the leading candidates and has therefore been studied extensively.
In studies of quantum spin liquids, a characteristic difficulty is to compute spin dynamics that can be probed experimentally.
In fact, available methods for calculating dynamical properties in disordered phases, such as those represented by quantum spin liquids, are limited. For example, quantum Monte Carlo methods~\cite{Sandvik-2010,Katzgraber-2011,Sandvik-2019} are among the most powerful numerical techniques for computing dynamics in quantum magnets, but their application to frustrated magnets is severely hampered by the negative sign problem~\cite{Troyer-2005}.
Against this background, the parton construction has become a powerful tool for the theoretical study of quantum spin liquids~\cite{Auerbach-2012,Savary-Balents-2016,Zhou-Kanoda-2017}. In this approach, spin operators are mapped onto bosonic or fermionic partons, and a mean-field theory is formulated in terms of these particles, which allows one to compute dynamical properties of disordered phases in a relatively straightforward manner.
Such approaches, often referred to as slave-particle methods, have long been used as analytical tools in the field of correlated electron systems, in particular in gauge-theoretical descriptions~\cite{Kogut-1979} of cuprate high-temperature superconductivity~\cite{Anderson-1987,Baskaran-1987,Kotliar-1988,Lee-Nagaosa-Wen-2006} and in the context of the Kondo effect in heavy-fermion systems~\cite{Barnes-1976,Coleman-1984,Read-1983}.

In addition to serving as a theoretical tool, the parton construction offers a physical picture for quantum spin liquids. Although rewriting a spin operator in terms of two auxiliary particles may at first seem to be merely a mathematical tool, it in fact captures the fractionalization of spin degrees of freedom. In conventional magnets, the elementary excitations are bosonic magnons, and a wide variety of magnetic systems can indeed be successfully understood in terms of magnon excitations~\cite{Manousakis-1991}. By contrast, in disordered phases such as quantum spin liquids, a description in terms of fluctuations around an ordered state is no longer appropriate. Instead, the elementary excitations are believed to be charge-neutral $S=1/2$ degrees of freedom carried by the partons introduced above, which may appear as either bosons (Schwinger bosons)~\cite{Arovas-Auerbach-1998,Sachdev-Read-1991,Read-Sachdev-1991,Auerbach-2012} or fermions (Abrikosov fermions)~\cite{Affleck-Marston-1988,Wen-1991,Wen-2002,Wen-2004,Zhou-Kanoda-2017}. This nonuniqueness reflects the fact that the same spin operator admits both bosonic and fermionic parton representations in an enlarged Hilbert space. Although these representations are formally equivalent when the local constraint is enforced exactly, they generally lead to distinct approximate theories after mean-field decoupling and can therefore yield substantially different predictions for observables such as the dynamical spin structure factor.
This raises a central question: how strongly does the choice of parton representation—bosonic or fermionic—affect the predicted spin dynamics, and can such differences be used to infer the nature of the low-energy quasiparticles in a given quantum spin liquid? In particular, qualitatively different continua, dispersive features, or spectral-weight distributions in $S(\bm{q},\omega)$ may provide useful clues as to whether the relevant fractionalized excitations are more naturally described in terms of bosonic or fermionic partons. Although both Abrikosov fermion and Schwinger boson mean-field theories have been extensively developed for a wide variety of frustrated magnets, they have usually been pursued separately rather than compared on equal footing within the same microscopic model. To our knowledge, a systematic discussion of how the two representations differ at the level of spin dynamics has remained lacking.
This issue is also directly relevant to experiment. In inelastic neutron-scattering experiments on candidate quantum spin liquids, one often observes a broad continuum in the dynamical spin structure factor. Such a continuum cannot be explained in terms of single-particle excitations such as magnons and is instead commonly interpreted as a signature of multi-parton excitations. The nature of the low-energy excitations therefore provides an important guideline for selecting an appropriate parton description. For instance, the Kitaev model~\cite{Motome-Nasu-2020,Trebst-Hickey-2022,Matsuda-2025} realizes a quantum spin liquid that is exactly described in terms of Majorana fermions~\cite{Kitaev-2006}, which justifies an Abrikosov fermion- (or Majorana-fermion-) based description~\cite{Rao-Moessner-Knolle-2025}. By contrast, for models without an exact solution, the character of the excitations must be inferred from both theoretical and experimental considerations. For example, recent theoretical studies have suggested that the $J_{1}$-$J_{2}$ Heisenberg model on a triangular lattice can be well described in terms of fermionic partons~\cite{Hu-2015,Iqbal-Hu-2016,Willsher-2025}.

In this context, the nature of the low-energy excitations in the $S=1/2$ kagome Heisenberg antiferromagnet remains an open problem. In particular, whether the ground state is gapped or gapless has been debated for many years without a definitive conclusion. From a theoretical perspective, this issue is often framed in terms of competing candidate spin-liquid states, most notably a gapped $\mathbb{Z}_{2}$ spin liquid and a gapless $\mathrm{U}(1)$ Dirac spin liquid.
In addition, the possibility of a chiral spin liquid (CSL), namely a topologically ordered spin liquid that spontaneously breaks time-reversal symmetry, has also been discussed in kagome models~\cite{Yang-Warman-Girvin-1993,Messio-Bernu-2012,Messio-Lhuillier-2013,Bauer-Cincio-2014,He-Sheng-Chen-2014,Gong-Zhu-Sheng-2014,Gong-Zhu-Balents-Sheng-2015,Wietek-Sterdyniak-Lauchli-2015,Hu-Zhu-Zhang-2015,Bieri-Messio-2015,Bieri-Lhuillier-2016}. In particular, a recent large-scale DMRG study combined with analytical analyses has suggested that the nearest-neighbor kagome Heisenberg antiferromagnet itself may lie in a gapped Kalmeyer-Laughlin-type CSL phase~\cite{Sun-Jin-Tu-Zhou-2024}.
At the mean-field level, this contrast is often reflected in the parton descriptions themselves, although the situation in the fermionic literature is more nuanced. Existing Schwinger boson mean-field studies of kagome antiferromagnets have often yielded gapped $\mathbb{Z}_{2}$ spin liquids~\cite{
Sachdev-1992,
Li-Su-Shen-2007,
Messio-Cepas-2010,
Huh-Punk-Sachdev-2011,
Fak-2012,
Messio-Bernu-2012,
Messio-Lhuillier-2013,
Punk-2014,
Halimeh-Punk-2016,
Mondal-2017,
Messio-Bieri-2017,
Halimeh-Singh-2019,
Mondal-2019,
Mondal-2021,
Lugan-Jaubert-2022,
Rossi-Motruk-2023}, whereas fermionic parton studies of the nearest-neighbor kagome Heisenberg model have discussed both gapless $\mathrm{U}(1)$ states, most notably the $\mathrm{U}(1)$ Dirac spin liquid, and candidate $\mathbb{Z}_{2}$ spin-liquid states~\cite{Lu-2011,Dodds-2013}. Thus, the two parton descriptions should not be regarded as forming a strict dichotomy, but rather as emphasizing different aspects of the problem.
Numerically, exact-diagonalization~\cite{Lecheminant-1997,Waldtmann-1998} and density-matrix renormalization-group (DMRG) studies~\cite{Jiang-Weng-Sheng-2008,Yan-Huse-White-2011,Depenbrock-2012,Nishimoto-2013} have reported a small spin gap of order $J/20$--$J/10$. Other DMRG works~\cite{He-Zaletel-Oshikawa-Pollmann-2017}, performed under different numerical conditions, instead argue for a gapless spectrum, while tensor-network~\cite{Liao-Xie-2017} and variational Monte Carlo studies~\cite{Iqbal-Becca-2013} likewise support gapless excitations. More recently, a machine-learning-assisted variational Monte Carlo study of the $S=1/2$ kagome Heisenberg antiferromagnet has been reported~\cite{Duric-2025}. Its results strongly suggest a spinon pair-density-wave ground state and provide evidence consistent with a gapless spinon spectrum; however, given the limited system sizes accessible, it remains difficult to distinguish a genuinely gapless state from one with an extremely small but finite gap.
Experimentally, the situation is similarly unsettled. In herbertsmithite, interpretations range from gapless or nearly gapless behavior \cite{Han-2012,Khuntia-2020} to a small but finite gap~\cite{Fu-2015}, while another study has also pointed to possible sensitivity to disorder and inhomogeneity~\cite{Han-2016}.
Related kagome materials such as Zn-barlowite~\cite{Feng-2017,Wang-2021,Breidenbach-2025} and kapellasite~\cite{Fak-2012,Ihara-2017} likewise exhibit low-energy behaviors that have been discussed in terms of either gapless excitations or a small finite gap.
Taken together, the current experimental situation does not yet provide a definitive conclusion regarding whether the kagome spin liquid is truly gapless or only very weakly gapped.
Overall, both numerical and experimental studies indicate that the low-energy spin excitations extend down to very small energies, while leaving unresolved whether the kagome spin liquid is truly gapless or only very weakly gapped. Although the gauge structure is not identical to the choice of bosonic or fermionic parton representation, these competing scenarios are expected to produce qualitatively different low-energy spin dynamics.

In this paper, we revisit the dynamical spin structure factor of kagome antiferromagnets with a $z$-directed Dzyaloshinskii-Moriya interaction within the framework of parton theories. We compute the dynamical spin structure factor using two approaches, namely the Abrikosov fermion mean-field theory (AFMFT) and the Schwinger boson mean-field theory (SBMFT).
Within AFMFT, we examine spin-liquid ans\"atze constructed solely from nearest-neighbor mean-field channels, including the previously considered $\mathrm{U}(1)$ ans\"atze as well as a $\mathbb{Z}_{2}$ ansatz.
We find that many $\mathrm{U}(1)$ spin-liquid states exhibit characteristic dome-shaped structures in the dynamical spin structure factor, whereas the $\mathbb{Z}_{2}$ spin-liquid state shows a qualitatively different high-energy response characterized by a relatively flat spectrum rather than a dome-like one. 
This suggests that, among spin-liquid ans\"atze constructed solely from nearest-neighbor mean-field channels, the dynamical spin structure factor can distinguish between $\mathrm{U}(1)$ and $\mathbb{Z}_{2}$ spin-liquid states.
We also investigate a $\mathbb{Z}_{2}$ spin-liquid state within SBMFT.
Although the corresponding mean-field dynamical spin structure factor exhibits gapped spin excitations, the inclusion of many-body effects beyond the mean-field approximation substantially reduces the spin gap and can even render the spectrum effectively gapless.
The resulting low-energy spectral structure is concave down and exhibits significant spectral weight near zero energy, in sharp contrast to the spectra obtained within AFMFT.
Moreover, the dynamical spin structure factor obtained by incorporating these many-body effects within SBMFT captures key features observed experimentally, suggesting that the Schwinger boson framework can capture the low-energy spin dynamics of kagome antiferromagnets once many-body effects between quasiparticles are properly taken into account.
Taken together, these results show that characteristic features of the dynamical spin response encode information about the nature and statistics of the underlying quasiparticles.

This paper is organized as follows.
In Sec.~\ref{sec:Method}, we present the methods employed in this study.
The Abrikosov fermion mean-field theory (AFMFT) and Schwinger boson mean-field theory (SBMFT) are described in Secs.~\ref{sec:Method_Abrikosov fermion mean-field theory} and \ref{sec:Method_Schwinger boson mean-field theory}, respectively.
We explain how to solve the mean-field Hamiltonians obtained in AFMFT and SBMFT, and how to compute the spin structure factors as observables, in Secs.~\ref{sec:Diagonalization of the BdG Hamiltonian} and \ref{sec:Calculation of the spin structure factor}.
Section~\ref{sec:Random phase approximation} briefly introduces the RPA framework used in this paper; further technical details are provided in Appendix~\ref{app:Detailed calculation of the random-phase approximation}.
In Sec.~\ref{sec:Application to kagome antiferromagnets}, we present the $S=1/2$ kagome antiferromagnetic Heisenberg model.
Section~\ref{sec:Results} reports the results obtained in this work.
For AFMFT and SBMFT, we describe the mean-field ansatz, the resulting spinon dispersions, and the static and dynamical structure factors in Secs.~\ref{sec:Results_Abrikosov fermion mean-field theory} and \ref{sec:Results_Schwinger boson mean-field theory}, respectively.
The dynamical structure factor incorporating many-body effects is presented in Sec.~\ref{sec:Random phase approximation beyond Schwinger boson mean-field}.
In Sec.~\ref{sec:Discussion}, we discuss the differences between the spin structure factors obtained in AFMFT and SBMFT, and examine how the spin structure factors computed within the Schwinger boson approach are related to the instability of the spin-liquid phase and the impact of many-body effects.
Finally, we summarize our findings in Sec.~\ref{sec:Summary}.

\section{Method}
\label{sec:Method}

\subsection{Abrikosov fermion mean-field theory}
\label{sec:Method_Abrikosov fermion mean-field theory}

In this section, we introduce the Abrikosov fermion method~\cite{Wen-1991,Affleck-Marston-1988,Zhou-Kanoda-2017,Wen-2004}, in which spin operators are rewritten in terms of fermionic operators, and AFMFT. In the following, we assume $S=1/2$ in the calculations based on the Abrikosov fermion method. Although in principle it is possible within the Abrikosov fermion formalism to describe an arbitrary spin quantum number $S$, the resulting expressions often become rather complicated~\cite{Zhou-Kanoda-2017}. 

In the Abrikosov fermion method, the spin operator $\bm{S}_{i}$ at lattice site $i$ is expressed in terms of a pair of fermionic operators
$\bm{f}_{i}=\left(f_{i\uparrow},f_{i\downarrow}\right)^{T}$ as
\begin{align}
    \label{Eq:fermionic-representation}
    S_{i}^{\gamma}=\frac{1}{2}\sum_{\mu,\nu}f_{i\mu}^{\dagger}\sigma_{\mu\nu}^{\gamma}f_{i\nu}.
\end{align}
Here, the fermionic operators introduced above satisfy the anticommutation relation
\begin{align}
    \label{Eq:fermionic-commutation-relation}
    \{f_{i\mu},f_{j\nu}^{\dagger}\}=\delta_{ij}\delta_{\mu\nu},
\end{align}
where $\delta_{ij}$ denotes the Kronecker delta.
Since this representation enlarges the Hilbert space, it is necessary to impose the local constraint
\begin{align}
    \label{Eq:local-number-constraint-fermion}
    n_{i}=\sum_{\mu}f_{i\mu}^{\dagger}f_{i\mu}=1 \, .
\end{align}
However, it is technically difficult to enforce this constraint exactly on every site. Therefore, within the mean-field framework discussed below, we introduce a uniform Lagrange multiplier and relax the constraint to
\begin{align}
    \label{Eq:average-number-constraint-fermion}
    \langle n\rangle=\frac{1}{N}\sum_{i}\langle n_{i}\rangle=1 \, ,
\end{align}
where $N$ is the total number of lattice sites.
The local constraint immediately implies, taking into account the fermionic nature of the operators, that
\begin{align}
    \label{Eq:local-constraint-fermion}
    f_{i\uparrow}f_{i\downarrow}=f_{i\uparrow}^{\dagger}f_{i\downarrow}^{\dagger}=0 \, .
\end{align}
Within the mean-field framework, this constraint is also implemented only on average.

With this rewriting of the spin operators, the interaction between spins can be expressed as
\begin{align}
    \label{Eq:interaction-fermion}
    S_{i}^{\gamma}S_{j}^{\gamma^{\prime}}=\frac{1}{4}\sum_{\mu,\nu,\rho,\lambda}\sigma^{\gamma}_{\mu\nu}\sigma^{\gamma^{\prime}}_{\rho\lambda}f_{i\mu}^{\dagger}f_{i\nu}f_{j\rho}^{\dagger}f_{j\lambda}
\end{align}
which takes the form of a quartic interaction term in the fermionic operators.

Since it is difficult to treat these interaction terms exactly, we now apply mean-field theory to this interaction. Namely, we decouple the product of four operators in a mean-field channel.
Because we are interested in spin-liquid phases, we assume that onsite decouplings associated with magnetic order, such as $\langle f_{i\mu}^{\dagger}f_{i\nu}\rangle$ and $\langle f_{j\rho}^{\dagger}f_{j\lambda}\rangle$, vanish. The spin–spin interaction above is then approximated as
\begin{align}
    \label{Eq:Heisenberg-mean-field-decouple-fermion}
    &S_{i}^{\gamma}S_{j}^{\gamma^{\prime}}=\frac{1}{4}\sum_{\mu,\nu,\rho,\lambda}\sigma^{\gamma}_{\mu\nu}\sigma^{\gamma^{\prime}}_{\rho\lambda}f_{i\mu}^{\dagger}f_{i\nu}f_{j\rho}^{\dagger}f_{j\lambda}\notag\\
    &\approx\frac{1}{4}\sum_{\mu,\nu,\rho,\lambda}\sigma^{\gamma}_{\mu\nu}\sigma^{\gamma^{\prime}}_{\rho\lambda}\left(\langle f_{i\nu}f_{j\rho}^{\dagger}\rangle f_{i\mu}^{\dagger}f_{j\lambda} +\langle f_{i\mu}^{\dagger}f_{j\lambda}\rangle f_{i\nu}f_{j\rho}^{\dagger}\right.\notag\\
    &\left. -\langle f_{i\nu}f_{j\lambda}\rangle f_{i\mu}^{\dagger}f_{j\rho}^{\dagger}-\langle f_{i\mu}^{\dagger}f_{j\rho}^{\dagger}\rangle f_{i\nu}f_{j\lambda}\right) + \text{const.}
\end{align}
We introduce the following bond operators:
\begin{align}
    \label{Eq:bond operators-fermion}
    \chi_{ij}
    &=\frac{1}{2}\sum_{\mu,\nu}f_{i\mu}^{\dagger}\sigma^{0}_{\mu\nu}f_{j\nu}=\frac{1}{2}\left(f_{i\uparrow}^{\dagger}f_{j\uparrow}+f_{i\downarrow}^{\dagger}f_{j\downarrow}\right),\\
    \eta_{ij}
    &=-\frac{i}{2}\sum_{\mu,\nu}f_{i\mu}\sigma^{y}_{\mu\nu}f_{j\nu}=-\frac{1}{2}\left(f_{i\uparrow}f_{j\downarrow}-f_{i\downarrow}f_{j\uparrow}\right),\\
    E_{ij}^{\gamma}
    &=\frac{1}{2}\sum_{\mu,\nu}f_{i\mu}^{\dagger}\sigma^{\gamma}_{\mu\nu}f_{j\nu},\\
    D_{ij}^{\gamma}
    &=\frac{i}{2}\sum_{\mu,\nu}f_{i\mu}\left(\sigma^{y}\sigma^{\gamma}\right)_{\mu\nu}f_{j\nu}.
\end{align}
Here, $\chi_{ij}$ and $\eta_{ij}$ are $\mathrm{SU}(2)$-invariant operators, i.e., they are invariant under global $\mathrm{SU}(2)$ transformations of the two-component spinon $\bm{f}_{i}=\left(f_{i\uparrow},f_{i\downarrow}\right)^{T}$. Accordingly, when the Hamiltonian explicitly preserves $\mathrm{SU}(2)$ symmetry and can be written solely in terms of Heisenberg interactions, the spin–spin interactions can be expressed in terms of the $\mathrm{SU}(2)$-invariant bond operators $\chi_{ij}$ and $\eta_{ij}$. On the other hand, in cases such as the Ising-type interactions, which explicitly break $\mathrm{SU}(2)$ symmetry, it becomes necessary to introduce the $\mathrm{SU}(2)$-breaking bond operators $E_{ij}^{\gamma}$ and $D_{ij}^{\gamma}$~\cite{Choi-Klein-Rosch-Kim-2018}. We note that these bond operators satisfy
$\chi_{ji}=\chi_{ij}^{\dagger},\eta_{ji}=\eta_{ij},E_{ji}^{\gamma}=E_{ij}^{\gamma\dagger},D_{ji}^{\gamma}=-D_{ij}^{\gamma}$.

Since the Heisenberg interaction manifestly preserves $\mathrm{SU}(2)$ symmetry, we may adopt a mean-field ansatz in which only the mean-field channels $\langle \chi_{ij}\rangle$ and $\langle\eta_{ij}\rangle$ are taken to be finite. In this case, we obtain
\begin{align}
    &\bm{S}_{i}\cdot\bm{S}_{j}=\frac{1}{4}\sum_{\mu,\nu,\rho,\lambda}\sum_{\gamma}\sigma^{\gamma}_{\mu\nu}\sigma^{\gamma}_{\rho\lambda}f_{i\mu}^{\dagger}f_{i\nu}f_{j\rho}^{\dagger}f_{j\lambda}\notag\\
    &\approx-\frac{3}{2}\left(\langle\chi_{ij}^{\dagger}\rangle\chi_{ij}+\langle\chi_{ij}
    \rangle\chi_{ij}^{\dagger}+\langle\eta_{ij}^{\dagger}\rangle\eta_{ij}+\langle\eta_{ij}\rangle\eta_{ij}^{\dagger}\right)+\text{const.}
\end{align}

In the presence of a Dzyaloshinskii-Moriya (DM) interaction, the spin Hamiltonian contains a term of the form
\begin{align}
    \label{eq:spin-Hamiltonian-DM}
    \bm{d}_{ij}\cdot\left(\bm{S}_{i}\times\bm{S}_{j}\right).
\end{align}
We now examine how this interaction can be treated within the present mean-field framework. 
Once the DM interaction is included, the $\mathrm{SU}(2)$ symmetry of the Hamiltonian is broken, and it becomes necessary to incorporate the $\mathrm{SU}(2)$-breaking bond operators $E_{ij}^{\gamma}$ and $D_{ij}^{\gamma}$. 
For example, let us consider the case in which the DM vector on the bond $\langle ij\rangle$ is oriented along the $z$ direction,
$\bm{d}_{ij}=\left(0,0,d_{ij}^{z}\right)^{T}.$
Applying the mean-field decoupling to the interaction above, we obtain
\begin{align}
    \label{Eq:DM-mean-field-decouple-fermion}
    &\bm{d}_{ij}\cdot\left(\bm{S}_{i}\times\bm{S}_{j}\right)=d_{ij}^{z}\left(S_{i}^{x}S_{j}^{y}-S_{i}^{y}S_{j}^{x}\right)\notag\\
    &\approx id_{ij}^{z}\left(\langle E_{ij}^{z\dagger}\rangle\chi_{ij}+\langle\chi_{ij}\rangle E_{ij}^{z\dagger}-\langle\chi_{ij}^{\dagger}\rangle E_{ij}^{z}-\langle E_{ij}^{z}\rangle\chi_{ij}^{\dagger}\right.\nonumber\\
    &\left.-\langle D_{ij}^{z\dagger}\rangle \eta_{ij}-\langle \eta_{ij}\rangle D_{ij}^{z\dagger}+\langle\eta_{ij}^{\dagger}\rangle D_{ij}^{z}+\langle D_{ij}^{z}\rangle\eta_{ij}^{\dagger}\right)+\text{const.}
\end{align}
Here, the mean-field channels $E_{ij}^{z}$ and $D_{ij}^{z}$ associated with the $\mathrm{SU}(2)$-breaking bond operators can in general acquire finite values. Consistently with this, the Heisenberg Hamiltonian is also modified as
\begin{align}
    &\bm{S}_{i}\cdot\bm{S}_{j}\notag\\
    &\approx-\frac{3}{2}\left(\langle\chi_{ij}^{\dagger}\rangle\chi_{ij}+\langle\chi_{ij}
    \rangle\chi_{ij}^{\dagger}+\langle\eta_{ij}^{\dagger}\rangle\eta_{ij}+\langle\eta_{ij}\rangle\eta_{ij}^{\dagger}\right)\notag\\
    &+\frac{1}{2}\left(\langle E_{ij}^{z\dagger}\rangle E_{ij}^{z}+\langle E_{ij}^{z}\rangle E_{ij}^{z\dagger}+\langle D_{ij}^{z\dagger}\rangle D_{ij}^{z}+\langle D_{ij}^{z}\rangle D_{ij}^{z\dagger}\right)+\text{const.}
\end{align}

Up to this point, we have introduced the procedure of the mean-field approximation within the Abrikosov fermion method. In the following, we reformulate the above mean-field theory using a convenient notation for the Abrikosov fermion representation~\cite{Wen-1991,Wen-2002,Wang-Normand-2019}.
To this end, we introduce the time-reversal and particle-hole partner of the two-component fermionic operator $\bm{f}_{i}=\left(f_{i\uparrow},f_{i\downarrow}\right)^{T}$, defined as $\bar{\bm{f}}_{i}=\left(f_{i\downarrow}^{\dagger},-f_{i\uparrow}^{\dagger}\right)^{T}$. Furthermore, by introducing the matrix of fermionic operators
\begin{align}
    \label{Eq:matrix of fermionic operators}
    \psi_{i}=\begin{pmatrix}
        \bm{f}_{i},\bar{\bm{f}}_{i}
    \end{pmatrix}=
    \begin{pmatrix}
        f_{i\uparrow}&f_{i\downarrow}^{\dagger}\\
        f_{i\downarrow}&-f_{i\uparrow}^{\dagger}
    \end{pmatrix},
\end{align}
we can formulate the redundancy inherent in the Abrikosov fermion representation in a more transparent way.
In terms of this notation, the spin operators can be written as
\begin{align}
    \label{Eq:fermionic-matrix-representation}
    S_{i}^{\gamma}=\frac{1}{4}\text{Tr}\left(\psi_{i}^{\dagger}\sigma^{\gamma}\psi_{i}\right) \, .
\end{align}

A characteristic feature of the Abrikosov fermion representation is the presence of a local $\mathrm{SU}(2)$ redundancy. To make this explicit, let us consider a local $\mathrm{SU}(2)$ transformation of the form
\begin{align}
    \label{eq:local-SU2-transformation}
    \psi_{i} \rightarrow \psi_{i} W_{i}.
\end{align}
Under this transformation, the spin operators remain invariant. This shows that the transformation generated by $W_{i}$ is not a physical spin rotation but a redundancy of the Abrikosov fermion representation.
To characterize this redundancy, we introduce the operators
\begin{align}
    \label{Eq:definition-Lambda}
    \Lambda_{i}^{\gamma}
    =
    \frac{1}{4}\mathrm{Tr}
    \left(
        \psi_{i} \tau^{\gamma}\psi_{i}^{\dagger}
    \right),
\end{align}
where $\tau^\gamma$ denote Pauli matrices acting in the internal SU(2) particle-hole space, in contrast to $\sigma^\gamma$, which act on the physical spin degrees of freedom. Under the local $\mathrm{SU}(2)$ transformation $\psi_i \rightarrow \psi_i W_i$, these operators transform as
\begin{align}
    \label{Eq:right-transformation-Lambda}
    \Lambda_{i}^{\gamma}\rightarrow
    \frac{1}{4}\mathrm{Tr}
    \left(
        \psi_{i}W_{i}\tau^{\gamma}W_{i}^{\dagger}\psi_{i}^{\dagger}
    \right)
    =
    \sum_{\rho} R_{\rho\gamma}(W_{i})\Lambda_{i}^{\rho},
\end{align}
where $R(W_i)$ denotes the three dimensional representation of $W_{i}$.
Since the original spin operators possess only the physical $\mathrm{SU}(2)$ degrees of freedom associated with spin rotations, the $\mathrm{SU}(2)$ redundancy associated with $W_i$ must be projected out in order to recover the physical Hilbert space. This is achieved by imposing the local constraint $\bm{\Lambda}_i=\bm{0}$, namely
\begin{align}
    \label{Eq:local-matrix-constraint-fermion}
    \mathrm{Tr}\left(\psi_{i}\tau^{\gamma}\psi_{i}^{\dagger}\right)=0,
\end{align}
on each site. When written out explicitly, this constraint reproduces the local constraints Eqs.~\eqref{Eq:local-number-constraint-fermion} and \eqref{Eq:local-constraint-fermion}.
It is difficult to enforce these constraints locally. We therefore introduce global Lagrange multipliers $a^{\gamma}$ and impose the constraints on average as
\begin{align}
    \label{Eq:global-number-constraint-fermion}
    \frac{1}{N}\sum_{i}\Bigl\langle\mathrm{Tr}\left(\psi_{i}\tau^{\gamma}\psi_{i}^{\dagger}\right)\Bigr\rangle=0,
\end{align}
and determine the Lagrange multipliers self-consistently so that Eq.~\eqref{Eq:global-number-constraint-fermion} is satisfied.

Here, we formulate AFMFT using the matrix $\psi_{i}$ defined by Eq.~\eqref{Eq:matrix of fermionic operators}. To this end, we introduce two matrices $u_{ij}^{0}$ and $u_{ij}^{z}$ composed of mean fields as
\begin{align}
    \label{Eq:mean-field-matrix}
    u_{ji}^{0}
    &=\frac{1}{2}\langle \psi_{j}^{\dagger}\psi_{i}\rangle
    =\begin{pmatrix}
    \langle\chi_{ij}^{\dagger}\rangle&\langle\eta_{ij}^{\dagger}\rangle\\
    \langle\eta_{ij}\rangle&-\langle\chi_{ij}\rangle
    \end{pmatrix},\\
    u_{ji}^{z}
    &=\frac{1}{2}\langle\psi_{j}^{\dagger}\sigma^{z}\psi_{i}\rangle
    =\begin{pmatrix}
    \langle E_{ij}^{z\dagger}\rangle&-\langle D_{ij}^{z\dagger}\rangle\\
    \langle D_{ij}^{z}\rangle&\langle E_{ij}^{z}\rangle
    \end{pmatrix}.
\end{align}
Note that these matrices satisfy $u_{ij}^{0}=u_{ji}^{0\dagger}$ and $u_{ij}^{z}=u_{ji}^{z\dagger}$.
In terms of these matrices, the mean-field Hamiltonian for the Heisenberg and DM interactions can be written as
\begin{align}
&\bm{S}_{i}\cdot\bm{S}_{j}\notag\\
&\approx -\frac{3}{8}\text{Tr}\left(u_{ji}^{0}\psi_{i}^{\dagger}\psi_{j}+\text{H.c.}\right)
+\frac{1}{8}\text{Tr}\left(u_{ji}^{z}\psi_{i}^{\dagger}\sigma^{z}\psi_{j}+\text{H.c.}\right)
+\text{const}, \label{Eq:Heisenberg-representation-fermion-matrix}\\
\shortintertext{and}
&d_{ij}^{z}\left(S_{i}^{x}S_{j}^{y}-S_{i}^{y}S_{j}^{x}\right)\notag\\
&\approx \frac{d_{ij}^{z}}{4}\text{Tr}\left(
iu_{ji}^{z}\psi_{i}^{\dagger}\psi_{j}
-iu_{ji}^{0}\psi_{i}^{\dagger}\sigma^{z}\psi_{j}
+\text{H.c.}\right)
+\text{const}. \label{Eq:DM-representation-fermion-matrix}
\end{align}
respectively.
This representation is useful for understanding how the mean-field parameters $u_{ji}^{0}$ and $u_{ji}^{z}$ transform under local gauge transformations $W_{i}$. Under a local $\mathrm{SU}(2)$ transformation $\psi_{i}\rightarrow \psi_{i}W_{i}$, the mean-field matrices transform as
\begin{align}
    \label{Eq:SU(2)transformation-mean-field-matrix}
    u_{ji}^{0}
    &\rightarrow \tilde{u}^{0}_{ji}=W_{j}u_{ji}^{0}W_{i}^{\dagger},\\
    u_{ji}^{z}
    &\rightarrow \tilde{u}^{z}_{ji}=W_{j}u_{ji}^{z}W_{i}^{\dagger}\, .
\end{align}
We define the invariant gauge group (IGG) as the set of local gauge transformations that leave the mean-field ansatz
$\{u^{0}_{ij}\}$ and $\{u^{z}_{ij}\}$ invariant:
\begin{align}
    \label{Eq:definition-IGG}
    \text{IGG}=\{W_{i}\mid W_{i}u_{ij}^{0}W_{j}^{\dagger}=u_{ij}^{0},~
    W_{i}u_{ij}^{z}W_{j}^{\dagger}=u_{ij}^{z},~
    W_{i}\in \text{SU}(2)\} \, .
\end{align}
The IGG is one of the key quantities used to classify quantum spin-liquid phases within the AFMFT framework~\cite{Wen-1991,Wen-2002,Wen-2004}.

We now consider a closed loop $C$ on the lattice,
\(
  i_0 \to i_1 \to \dots \to i_{n-1} \to i_n=i_0 \, ,
\)
and define along this loop the ordered product of bond matrices $U^{\gamma}(C)$ as
\begin{align}
    \label{Eq:definition-U(C)}
  U^{\gamma}(C)=u^{\gamma}_{i_0 i_1}u^{\gamma}_{i_1 i_2}\cdots u^{\gamma}_{i_{n-1} i_0}.
\end{align}
Under a local gauge transformation $\psi_{i}\rightarrow\psi_{i}W_{i}$, this matrix transforms as
\begin{align}
    \label{Eq:SU(2)transformation-U(C)}
  U^{\gamma}(C)\rightarrow W_{i_{0}}U^{\gamma}(C)W_{i_{0}}^{\dagger},
\end{align}
and is therefore gauge dependent. On the other hand, the trace of this quantity,
\begin{align}
    \label{Eq:definition-Wilson-loop}
  \widetilde{P}^{\gamma}(C)=\frac{1}{2}\text{Tr}U^{\gamma}(C),
\end{align}
is gauge invariant and is referred to as the Wilson loop.
However, since $u^{\gamma}_{ij}$ obtained in the mean-field theory is not necessarily unitary, $\widetilde{P}^{\gamma}(C)$ generally contains an overall amplitude.
To extract the flux information, we define the normalized Wilson loop as
\begin{align}
    P^{\gamma}(C)\equiv \frac{\widetilde{P}^{\gamma}(C)}{\sqrt{\det U^{\gamma}(C)}}.
\end{align}
Both $\widetilde{P}^{\gamma}(C)$ and $P^{\gamma}(C)$ are gauge invariant, since $\mathrm{Tr}$ and $\det$ are invariant under conjugation.
For an ansatz with IGG $\mathrm{U}(1)$, one may choose a gauge in which the loop operator takes the form
$U^{\gamma}(C)=\rho^{\gamma}(C)\,e^{i\Phi^{\gamma}(C)\tau^{z}}$ with $\rho^{\gamma}(C)>0$,
so that the normalized Wilson loop can be parameterized as
\begin{align}
  P^{\gamma}(C)=\cos\Phi^{\gamma}(C)\, ,
\end{align}
thereby defining the $\mathrm{U}(1)$ gauge flux $\Phi^{\gamma}(C)$ modulo $2\pi$.
In particular, when the flux is restricted to $\Phi^{\gamma}(C)=0$ or $\pi$, the sign of $P^{\gamma}(C)$ distinguishes the $0$- and $\pi$-flux sectors.
For an ansatz with IGG $\mathbb{Z}_{2}$, the normalized Wilson loop is restricted to $P^{\gamma}(C)=\pm 1$, which can likewise be interpreted as a $\mathbb{Z}_{2}$ flux ($0$ or $\pi$) piercing the loop.
In the classification of quantum spin-liquid phases within AFMFT, it is thus important to characterize a phase by combining the gauge structure encoded in the IGG with the flux pattern encoded in the Wilson loops.

\subsection{Schwinger boson mean-field theory}
\label{sec:Method_Schwinger boson mean-field theory}
In this section, we describe the Schwinger boson representation~\cite{Arovas-Auerbach-1998,Read-Sachdev-1991,Auerbach-2012,Sachdev-Read-1991}, in which spin operators are rewritten in terms of bosonic operators, and SBMFT. In analogy with the Abrikosov fermion case, the spin operator is expressed in terms of a two-component bosonic operator $\bm{b}_{i}=\left(b_{i\uparrow},b_{i\downarrow}\right)^{T}$ as
\begin{align}
    \label{Eq:bosonic-representation}
    S_{i}^{\gamma}=\frac{1}{2}\sum_{\mu,\nu}b_{i\mu}^{\dagger}\sigma_{\mu\nu}^{\gamma}b_{i\nu} \, .
\end{align}
Here, the bosonic operators satisfy the commutation relations
\begin{align}
    \label{Eq:bosonic-commutation-relation}
    [b_{i\mu},b_{j\nu}^{\dagger}]=\delta_{ij}\delta_{\mu\nu}.
\end{align}
Since this representation enlarges the Hilbert space, one has to impose the local constraint
\begin{align}
    \label{Eq:local-number-constraint-boson}
    n_{i}=\sum_{\mu}b_{i\mu}^{\dagger}b_{i\mu}=2S.
\end{align}
However, it is difficult to enforce this constraint locally. We therefore introduce a global Lagrange multiplier and implement the constraint on average as
\begin{align}
    \label{Eq:global-number-constraint-boson}
    \frac{1}{N}\sum_{i}\langle n_{i}\rangle=\kappa,
\end{align}
and determine the Lagrange multiplier such that $\kappa=2S$, thereby incorporating the constraint.

A key difference from the Abrikosov fermion representation is that, in the Schwinger boson approach, an arbitrary spin quantum number $S$ can be included simply through the constraint $\kappa=2S$. As will become clear in the following, as long as we work within the SBMFT framework, the information about the spin quantum number $S$ appears only in the constraint Eq.~\eqref{Eq:global-number-constraint-boson}. Consequently, the SBMFT has the advantage that it can be extended to arbitrary $S$ in a transparent way. Furthermore, the parameter $\kappa$ serves as a tuning parameter that controls the strength of quantum fluctuations. In particular, the limit of small $\kappa$ corresponds to the quantum limit. This provides the advantage that one can analyze how magnetic order melts as a consequence of quantum fluctuations.
We note, however, that this mean-field treatment does not enforce the local boson-number constraint exactly. Consequently, the spin sum rule for the momentum- and frequency-integrated spin structure factor is not exactly satisfied at the SBMFT level. In the conventional mean-field evaluation, the integrated spectral weight can be larger than the exact value $S(S+1)$ by a factor of $3/2$~\cite{Auerbach-Arovas-1988,Messio-Cepas-2010,Mezio-2011,Mezio-2012}. In this sense, changing $\kappa$ can partly compensate for the sum-rule violation, but in the present work we use $\kappa$ primarily as a parameter controlling the strength of quantum fluctuations rather than as an exact sum-rule correction.

As in the Abrikosov fermion case, we introduce the following bond operators in the Schwinger boson representation:
\begin{align}
    \label{Eq:bond operators-boson}
    \mathcal{A}_{ij}
    &= \frac{i}{2}\sum_{\mu,\nu}
        b_{i\mu}\sigma^{y}_{\mu\nu} b_{j\nu}
      = \frac{1}{2}\left(
        b_{i\uparrow}b_{j\downarrow}
        - b_{i\downarrow}b_{j\uparrow}
      \right),\\
    \mathcal{B}_{ij}
    &= \frac{1}{2}\sum_{\mu,\nu}
        b_{i\mu}^{\dagger}\sigma^{0}_{\mu\nu} b_{j\nu}
      = \frac{1}{2}\left(
        b_{i\uparrow}^{\dagger}b_{j\uparrow}
        + b_{i\downarrow}^{\dagger}b_{j\downarrow}
      \right),\\
    \mathcal{C}_{ij}^{\gamma}
    &= \frac{1}{2}\sum_{\mu,\nu}
        b_{i\mu}^{\dagger}\sigma^{\gamma}_{\mu\nu}b_{j\nu},\\
    \mathcal{D}_{ij}^{\gamma}
    &= \frac{i}{2}\sum_{\mu,\nu}
        b_{i\mu}\left(\sigma^{y}\sigma^{\gamma}\right)_{\mu\nu}b_{j\nu}.
\end{align}
Here, $\mathcal{A}_{ij}$ and $\mathcal{B}_{ij}$ are $\mathrm{SU}(2)$-invariant operators, i.e., they are invariant under global $\mathrm{SU}(2)$ transformations of the two-component spinon $\bm{b}_{i}=\left(b_{i\uparrow},b_{i\downarrow}\right)^{T}$, and they are the bond operators used to describe Heisenberg interactions. When the $\mathrm{SU}(2)$ symmetry is broken, as in the case of Ising-type interactions, it becomes necessary to introduce the $\mathrm{SU}(2)$-breaking bond operators $\mathcal{C}_{ij}^{\gamma}$ and $\mathcal{D}_{ij}^{\gamma}$. Finally, we note that these bond operators satisfy
$\mathcal{B}_{ji}=\mathcal{B}_{ij}^{\dagger},\mathcal{A}_{ji}=-\mathcal{A}_{ij},\mathcal{C}_{ji}^{\gamma}=\mathcal{C}_{ij}^{\gamma\dagger},\mathcal{D}_{ji}^{\gamma}=\mathcal{D}_{ij}^{\gamma}$.

We now discuss the representation of the interactions in the Schwinger boson formalism. We first consider the Heisenberg interaction,
\begin{align}
    \label{Eq:Heisenberg-interaction-boson-representation}
    \bm{S}_{i}\cdot\bm{S}_{j}
    = \frac{1}{4}\sum_{\mu,\nu,\rho,\lambda}\sum_{\gamma}
      \sigma^{\gamma}_{\mu\nu}\sigma^{\gamma}_{\rho\lambda}
      b_{i\mu}^{\dagger}b_{i\nu}b_{j\rho}^{\dagger}b_{j\lambda} \, .
\end{align}
Using the identities
\begin{align}
    \label{Eq:identity-Heinsenberg}
    \sum_{\gamma}\sigma^{\gamma}_{\mu\nu}\sigma^{\gamma}_{\rho\lambda}
    &= 2\sigma_{\mu\lambda}^{0}\sigma_{\nu\rho}^{0}
       -\sigma_{\mu\nu}^{0}\sigma_{\rho\lambda}^{0},\\
    &= \sigma_{\mu\nu}^{0}\sigma_{\rho\lambda}^{0}
       +2\sigma_{\mu\rho}^{y}\sigma_{\nu\lambda}^{y},
\end{align}
we obtain two different representations of the interaction:
\begin{align}
    \bm{S}_{i}\cdot\bm{S}_{j}
    &= 2 {} :\mathcal{B}_{ij}^{\dagger}\mathcal{B}_{ij}: - S^{2},\label{Eq:Ferro-representation}\\
    &= S^{2}-2\mathcal{A}_{ij}^{\dagger}\mathcal{A}_{ij}\label{Eq:Antierro-representation}.
\end{align}
Here, $:\mathcal{O}_{1}\mathcal{O}_{2}:$ denotes normal ordering, i.e., rearranging the operators so that all creation operators stand to the left of the annihilation operators.
Originally, in the Schwinger boson approach, Eq.~\eqref{Eq:Ferro-representation} was used for ferromagnetic interactions and Eq.~\eqref{Eq:Antierro-representation} for antiferromagnetic interactions~\cite{Arovas-Auerbach-1998}. Subsequent studies~\cite{Gazza-1993,Mattsson-1995,Flint-Coleman-2009,Messio-Cepas-2010,Mezio-2011,Messio-Bernu-2012,Messio-Lhuillier-2013,Merino-Holt-Powell-2014,Halimeh-Punk-2016,Mondal-2017,Ghioldi-Gonzalez-2018,Zhang-Ghioldi-2019,Mondal-2019,Halimeh-Singh-2019,Mondal-2021,Zhang-2021} have shown, however, that in order to obtain more accurate ground-state energies and spin structure factors, it is advantageous to treat the two representations on an equal footing and employ the averaged form
\begin{align}
    \label{Eq:Heisenberg-representation-boson}
    \bm{S}_{i}\cdot\bm{S}_{j}
    = {} :\mathcal{B}_{ij}^{\dagger}\mathcal{B}_{ij}:
      -\mathcal{A}_{ij}^{\dagger}\mathcal{A}_{ij}.
\end{align}
In this paper, we also adopt Eq.~\eqref{Eq:Heisenberg-representation-boson} as the representation of the Heisenberg interaction.

We next consider the representation of the same DM term discussed in Sec.~\ref{sec:Method_Abrikosov fermion mean-field theory},
$d_{ij}^{z}\left(S_{i}^{x}S_{j}^{y}-S_{i}^{y}S_{j}^{x}\right)$.
To this end, we first write
\begin{align}
    \label{Eq:DM-interaction-boson-representation}
    S_{i}^{x}S_{j}^{y}
    = \frac{1}{4}\sum_{\mu,\nu,\rho,\lambda}
      \sigma^{x}_{\mu\nu}\sigma^{y}_{\rho\lambda}
      b_{i\mu}^{\dagger}b_{i\nu}b_{j\rho}^{\dagger}b_{j\lambda} \, .
\end{align}
Using the identity
\begin{align}
    \label{Eq:identity-DM}
    \sigma_{\mu\nu}^{x}\sigma_{\rho\lambda}^{y}  
    &=-\sigma_{\mu\lambda}^{x}\sigma_{\nu\rho}^{y}
      +\sigma_{\mu\rho}^{x}\sigma_{\nu\lambda}^{y}  
      =\sigma_{\mu\lambda}^{y}\sigma_{\nu\rho}^{x}
      -\sigma_{\mu\rho}^{y}\sigma_{\nu\lambda}^{x},\\  
    &=-i\sigma_{\mu\rho}^{z}\sigma_{\nu\lambda}^{0}
      +i\sigma_{\mu\lambda}^{z}\sigma_{\nu\rho}^{0}
      =i\sigma_{\mu\rho}^{0}\sigma_{\nu\lambda}^{z}
      -i\sigma_{\mu\lambda}^{0}\sigma_{\nu\rho}^{z},
\end{align}
we obtain
\begin{align}
    \label{Eq:DM-xy-two-representations-boson}
    S_{i}^{x}S_{j}^{y}
    &=\frac{1}{2}\left[
      :\mathcal{C}_{ij}^{x\dagger}\mathcal{C}_{ij}^{y}:
      + {} :\mathcal{C}_{ij}^{y\dagger}\mathcal{C}_{ij}^{x}:
      +i\left(\mathcal{D}_{ij}^{z\dagger}\mathcal{A}_{ij}
      -\mathcal{A}_{ij}^{\dagger}\mathcal{D}_{ij}^{z}\right)\right]\\
    &=-\frac{1}{2}\left[
      \mathcal{D}_{ij}^{x\dagger}\mathcal{D}_{ij}^{y}
      +\mathcal{D}_{ij}^{y\dagger}\mathcal{D}_{ij}^{x}
      +i\left(:\mathcal{C}_{ij}^{z\dagger}\mathcal{B}_{ij}:
      - {} :\mathcal{B}_{ij}^{\dagger}\mathcal{C}_{ij}^{z}:\right)\right] \, .
\end{align}
Here, the above expressions for the interaction have been derived by exploiting the hermiticity of $S_{i}^{x}S_{j}^{y}$, namely $(S_{i}^{x}S_{j}^{y})^{\dagger}=S_{i}^{x}S_{j}^{y}$.

Similarly, we obtain
\begin{align}
    \label{Eq:DM-yx-two-representations-boson}
    S_{i}^{y}S_{j}^{x}
    &=\frac{1}{2}\left[
      :\mathcal{C}_{ij}^{x\dagger}\mathcal{C}_{ij}^{y}:
      + {} :\mathcal{C}_{ij}^{y\dagger}\mathcal{C}_{ij}^{x}:
      -i\left(\mathcal{D}_{ij}^{z\dagger}\mathcal{A}_{ij}
      -\mathcal{A}_{ij}^{\dagger}\mathcal{D}_{ij}^{z}\right)\right]\\
    &=-\frac{1}{2}\left[
      \mathcal{D}_{ij}^{x\dagger}\mathcal{D}_{ij}^{y}
      +\mathcal{D}_{ij}^{y\dagger}\mathcal{D}_{ij}^{x}
      -i\left(:\mathcal{C}_{ij}^{z\dagger}\mathcal{B}_{ij}:
      - {} :\mathcal{B}_{ij}^{\dagger}\mathcal{C}_{ij}^{z}:\right)\right].
\end{align}
Combining these results, we finally obtain
\begin{align}
    \label{Eq:DM-representation-boson}
	    S_{i}^{x}S_{j}^{y}-S_{i}^{y}S_{j}^{x}
	    &= \frac{i}{2}\left[
	      \mathcal{D}_{ij}^{z\dagger}\mathcal{A}_{ij}
	      -\mathcal{A}_{ij}^{\dagger}\mathcal{D}_{ij}^{z}\right.\notag\\
	    &\left.\qquad
	      - {} :\mathcal{C}_{ij}^{z\dagger}\mathcal{B}_{ij}:
	      + {} :\mathcal{B}_{ij}^{\dagger}\mathcal{C}_{ij}^{z}:\right].
\end{align}
Here again, we adopt the representation obtained by averaging over the possible decompositions of the interaction.

Up to this point, we have shown that, within the Schwinger boson representation, spin–spin interactions can be written in a quadratic form of bond operators. In fact, by using the bond operators defined above, an arbitrary spin interaction can be expressed in a quadratic form of bond operators. We then apply the following mean-field approximation to this quadratic form:
\begin{align}
    &S_{i}^{\alpha}S_{j}^{\beta}=\sum_{p,q}A_{pq}^{\alpha\beta} {} :\mathcal{Q}_{ij}^{p\dagger}\mathcal{Q}_{ij}^{q}:\nonumber\\
    &\approx\sum_{p,q}A_{pq}^{\alpha\beta}\left(\langle \mathcal{Q}_{ij}^{p\dagger}\rangle\mathcal{Q}_{ij}^{q}+\langle\mathcal{Q}_{ij}^{q}\rangle\mathcal{Q}_{ij}^{p\dagger}-\langle\mathcal{Q}_{ij}^{p\dagger}\rangle\langle\mathcal{Q}_{ij}^{q}\rangle\right),
\end{align}
where $\mathcal{Q}_{ij}^{p}$ denotes the $p$-th component of the bond operator vector $\bm{\mathcal{Q}}_{ij}$ defined as $\left(\mathcal{A}_{ij},\mathcal{B}_{ij},\mathcal{C}_{ij}^{x},\mathcal{C}_{ij}^{y},\mathcal{C}_{ij}^{z},\mathcal{D}_{ij}^{x},\mathcal{D}_{ij}^{y},\mathcal{D}_{ij}^{z}\right)^{T}.$
With this mean-field decoupling, the problem is reduced to a bosonic Bogoliubov--de Gennes (BdG) form.

\subsection{Diagonalization of the BdG Hamiltonian}
\label{sec:Diagonalization of the BdG Hamiltonian}

We consider a generic quantum spin model described by the following Hamiltonian
\begin{align}
    \label{Eq:quantum-spin-Hamiltonian}
    \mathcal{H}=\frac{1}{2}\sum_{i,j}\sum_{\alpha,\beta}J_{ij}^{\alpha\beta}S_{i}^{\alpha}S_{j}^{\beta}. 
\end{align}
Here, $\bm{S}_{i}$ denotes the spin operator defined at site $i$ on an $N$-site lattice. The coefficient $J_{ij}^{\alpha\beta}$ represents the exchange interaction between the spin component $S_{i}^{\alpha}$ at site $i$ and the spin component $S_{j}^{\beta}$ at site $j$, and satisfies the relation $J_{ij}^{\alpha\beta}=J_{ji}^{\beta\alpha}$.
The Heisenberg model with a $z$-directed DM interaction discussed above is expressed as a special case of Eq.~\eqref{Eq:quantum-spin-Hamiltonian}.
In the previous two sections, we showed that the interaction terms written in terms of these spin operators can be expressed, using the Abrikosov fermion and Schwinger boson representations, as products of two bond operators in the form
\begin{align}
    \mathcal{H}=\frac{1}{2}\sum_{i,j}\sum_{\alpha,\beta}\sum_{p,q}J_{ij}^{\alpha\beta} A_{pq}^{\alpha\beta} {} :\mathcal{Q}_{ij}^{p\dagger}\mathcal{Q}_{ij}^{q}: .
\end{align}
Here, $\mathcal{Q}_{ij}^{p}$ denotes the $p$-th component of the bond operator vector $\bm{\mathcal{Q}}_{ij}$. In the Abrikosov fermion representation, it is defined as
$\bm{\mathcal{Q}}_{ij}=\left(\chi_{ij},\eta_{ij},E_{ij}^{x},E_{ij}^{y},E_{ij}^{z},D_{ij}^{x},D_{ij}^{y},D_{ij}^{z}\right)^{T}$,
whereas in the Schwinger boson representation, it is defined as
$\bm{\mathcal{Q}}_{ij}=\left(\mathcal{A}_{ij},\mathcal{B}_{ij},\mathcal{C}_{ij}^{x},\mathcal{C}_{ij}^{y},\mathcal{C}_{ij}^{z},\mathcal{D}_{ij}^{x},\mathcal{D}_{ij}^{y},\mathcal{D}_{ij}^{z}\right)^{T}$.
Applying a mean-field approximation, we can write
\begin{align}
    \mathcal{H}^{\text{MF}}=\sum_{i,j}\mathcal{H}^{\text{MF}}_{ij}+\mathcal{H}_{\text{c}} \, .
\end{align}
Here, $\mathcal{H}_{ij}^{\text{MF}}$ is defined as
\begin{align}
    \mathcal{H}^{\text{MF}}_{ij}=\frac{1}{2}\sum_{\alpha,\beta}\sum_{p,q}J_{ij}^{\alpha\beta} A_{pq}^{\alpha\beta}\left(\langle \mathcal{Q}_{ij}^{p\dagger}\rangle\mathcal{Q}_{ij}^{q}+\langle\mathcal{Q}_{ij}^{q}\rangle\mathcal{Q}_{ij}^{p\dagger}-\langle\mathcal{Q}_{ij}^{p\dagger}\rangle\langle\mathcal{Q}_{ij}^{q}\rangle\right).
\end{align}

Furthermore, $\mathcal{H}_{\text{c}}$ denotes the term that enforces the constraint, and is defined as
\begin{align}
\label{eq:mean-field-constraint-term}
\mathcal{H}_{\text{c}}
=
\left\{
\begin{aligned}
& \displaystyle \sum_{i}\left\{a^{z}\left(n_{i}-1\right)
+\left[\left(a^{x}+ia^{y}\right)f_{i\uparrow}^{\dagger}f_{i\downarrow}^{\dagger}+\text{H.c.}\right]\right\},\\
& \displaystyle \lambda\sum_{i}\left(n_{i}-\kappa\right).
\end{aligned}
\right.
\end{align}
The upper (lower) line corresponds to the constraint in AFMFT (SBMFT).

Since $\bm{\mathcal{Q}}_{ij}$ is expressed in terms of either the Abrikosov fermions $\bm{f}_{i},\bm{f}_{j}$ or the Schwinger bosons $\bm{b}_{i},\bm{b}_{j}$, the mean-field Hamiltonian $\mathcal{H}^{\text{MF}}$ can be written in a bilinear form of these operators as
\begin{align}
    \mathcal{H}^{\text{MF}}=\frac{1}{2}\sum_{l,l^{\prime}}C_{l}^{\dagger}\mathcal{M}_{ll^{\prime}}C_{l^{\prime}}+\text{const},
\end{align}
where the site $i$ is labeled as $i=(l,m)$ by a unit-cell index $l=1,2,\dots,N/M$ and a sublattice index $m=1,2,\dots,M$. Here, $\mathcal{M}_{l,l^{\prime}}$ is a $4M\times 4M$ Hermitian matrix, and $C_{l}^{\dagger}$ is a $4M$-dimensional vector defined by
\begin{align}
    C_{l}^{\dagger}=\Bigl(
    c_{\left(l,1\right),\uparrow}^{\dagger}\cdots c_{\left(l,M\right),\uparrow}^{\dagger},c_{\left(l,1\right),\downarrow}^{\dagger}\cdots c_{\left(l,M\right),\downarrow}^{\dagger},\notag\\
    c_{\left(l,1\right),\uparrow}\cdots c_{\left(l,M\right),\uparrow},c_{\left(l,1\right),\downarrow}\cdots c_{\left(l,M\right),\downarrow} 
    \Bigr)
\end{align}
In this notation, the operator $c$ denotes the fermionic operator $f$ in AFMFT, while it denotes the bosonic operator $b$ in SBMFT.
In what follows, we drop the constant terms in the above expression, since they only produce an overall energy shift and do not affect the Heisenberg time evolution or spin correlation functions.

By performing a Fourier transform of the operators $C_{l}$, we obtain
\begin{align}
    \mathcal{H}^{\text{MF}}=\frac{1}{2}\sum_{\bm{k}}^{\text{BZ}}C_{\bm{k}}^{\dagger}\mathcal{M}_{\bm{k}}C_{\bm{k}},
\end{align}
where the sum over $\bm{k}$ is taken within the first Brillouin zone of the superlattice, and $C_{\bm{k}}^{\dagger}$ is the Fourier-transformed vector defined as
\begin{align}
    C_{\bm{k}}^{\dagger}&=\Bigl(c_{\bm{k},1,\uparrow}^{\dagger}\cdots c_{\bm{k},M,\uparrow}^{\dagger},c_{\bm{k},1,\downarrow}^{\dagger}\cdots c_{\bm{k},M,\downarrow}^{\dagger},\notag\\
    &c_{-\bm{k},1,\uparrow}\cdots c_{-\bm{k},M,\uparrow},c_{-\bm{k},1,\downarrow}\cdots c_{-\bm{k},M,\downarrow}\Bigr)
\end{align}
Here,
\begin{align}
    c^{\dagger}_{\bm{k},m}=\sqrt{\frac{M}{N}}\sum_{l}c_{\left(l,m\right)}^{\dagger}e^{i\bm{k}\cdot\bm{R}_{l}}.
\end{align}
In this expression, $\bm{R}_{l}$ denotes the representative position of the unit cell $l$, and $\mathcal{M}_{\bm{k}}$ is a $4M\times 4M$ matrix defined as the Fourier transform of the real-space matrix $\mathcal{M}_{ll^{\prime}}$,
\begin{align}
    \mathcal{M}_{\bm{k}}=\sum_{l^{\prime}}\mathcal{M}_{ll^{\prime}}e^{-i\bm{k}\cdot\left(\bm{R}_{l}-\bm{R}_{l^{\prime}}\right)},
\end{align}
where $\mathcal{M}_{\bm{k}}$ does not depend on $l$ due to translational symmetry, i.e., $\mathcal{M}_{ll^{\prime}}$ depends only on the relative coordinate $\bm{R}_{l}-\bm{R}_{l^{\prime}}$.

We diagonalize $\mathcal{M}_{\bm{k}}$ by a Bogoliubov transformation,
$\mathcal{E}_{\bm{k}}=\mathcal{T}_{\bm{k}}^{\dagger}\mathcal{M}_{\bm{k}}\mathcal{T}_{\bm{k}}$.
Here, $\mathcal{T}_{\bm{k}}$ is a $4M\times 4M$ matrix. When $c$ represents fermionic operators, $\mathcal{T}_{\bm{k}}$ is a unitary matrix satisfying
$\mathcal{T}_{\bm{k}}^{\dagger}\mathcal{T}_{\bm{k}}=\mathcal{T}_{\bm{k}}\mathcal{T}_{\bm{k}}^{\dagger}=\bm{1}_{4M\times4M}$,
whereas when $c$ represents bosonic operators, $\mathcal{T}_{\bm{k}}$ is a paraunitary matrix satisfying
$\mathcal{T}_{\bm{k}}^{\dagger}\sigma^{3}\mathcal{T}_{\bm{k}}=\mathcal{T}_{\bm{k}}\sigma^{3}\mathcal{T}_{\bm{k}}^{\dagger}=\sigma^{3}$.
Here, $\bm{1}_{4M\times 4M}$ is the $4M\times 4M$ unit matrix and $\sigma^{3}$ is defined as 
$\sigma^{3}=\begin{pmatrix}
    \bm{1}_{2M\times2M}&0\\
    0&-\bm{1}_{2M\times2M}
\end{pmatrix}$.

The matrix $\mathcal{E}_{\bm{k}}=\operatorname{diag}\{\varepsilon_{\bm{k},1},\cdots,\varepsilon_{\bm{k},2M},\mp\varepsilon_{-\bm{k},1},\cdots,\mp\varepsilon_{-\bm{k},2M}\}$ is diagonal, where the upper (lower) sign corresponds to AFMFT (SBMFT).
Using this Bogoliubov transformation, the mean-field Hamiltonian can be written in a diagonalized form as
\begin{align}
    \label{eq:diagonalized-quasiparticle-Hamiltonian}
    \mathcal{H}^{\text{MF}}=\frac{1}{2}\sum_{\bm{k}}^{\text{BZ}}\Gamma_{\bm{k}}^{\dagger}\mathcal{E}_{\bm{k}}\Gamma_{\bm{k}}.
\end{align}
Here, the basis that diagonalizes the Hamiltonian is given by $\Gamma_{\bm{k}}=\mathcal{T}_{\bm{k}}^{-1}C_{\bm{k}}$, which is defined as
\begin{align}
    \Gamma_{\bm{k}}=\Bigl(\gamma_{\bm{k},1},\cdots,\gamma_{\bm{k},2M},\gamma_{-\bm{k},1}^{\dagger},\cdots,\gamma_{-\bm{k},2M}^{\dagger}\Bigr)^{T} \, .
\end{align}
The operators $\gamma_{\bm{k},\eta}$ and $\gamma_{\bm{k},\eta}^{\dagger}$ are the annihilation and creation operators of quasiparticles, and they satisfy the same algebraic relations as the original $c_{\bm{k}\eta}$ and $c_{\bm{k}\eta}^{\dagger}$. Accordingly, in AFMFT the $\gamma$ operators are fermionic, while in SBMFT they are bosonic.
Therefore, the distribution function $g(\varepsilon_{\bm{k},\eta})=\langle \gamma_{\bm{k},\eta}^{\dagger}\gamma_{\bm{k},\eta}\rangle$ at temperature $T$ is given by
\begin{align}
    \label{eq:quasiparticle-distribution-function}
    g(\varepsilon)
=\left\{
\begin{array}{ll}
\dfrac{1}{e^{\varepsilon/T}+1}, & \text{AFMFT}, \\
\dfrac{1}{e^{\varepsilon/T}-1}, & \text{SBMFT},
\end{array}
\right.
\end{align}
which yields the Fermi distribution function for AFMFT and the Bose distribution function for SBMFT. Here, the Boltzmann constant $k_{\text{B}}$ is set to unity. With the above diagonalization, one can determine the mean-field parameters and the Lagrange multipliers self-consistently.
Using the diagonalized quasiparticle Hamiltonian in Eq.~\eqref{eq:diagonalized-quasiparticle-Hamiltonian} and the distribution function in Eq.~\eqref{eq:quasiparticle-distribution-function}, we evaluate the expectation values of the bond operators and the constraint equations. Since the mean-field Hamiltonian depends on the mean-field parameters and Lagrange multipliers, the bond fields are determined self-consistently. The Lagrange multipliers in Eq.~\eqref{eq:mean-field-constraint-term} are fixed so that the corresponding averaged constraints, such as Eqs.~\eqref{Eq:global-number-constraint-fermion} and \eqref{Eq:global-number-constraint-boson}, are satisfied. Enlarging the unit cell changes the number of independent bond fields and the dimension $M$ of the BdG matrix, but the same self-consistency procedure is used.

\subsection{Calculation of the spin structure factor}
\label{sec:Calculation of the spin structure factor}

We describe the methodology for evaluating the spin structure factor within the AFMFT or SBMFT framework. The dynamical structure factor is defined as the Fourier transform of the real-time, real-space spin correlation function:
\begin{align}
    \label{eq:definition-dynamical-structure-factor}
    S^{\alpha\alpha^{\prime}}(\bm{q},\omega)
    =\int_{-\infty}^{\infty}\frac{dt}{2\pi} \, e^{i\omega t}\frac{1}{N}\sum_{i,j}
    e^{-i\bm{q}\cdot\left(\bm{r}_{i}-\bm{r}_{j}\right)}
    \langle S_{i}^{\alpha}(t)S_{j}^{\alpha^{\prime}}\rangle.
\end{align}
For the actual calculation, it is convenient to formulate the same response in imaginary time. We introduce
\begin{align}
    \label{eq:imaginary-time-spin-susceptibility}
    \chi_{ij}^{\alpha\alpha^{\prime}}(\tau)
    =\left\langle T_{\tau}S_{i}^{\alpha}(\tau)S_{j}^{\alpha^{\prime}}(0)\right\rangle,\qquad
    0\leq \tau<\beta,
\end{align}
where $T_{\tau}$ is the imaginary-time-ordering operator. The imaginary-time spin operator is written as
\begin{align}
    \label{eq:imaginary-time-spin-operator}
    S_{i}^{\alpha}(\tau)
    =\frac{1}{2}\sum_{\mu,\nu}\bar{c}_{i\mu}(\tau)\sigma_{\mu\nu}^{\alpha}c_{i\nu}(\tau),
\end{align}
with
\begin{align}
    c_{i\mu}(\tau)=e^{\tau\mathcal{H}^{\text{MF}}}c_{i\mu}e^{-\tau\mathcal{H}^{\text{MF}}},\qquad
    \bar{c}_{i\mu}(\tau)=e^{\tau\mathcal{H}^{\text{MF}}}c_{i\mu}^{\dagger}e^{-\tau\mathcal{H}^{\text{MF}}}.
\end{align}
The bar denotes the imaginary-time-evolved creation operator. The corresponding Matsubara susceptibility is
\begin{align}
    \label{eq:matsubara-susceptibility}
    \chi^{\alpha\alpha^{\prime}}(\bm{q},i\omega_{n})
    =\frac{1}{N}\sum_{i,j}e^{-i\bm{q}\cdot\left(\bm{r}_{i}-\bm{r}_{j}\right)}
    \int_{0}^{\beta}d\tau\,e^{i\omega_{n}\tau}\chi_{ij}^{\alpha\alpha^{\prime}}(\tau),
\end{align}
where $\omega_{n}=2\pi n/\beta$ is a bosonic Matsubara frequency. The retarded susceptibility is obtained by the analytic continuation $i\omega_{n}\to\omega+i\delta$, where the positive infinitesimal $\delta$ is replaced by a finite broadening in numerical calculations. With the convention used in Eq.~\eqref{eq:definition-dynamical-structure-factor},
\begin{align}
    \label{eq:dssf-from-matsubara-susceptibility}
    S^{\alpha\alpha^{\prime}}(\bm{q},\omega)
    =\frac{1}{\pi}\frac{1}{1-e^{-\beta\omega}}
    \operatorname{Im}\chi^{\alpha\alpha^{\prime}}(\bm{q},\omega+i\delta).
\end{align}
Although the above expressions are written in the standard finite-temperature Matsubara form, all spectra shown in this work are evaluated in the zero-temperature limit $T\to0$, or equivalently $\beta\to\infty$. In this limit, the distribution functions and the fluctuation--dissipation prefactor are understood in their zero-temperature forms.
Furthermore, the static structure factor is defined as the equal-time correlation in momentum space, obtained by integrating the dynamical structure factor over frequency:
\begin{align}
    \label{eq:definition-static-structure-factor}
    S^{\alpha\alpha^{\prime}}(\bm{q})
    =\int_{-\infty}^{\infty}d\omega\:S^{\alpha\alpha^{\prime}}(\bm{q},\omega)
    =\frac{1}{N}\sum_{i,j}e^{-i\bm{q}\cdot\left(\bm{r}_{i}-\bm{r}_{j}\right)}
    \langle S_{i}^{\alpha}S_{j}^{\alpha^{\prime}}\rangle.
\end{align}
Thus, the central task is to evaluate $\chi_{ij}^{\alpha\alpha^{\prime}}(\tau)$ consistently with the bond-operator mean-field ansatz. We start from the operator identity
\begin{align}
    \label{eq:evaluation_spin_correlator}
    S_{i}^{\alpha}S_{j}^{\alpha^{\prime}}
    =\sum_{p,q}A_{pq}^{\alpha\alpha^{\prime}}{}:\mathcal{Q}_{ij}^{p\dagger}\mathcal{Q}_{ij}^{q}:,
\end{align}
and evaluate the imaginary-time spin correlation by decomposing the resulting four-spinon expectation value into products of two-point contractions,
\begin{align}
    \label{eq:imaginary-time-spin-correlator-bond}
    \chi_{ij}^{\alpha\alpha^{\prime}}(\tau)
    \simeq\sum_{p,q}A_{pq}^{\alpha\alpha^{\prime}}\overline{\mathcal{F}}_{ij}^{p}(\tau)\mathcal{F}_{ij}^{q}(\tau).
\end{align}
Here, $\mathcal{F}_{ij}^{q}(\tau)$ and $\overline{\mathcal{F}}_{ij}^{p}(\tau)$ denote the imaginary-time two-point contractions associated with $\mathcal{Q}_{ij}^{q}$ and $\mathcal{Q}_{ij}^{p\dagger}$, respectively. This expression is common to AFMFT and SBMFT; the difference between the two approaches enters through the statistics of $c$ and through the matrices defining each bond operator. For a hopping-type bond operator
\begin{align}
    \mathcal{Q}_{ij}^{p}
    =\frac{1}{2}\sum_{\mu,\nu}X_{\mu\nu}^{p}c_{i\mu}^{\dagger}c_{j\nu},
\end{align}
the corresponding contractions are
\begin{subequations}
\label{eq:hopping-type-Matsubara-contractions}
\begin{align}
    \mathcal{F}_{ij}^{p}(\tau)
    &=\frac{1}{2}\sum_{\mu,\nu}X_{\mu\nu}^{p}
    \left\langle T_{\tau}\bar{c}_{i\mu}(\tau)c_{j\nu}(0)\right\rangle,\\
    \overline{\mathcal{F}}_{ij}^{p}(\tau)
    &=\frac{1}{2}\sum_{\mu,\nu}X_{\mu\nu}^{p*}
    \left\langle T_{\tau}c_{i\mu}(\tau)\bar{c}_{j\nu}(0)\right\rangle.
\end{align}
\end{subequations}
For a pairing-type bond operator
\begin{align}
    \mathcal{Q}_{ij}^{p}
    =\frac{1}{2}\sum_{\mu,\nu}Y_{\mu\nu}^{p}c_{i\mu}c_{j\nu},
\end{align}
we use
\begin{subequations}
\label{eq:pairing-type-Matsubara-contractions}
\begin{align}
    \mathcal{F}_{ij}^{p}(\tau)
    &=\frac{1}{2}\sum_{\mu,\nu}Y_{\mu\nu}^{p}
    \left\langle T_{\tau}c_{i\mu}(\tau)c_{j\nu}(0)\right\rangle,\\
    \overline{\mathcal{F}}_{ij}^{p}(\tau)
    &=\frac{1}{2}\sum_{\mu,\nu}Y_{\mu\nu}^{p*}
    \left\langle T_{\tau}\bar{c}_{i\mu}(\tau)\bar{c}_{j\nu}(0)\right\rangle.
\end{align}
\end{subequations}
For example, the SU(2)-invariant hopping channel has $X^{p}=\sigma^{0}$ in both AFMFT and SBMFT, while the singlet-pairing matrix is $Y^{p}=-i\sigma^{y}$ for the AFMFT $\eta_{ij}$ channel and $Y^{p}=i\sigma^{y}$ for the SBMFT $\mathcal{A}_{ij}$ channel. The signs specific to fermionic Wick contractions are included when evaluating the corresponding two-point functions. All contractions are selected from the Nambu two-point function
\begin{align}
    \label{eq:evaluation_two_point_correlator}
    \mathcal{G}_{mm^{\prime}}(\bm{k},\tau)
    =\left\langle T_{\tau}C_{\bm{k},m}(\tau)C_{\bm{k},m^{\prime}}^{\dagger}(0)\right\rangle.
\end{align}
Using $C_{\bm{k}}=\mathcal{T}_{\bm{k}}\Gamma_{\bm{k}}$, this two-point function is expressed in the diagonal quasiparticle basis as
\begin{align}
    \label{eq:evaluation_two_point_correlator_diagonal}
    \mathcal{G}_{mm^{\prime}}(\bm{k},\tau)
    =\sum_{n,n^{\prime}}\mathcal{T}_{mn}
    \left\langle T_{\tau}\Gamma_{\bm{k}n}(\tau)\Gamma_{\bm{k}n^{\prime}}^{\dagger}(0)\right\rangle
    \mathcal{T}^{\dagger}_{n^{\prime}m^{\prime}}.
\end{align}
Since the $\Gamma$ basis diagonalizes $\mathcal{H}^{\text{MF}}$, the correlator is diagonal in the quasiparticle index. For $0\leq\tau<\beta$,
\begin{align}
    \label{eq:distribution_function}
    \left\langle T_{\tau}\Gamma_{\bm{k}n}(\tau)\Gamma_{\bm{k}n}^{\dagger}(0)\right\rangle
    =\left\{
    \begin{array}{ll}
    e^{-\varepsilon_{\bm{k},n}\tau}\left(1\mp g\left(\varepsilon_{\bm{k},n}\right)\right), & n\leq 2M,\\
    e^{\varepsilon_{-\bm{k},n-2M}\tau}g\left(\varepsilon_{-\bm{k},n-2M}\right), & n>2M,
    \end{array}
\right.
\end{align}
where the upper (lower) sign in \(\mp\) corresponds to AFMFT (SBMFT).
By Fourier transforming Eq.~\eqref{eq:evaluation_two_point_correlator} back to real space and selecting the appropriate particle-hole, spin, and sublattice components, one obtains the contractions in Eq.~\eqref{eq:imaginary-time-spin-correlator-bond}. Equivalently, after carrying out the internal Matsubara-frequency summation, the bare susceptibility can be written in the generic pole form
\begin{align}
    \label{eq:generic-pole-form-susceptibility}
    \chi^{\alpha\alpha^{\prime}}(\bm{q},i\omega_{n})
    =\frac{1}{N}\sum_{\bm{k}}^{\mathrm{BZ}}\sum_{\eta,\eta^{\prime}}
    \mathcal{W}_{\eta\eta^{\prime}}^{\alpha\alpha^{\prime}}(\bm{k},\bm{q})
    \frac{g\left(E_{\bm{k},\eta}\right)
    -g\left(E_{\bm{k}+\bm{q},\eta^{\prime}}\right)}
    {i\omega_{n}+E_{\bm{k},\eta}-E_{\bm{k}+\bm{q},\eta^{\prime}}},
\end{align}
The retarded response is obtained by analytic continuation $i\omega_{n}\to\omega+i\delta$. The matrix-element weight $\mathcal{W}_{\eta\eta^{\prime}}^{\alpha\alpha^{\prime}}(\bm{k},\bm{q})$ contains the spin matrices, sublattice wave functions, the Bogoliubov or unitary transformation matrices, and the statistics-dependent signs from Wick contractions. The signed quasiparticle energies are defined as $E_{\bm{k},\eta}=\varepsilon_{\bm{k},\eta}$ for $\eta\leq2M$ and $E_{\bm{k},\eta}=-\varepsilon_{-\bm{k},\eta-2M}$ for $\eta>2M$; for negative-energy branches, $g(E)$ is understood by analytic continuation of the corresponding Fermi or Bose distribution defined in Eq.~\eqref{eq:quasiparticle-distribution-function}.
This formulation also clarifies how the spin structure factor should be interpreted in parton language. A single-spinon mean-field dispersion is gauge dependent in the parton representation and is therefore not a directly observable neutron-scattering spectrum. The physical spin operator, by contrast, defines a gauge-invariant spin correlation function and is bilinear in spinon operators. Therefore, within the mean-field approximation, $S(\bm{q},\omega)$ is obtained from a convolution of two spinon propagators, with form factors determined by the spin matrices, sublattice wave functions, and the Bogoliubov or unitary transformation matrices. Schematically, the spectral support is controlled by two-spinon processes with energies approximately given by $\varepsilon_{\bm{k},n}+\varepsilon_{\bm{k}+\bm{q},m}$, or by the corresponding particle-hole/anomalous channels in the Nambu representation, while the intensity is weighted by coherence factors. Thus, the lower boundary of the continuum at fixed $\bm{q}$ is governed by minimizing such two-spinon energies over the internal momentum $\bm{k}$ and band indices, but the visible high-intensity features need not coincide with either a single-spinon band or even with the continuum boundary if the relevant matrix elements are small. In this sense, a broad continuum in $S(\bm{q},\omega)$ is the gauge-invariant spin response associated with fractionalized spinons, whereas a sharp pole generated by additional interactions, such as in an RPA treatment, should be regarded as a collective or bound-state-like feature in the same gauge-invariant spin channel.

\subsection{Random phase approximation}
\label{sec:Random phase approximation}
\begin{figure}[t]
  \centering
      \includegraphics[width=\columnwidth,clip]{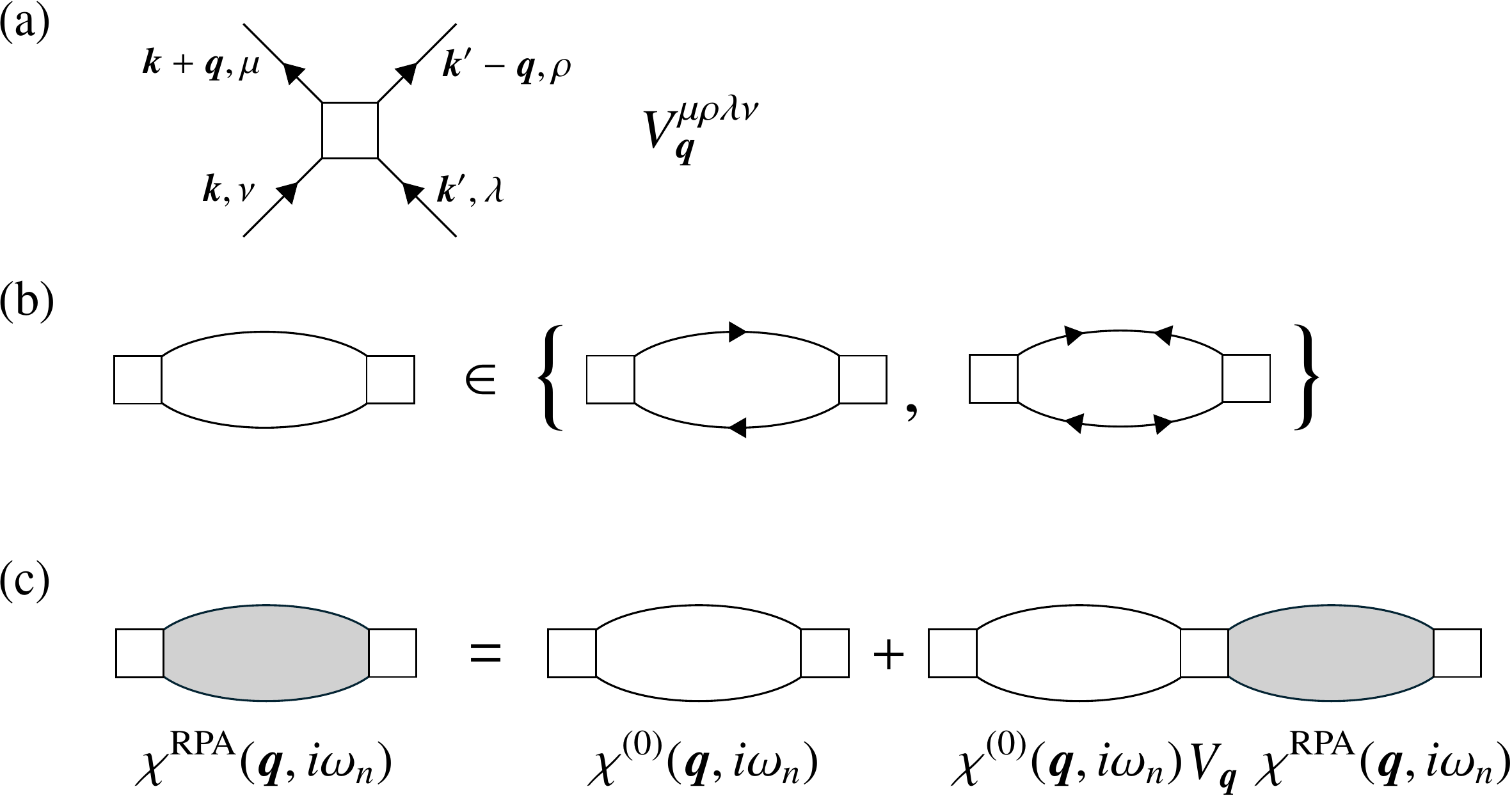}
      \caption{
      (a)
      Four-point vertex $V_{\bm{q}}^{\mu\nu\rho\lambda}$ defined in Eq.~\eqref{eq:definition-vertex-function}.
      (b)
      Bubble diagrams contributing to the bare susceptibility $\chi^{(0)}(\bm{q},i\omega_{n})$. In addition to the conventional particle-number-conserving bubble, anomalous bubbles (right) also contribute, reflecting the particle-number-nonconserving structure of the Schwinger boson mean-field theory.
      (c)
      Feynman-diagram representation of the Dyson equation for the RPA susceptibility. The shaded bubble denotes the RPA-dressed susceptibility $\chi_{\text{RPA}}(\bm{q},i\omega_{n})$, obtained by connecting the bare bubbles in (b) with the four-point vertex in (a).}
      \label{fig:RPA_diagram}
\end{figure}

In this section, we briefly review the random phase approximation (RPA) as one of the theoretical approaches that incorporate the effects of many-body interactions beyond mean-field theory.
Because an RPA scheme is built upon a mean-field description, the reliability of this scheme hinges on that of the underlying mean-field theory, as noted in previous studies~\cite{Rao-Moessner-Knolle-2025}.
As discussed in Sec.~\ref{sec:SB_Mean-field ansatz}, we therefore take as the starting point of the present perturbative treatment a Schwinger boson mean-field ansatz, motivated by previous theoretical studies suggesting that SBMFT can account for low-energy spectra consistent with inelastic neutron-scattering experiments~\cite{Punk-2014,Messio-Bieri-2017}.
By contrast, as explained in Sec.~\ref{sec:Discussion}, the Abrikosov fermion mean-field states considered in this work do not reproduce the experimentally relevant low-energy spectral features, and we therefore do not pursue an RPA analysis within that framework.
In this section, we only present the overall structure of the calculation; the detailed derivations are explained in Appendix~\ref{app:Detailed calculation of the random-phase approximation}.

In the scheme proposed in Ref.~\cite{Rao-Moessner-Knolle-2025,Willsher-2025}, the RPA interaction
vertex can be constructed from the original Hamiltonian in Eq.~\eqref{Eq:quantum-spin-Hamiltonian}.
Expressing this Hamiltonian in the Schwinger boson representation, we obtain
\begin{align}
    \label{eq:Perturbation-Hamiltonian-bosonic-representation-r-space}
    \mathcal{H}
    =\frac{1}{2}\frac{1}{4}\sum_{i,j}\sum_{\alpha,\beta}\sum_{\mu,\nu,\rho,\lambda}
    J_{ij}^{\alpha\beta}\sigma^{\alpha}_{\mu\nu}\sigma^{\beta}_{\rho\lambda}
    b_{i\mu}^{\dagger}b_{i\nu}b_{j\rho}^{\dagger}b_{j\lambda}.
\end{align}
Performing a Fourier transform, this becomes
\begin{align}
    \label{eq:Perturbation-Hamiltonian-bosonic-representation-k-space}
    \mathcal{H}
    =\frac{1}{2}\frac{M}{N}\sum_{\bm{k},\bm{k}^{\prime},\bm{q}}\sum_{\mu,\nu,\rho,\lambda}
    V_{\bm{q}}^{\mu\rho\lambda\nu}
    b_{\bm{k}+\bm{q}\mu}^{\dagger}b_{\bm{k}^{\prime}-\bm{q}\rho}^{\dagger}b_{\bm{k}^{\prime}\lambda}b_{\bm{k}\nu},
\end{align}
where $V_{\bm{q}}^{\mu\rho\lambda\nu}$ is the vertex function, defined as
\begin{align}
    \label{eq:definition-vertex-function}
    V_{\bm{q}}^{\mu\rho\lambda\nu}
    =\frac{1}{4}\frac{M}{N}\sum_{i,j}\sum_{\alpha,\beta}J_{ij}^{\alpha\beta}\sigma^{\alpha}_{\mu\nu}\sigma^{\beta}_{\rho\lambda}
    e^{-i\bm{q}\cdot\left(\bm{r}_{i}-\bm{r}_{j}\right)}.
\end{align}
Note that, since spatial translational symmetry in real space is defined with respect to the unit cell, the normalization factor is evaluated using $N/M$.
In a Feynman-diagram representation, this quantity corresponds to a four-point vertex as shown in Fig.~\ref{fig:RPA_diagram}(a). We emphasize that the two external legs entering the vertex and the two external legs leaving the vertex each carry an index, so that a single vertex is associated with four indices $\mu,\nu,\rho,\lambda$, and that the value of the vertex depends on the choice of these indices. In this work, we choose the Heisenberg interaction Hamiltonian as the perturbation Hamiltonian. For the corresponding four-point vertices with six distinct index structures, the explicit values are summarized in Appendix~\ref{app:Detailed calculation of the random-phase approximation}.

We now introduce the bare Matsubara spin susceptibility $\chi^{\alpha\alpha^{\prime}(0)}(\bm{q},i\omega_{n})$ as
\begin{align}
    \label{eq:definition_susceptibility}
    \chi^{\alpha\alpha^{\prime}(0)}(\bm{q},i\omega_{n})
    &=\frac{M}{N}\sum_{i,j}e^{-i\bm{q}\cdot\left(\bm{r}_{i}-\bm{r}_{j}\right)}
    \int_{0}^{\beta}d\tau\,e^{i\omega_{n}\tau}\notag\\
    &\qquad\times
    \left\langle T_{\tau}S_{i}^{\alpha}(\tau)S_{j}^{\alpha^{\prime}}(0)\right\rangle_{0}.
\end{align}
The subscript 0 indicates that the expectation value is evaluated with the quadratic SBMFT Hamiltonian. As discussed in Sec.~\ref{sec:Calculation of the spin structure factor}, this quantity is obtained from products of bosonic two-point Matsubara functions and therefore corresponds to the bare bubble diagrams shown in Fig.~\ref{fig:RPA_diagram}(b). In addition to the conventional particle-number-conserving bubble, anomalous bubble contributions also appear because the Schwinger boson mean-field Hamiltonian contains pairing terms. The retarded bare susceptibility used for the spectral function is obtained by analytic continuation, $\chi^{\alpha\alpha^{\prime}(0)}(\bm{q},\omega+i\delta)=\chi^{\alpha\alpha^{\prime}(0)}(\bm{q},i\omega_{n}\to\omega+i\delta)$.

We then dress these bare bubble diagrams by summing the Dyson-type geometric series to all orders, as illustrated in Fig.~\ref{fig:RPA_diagram}(c). The shaded bubble in Fig.~\ref{fig:RPA_diagram}(c) represents the RPA-corrected susceptibility $\chi_{\text{RPA}}(\bm{q},i\omega_{n})$. In constructing these diagrams, one must carefully keep track of the indices associated with each vertex. The corresponding diagrams and the detailed calculations are summarized in Appendix~\ref{app:Detailed calculation of the random-phase approximation}.
Formally, Fig.~\ref{fig:RPA_diagram}(c) leads to
\begin{align}
    \label{eq:RPA-Dyson}
    \chi_{\text{RPA}}(\bm{q},i\omega_{n})
    =\chi^{(0)}(\bm{q},i\omega_{n})
    -\chi^{(0)}(\bm{q},i\omega_{n})V_{\bm{q}}\chi_{\text{RPA}}(\bm{q},i\omega_{n}).
\end{align}
In practice, $V_{\bm{q}}$, $\chi^{(0)}(\bm{q},i\omega_{n})$, and $\chi_{\text{RPA}}(\bm{q},i\omega_{n})$ are matrices labeled by the spin-component indices $(\alpha,\alpha^{\prime})$ and the sublattice indices $(k,l)$, and Eq.~\eqref{eq:RPA-Dyson} should also be understood as a matrix equation. The explicit matrix formulation used in this work is described in Appendix~\ref{app:Detailed calculation of the random-phase approximation}.
By solving Eq.~\eqref{eq:RPA-Dyson}, we obtain
\begin{align}
    \label{eq:solution-RPA-Dyson}
    \chi_{\text{RPA}}(\bm{q},i\omega_{n})=\left(\bm{I}+\chi^{(0)}(\bm{q},i\omega_{n})V(\bm{q})\right)^{-1}\chi^{(0)}(\bm{q},i\omega_{n}),
\end{align}
where $\bm{I}$ is the identity matrix.
After solving the matrix equation in Matsubara space, we perform the analytic continuation $i\omega_{n}\to\omega+i\delta$ and obtain the retarded RPA susceptibility $\chi_{\text{RPA}}^{\alpha\alpha^{\prime}}(\bm{q},\omega+i\delta)$. Accordingly, once a particular component is evaluated, the dynamical spin structure factor corrected by the RPA, $S_{\text{RPA}}^{\alpha\alpha^{\prime}}(\bm{q},\omega)$, can be obtained via the fluctuation-dissipation theorem as
\begin{align}
    \label{eq:fluctuation-dissipation-theorem}
    S_{\text{RPA}}^{\alpha\alpha^{\prime}}(\bm{q},\omega)=\frac{1}{\pi}\frac{1}{1-e^{-\beta\omega}}\operatorname{Im}\chi_{\text{RPA}}^{\alpha\alpha^{\prime}}(\bm{q},\omega+i\delta).
\end{align}

\section{Application to kagome antiferromagnets}
\label{sec:Application to kagome antiferromagnets}

\begin{figure}[t]
  \centering
      \includegraphics[width=\columnwidth,clip]{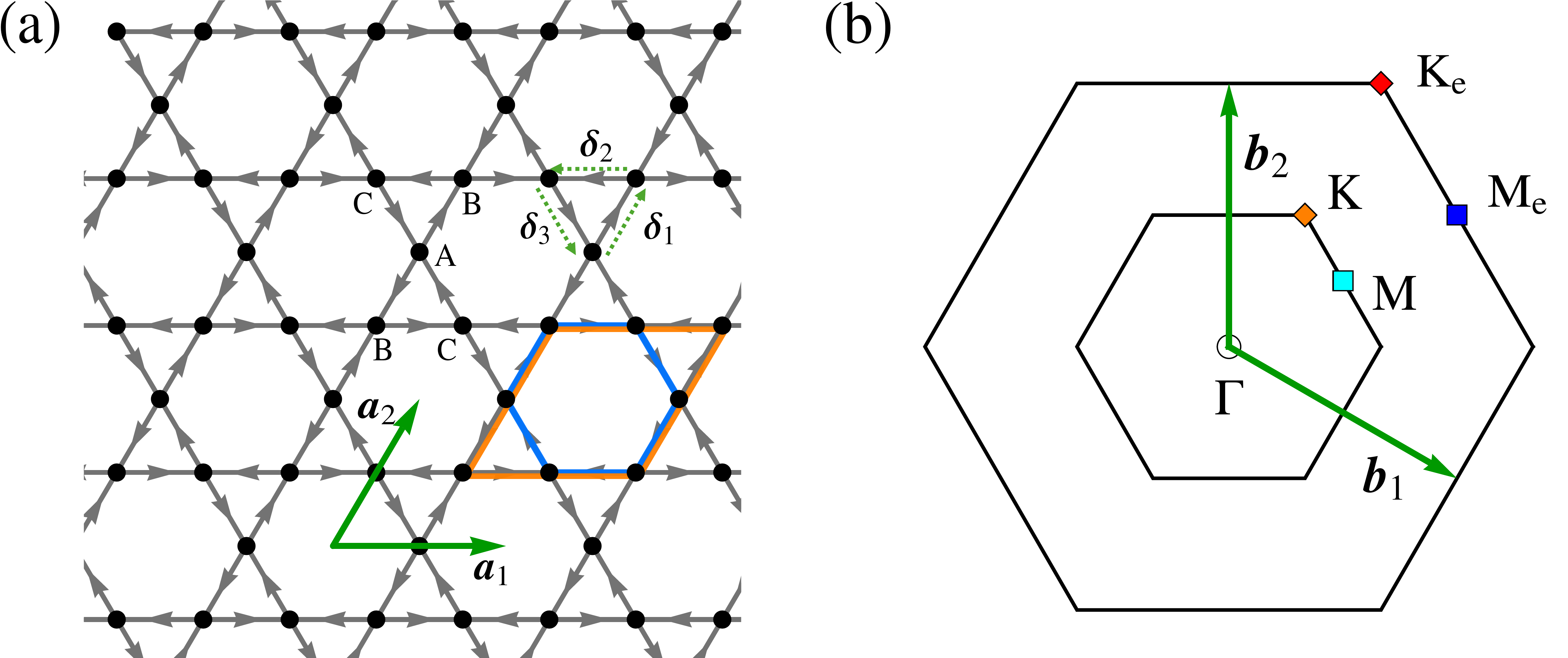}
      \caption{
        (a)
        Schematic picture of the kagome lattice on which the $S=1/2$ kagome antiferromagnetic Heisenberg model with a DM interaction is defined.
        The green arrows indicate the primitive translation vectors $\bm{a}_{1}$ and $\bm{a}_{2}$ of the unit cell, which are mainly used in the present calculations.
        The gray arrows on the bonds indicate the orientation of the DM vectors $\bm{d}_{ij}=(0,0,d_{ij}^{z})$, corresponding to the convention $d_{ij}^{z}>0$.
        The hexagon highlighted in blue and the parallelogram highlighted in orange indicate the closed loops on which the Wilson loops are defined in AFMFT to distinguish spin-liquid phases.
        (b)
        First and extended Brillouin zones of the kagome lattice.
        Filled symbols denote the high-symmetry points.
        The green arrows indicate the primitive reciprocal lattice vectors $\bm{b}_{1}$ and $\bm{b}_{2}$ corresponding to $\bm{a}_{1}$ and $\bm{a}_{2}$ in the real space.
      }
      \label{fig:kagome_lattice}
\end{figure}

We study a $S=1/2$ antiferromagnet on the two-dimensional kagome lattice with the nearest-neighbor Heisenberg exchange and the DM interaction. The Hamiltonian is
\begin{align}
\label{eq:kagome-Hamiltonian}
\mathcal{H}
=J\sum_{\langle i,j\rangle}\bm{S}_{i}\cdot\bm{S}_{j}
+\sum_{\langle i,j\rangle}\bm{d}_{ij}\cdot\left(\bm{S}_{i}\times \bm{S}_{j}\right),
\end{align}
where $\bm{S}_i$ is the $S=1/2$ operator at site $i$, $J>0$ is the nearest-neighbor antiferromagnetic exchange coupling, $\langle i,j\rangle$ denotes nearest-neighbor bonds, and energies are measured in units of $J$ throughout this paper.
The second term is the DM interaction, which originates from spin--orbit coupling and is allowed when the bond center lacks an inversion symmetry.
Such a situation naturally occurs in many materials where $\mathrm{Cu}^{2+}$ ions form a kagome lattice, making Eq.~\eqref{eq:kagome-Hamiltonian} a minimal and broadly relevant starting point for kagome quantum magnets.

In general, the DM vector $\bm{d}_{ij}$ can have both an out-of-plane component $d^z_{ij}$ (normal to the kagome plane) and in-plane components $(d^x_{ij},d^y_{ij})$.
Motivated by experiments on several kagome antiferromagnets where the out-of-plane component is dominant, we restrict ourselves to the perpendicular DM interaction,
\begin{align}
\label{eq:dm_pattern}
\bm{d}_{ij}=\left(0,0,d^{z}_{ij}\right)\equiv{d}^{z}\nu_{ij}\hat{\bm{z}},
\end{align}
with $\hat{\bm{z}}$ being the unit vector perpendicular to the kagome plane.
Here, $\nu_{ij}=\pm1$ is a bond-dependent sign satisfying $\nu_{ij}=-\nu_{ji}$, fixed by lattice symmetries; we follow the convention indicated in Fig.~\ref{fig:kagome_lattice}(a).
With this choice, changing the sign of $d^{z}$ corresponds to reversing the vector chirality of coplanar ordered states, and thus we focus on $d^{z}\ge 0$ without loss of generality.
For the present out-of-plane DM term, the spin-rotation symmetry of the Hamiltonian is lowered from $\mathrm{SU}(2)$ to $\mathrm{U}(1)$, which generally reduces quantum fluctuations and can promote long-range order.

The kagome lattice is a non-Bravais lattice built on a triangular Bravais lattice with three sublattices A, B, and C in the unit cell [see Fig.~\ref{fig:kagome_lattice}(a)].
We take the primitive translation vectors [green arrows in Fig.~\ref{fig:kagome_lattice}(a)] as
\begin{align}
\label{eq:realspace_vectors}
\bm{a}_1 = a(1,0),\qquad
\bm{a}_2 = a\left(\frac{1}{2},\frac{\sqrt{3}}{2}\right).
\end{align}
The corresponding reciprocal lattice vectors $\bm{b}_1$ and $\bm{b}_2$ [see Fig.~\ref{fig:kagome_lattice}(b)] are given by
\begin{align}
\label{eq:reciprocal_vectors}
\bm{b}_{1} = \frac{2\pi}{a}\left(1,-\frac{1}{\sqrt{3}}\right),\qquad
\bm{b}_{2} = \frac{2\pi}{a}\left(0,\frac{2}{\sqrt{3}}\right).
\end{align}
Throughout this paper we set the length of the primitive translation vectors to unity.
With this convention, the nearest-neighbor kagome bond length is $a/2=1/2$.

At $d^{z}=0$, the nearest-neighbor kagome antiferromagnetic Heisenberg model is highly frustrated and is widely discussed as a candidate for a QSL ground state.
The existence and magnitude of a spin gap remain actively debated in the literature.
Once a finite $d^{z}$ is introduced, the spin-rotation symmetry is reduced from $\mathrm{SU}(2)$ to $\mathrm{U}(1)$. Increasing $d^{z}$ then tends to reinforce antiferromagnetic correlations, eventually destabilizing the spin-liquid regime.
For the $S=1/2$ kagome antiferromagnet, previous numerical studies suggest a quantum phase transition from a magnetically disordered phase to an antiferromagnetically ordered phase at a critical value $d^{z}_{\text{c}}\simeq 0.1$~\cite{Cepas-2008,Messio-Cepas-2010}.

It is also worth emphasizing that kagome antiferromagnets form a rich materials platform, with many compounds that are close to the ideal kagome geometry and host sizable spin--orbit effects.
Representative examples include herbertsmithite $\mathrm{ZnCu}_{3}\mathrm{(OH)}_{6}\mathrm{Cl}_{2}$~\cite{
Helton-2007,
Han-2012,
Fu-2015,
Han-2016,
Norman-2016,
Khuntia-2020}, Zn-barlowite $\mathrm{ZnCu}_{3}\mathrm{(OH)}_{6}\mathrm{FBr}$~\cite{Jeschke-2015,Tustain-2020,Fu-2021,Smaha-2023}, and vesignieite $\mathrm{BaCu}_{3}\mathrm{V}_{2}\mathrm{O}_{8}\mathrm{(OH)}_{2}$~\cite{Okamoto-Yoshida-2009,Boldrin-2018}, among others~\cite{Li-2014,Orain-2017}.
In addition, the kapellasite family such as Ca-kapellasite provides further realizations of kagome $\mathrm{Cu}^{2+}$ networks~\cite{Ihara-2017,Ihara-2020}.
While real materials may involve additional ingredients such as further-neighbor exchange, weak interlayer coupling and structural disorder, the nearest-neighbor model Eq.~\eqref{eq:kagome-Hamiltonian} captures the minimal competition between strong frustration and spin-orbit-induced anisotropy, and thus serves as a useful reference point for understanding magnetic instabilities out of the kagome spin-liquid regime.
More explicitly, this minimal model captures the qualitative trend that a highly frustrated kagome antiferromagnet remains magnetically disordered for sufficiently small $d^{z}$, whereas increasing $d^{z}$ enhances antiferromagnetic correlations and pushes the system toward magnetic instability, in line with previous estimates of an instability scale around $d^{z}_{\mathrm{c}}\sim0.1$. It also provides a controlled setting for examining how spin-orbit-induced anisotropy and many-body spinon interactions reshape the low-energy neutron-scattering response. Within a given calculation, the ordering wave vector associated with an instability can be inferred from the momentum at which the low-energy response is most strongly enhanced or the gap closes. In the present RPA/SBMFT analysis, this instability occurs near the $\mathrm{M}_{\mathrm{e}}$ point, which is equivalent to the $\mathrm{M}$ point modulo a reciprocal lattice vector, and therefore points to an $\mathrm{M}$-channel magnetic tendency. However, we do not claim that this minimal Hamiltonian quantitatively determines the full material-specific phase diagram of herbertsmithite. In particular, the $\mathrm{M}_{\mathrm{e}}$ instability found here should be distinguished from the possible $q=0$ tendency discussed in the literature for herbertsmithite; the selection between such ordering channels can be affected by further-neighbor exchange, symmetric or off-diagonal anisotropies, interlayer coupling, structural disorder, and higher-order non-local exchange terms generated from microscopic Hubbard- or Anderson-type models. Such additional interactions may shift the instability scale, select a different ordering wave vector, or further stabilize the disordered regime, and their inclusion remains an important direction for future material-specific modeling.

To characterize which magnetic order is selected when the disordered phase becomes unstable, it is convenient to monitor the ordering wave vector through the spin structure factor.
A Bragg peak at the $\Gamma$ point, $\bm{Q}=\Gamma=(0,0)$, corresponds to the so-called $q=0$ coplanar $120^{\circ}$ antiferromagnetic order, in which the magnetic unit cell coincides with the crystallographic unit cell.
A Bragg peak at the $\mathrm{K}$ point, $\bm{Q}=\mathrm{K}=(\bm{b}_1+2\bm{b}_2)/3$, signals the $\sqrt{3}\times\sqrt{3}$ order, whose magnetic unit cell is enlarged by a factor of three along both primitive directions, with nine spins per magnetic unit cell.
A Bragg peak at the $\mathrm{M}$ point, $\bm{Q}=\mathrm{M}=(\bm{b}_{1}+\bm{b}_{2})/2$, indicates an ordering with doubled periodicity along one primitive direction; on the kagome lattice such $q=\mathrm{M}$ tendencies are commonly associated with stripe-type magnetic patterns, depending on the relative phases among the three sublattices.
In the extended-zone representation shown in Fig.~\ref{fig:kagome_lattice}(b), the points $\mathrm{K}_{\mathrm{e}}$ and $\mathrm{M}_{\mathrm{e}}$ denote wave vectors that are equivalent to $\mathrm{K}$ and $\mathrm{M}$ modulo reciprocal lattice vectors; Bragg peaks at $\mathrm{K}_{\mathrm{e}}$ or $\mathrm{M}_{\mathrm{e}}$ therefore correspond to the same ordering wave vectors as those at $\mathrm{K}$ or $\mathrm{M}$, and are useful when discussing the momentum dependence of scattering intensities in an extended Brillouin-zone scheme.

\section{Results}
\label{sec:Results}

In the following, we present our analysis of the kagome-lattice antiferromagnetic Heisenberg model with a DM interaction within AFMFT and SBMFT. In Sec.~\ref{sec:Results_Abrikosov fermion mean-field theory}, we present the AFMFT results in the order of mean-field ans\"atze, spinon dispersions, and the corresponding static and dynamical spin structure factors. We then present the SBMFT results in Sec.~\ref{sec:Results_Schwinger boson mean-field theory}.

All calculations presented in this section are performed at zero temperature. Unless otherwise noted, the Brillouin-zone momentum sums are evaluated on a uniform $50\times50$ mesh.
We have checked the mesh convergence of the spinon dispersions
and the spinon gap; they are almost unchanged for calculations up to a $200\times200$ mesh.
In calculating the dynamical spin structure factor, we use the broadening parameter $\delta=0.03$ for AFMFT and $\delta=0.01$ for SBMFT in Eq.~\eqref{eq:dssf-from-matsubara-susceptibility}.

In what follows, the static spin structure factor $S(\bm{q})$ and the dynamical spin structure factor $S(\bm{q},\omega)$ shown below correspond to the trace over spin components defined as
\begin{align}
    \label{eq:trace spin structure factor}
    S(\bm{q},\omega)
    &=\sum_{\gamma=x,y,z}S^{\gamma\gamma}(\bm{q},\omega),\\
    S(\bm{q})
    &=\sum_{\gamma=x,y,z}S^{\gamma\gamma}(\bm{q}).
\end{align}
Before presenting the results, we emphasize how the spinon dispersions shown below should be read together with the spin structure factors. The spinon bands provide the mean-field building blocks for fractionalized $S=1/2$ excitations, but the neutron-scattering response is a gauge-invariant spin correlation function. Therefore, the spectra in $S(\bm{q},\omega)$ should not be expected to reproduce the single-spinon dispersion one-to-one. Rather, as explained in Sec.~\ref{sec:Calculation of the spin structure factor}, they represent two-spinon continua whose lower edges, spectral weights, and possible flat or dome-like features are determined by convolutions of spinon bands together with coherence factors. Thus, when a feature in $S(\bm{q},\omega)$ overlaps with a feature in the spinon dispersion, we interpret this as the corresponding spinon band providing strong phase space or matrix elements for a gauge-invariant two-spinon excitation. Conversely, when the two do not overlap, this does not indicate an inconsistency; it reflects the fact that the observable spin response is a two-particle, gauge-invariant quantity rather than a direct image of a gauge-dependent single-spinon dispersion.
The dynamical structure factor $S(\bm{q},\omega)$ is evaluated along the high-symmetry path
$\Gamma$--$\mathrm{M}$--$\mathrm{M}_{\mathrm{e}}$--$\mathrm{K}_{\mathrm{e}}$--$\mathrm{K}$--$\Gamma$.
In the plots of the static spin structure factor $S(\bm{q})$, we also show two hexagons drawn by cyan dotted lines: the inner hexagon indicates the boundary of the first Brillouin zone, while the outer hexagon corresponds to the boundary of an extended Brillouin zone.

\subsection{Abrikosov fermion mean-field theory}
\label{sec:Results_Abrikosov fermion mean-field theory}

\subsubsection{Mean-field ansatz}
\label{sec:AF_Mean-field ansatz}

\begin{figure*}[t]
  \centering
      \includegraphics[width=0.95\textwidth,clip]{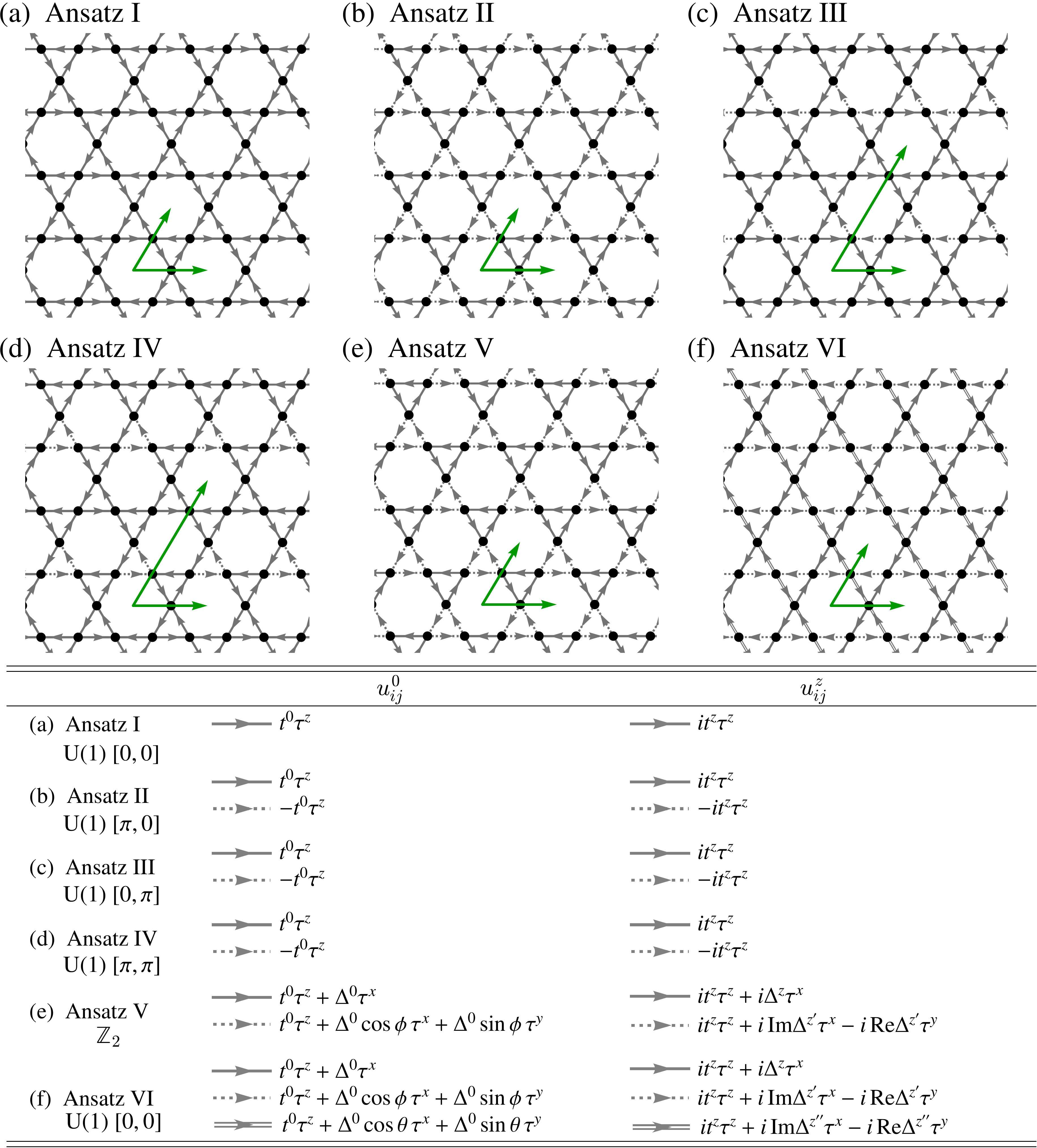}
      \caption{
      AFMFT mean-field ans\"atze studied in this work. 
      The figure specifies the pattern of the mean-field matrices $u_{ij}^{0}$ and $u_{ij}^{z}$ defined on solid, dashed, and open bonds. 
      In all ans\"atze the Lagrange multipliers $a_i^{\gamma}$ for the local constraints are taken to be uniform in site and sublattice.
      The labels below each ansatz indicate the IGG; for the $\mathrm{U}(1)$ states, $[\Phi_{\mathrm{Hex}},\Phi_{\mathrm{Para}}]$ specifies the flux sector.
      The green arrows indicate the primitive translation vectors. 
      Note that ans\"atze III and IV are analyzed in an enlarged kagome unit cell, doubled relative to the conventional primitive unit cell, and hence the unit-cell area is twice as large. 
      The arrow direction on each bond indicates the orientation from site $i$ to site $j$.
      }
      \label{fig:mean_field_ansatz_AFMFT}
\end{figure*}

\begin{table}[t]
  \centering
  \caption{Gauge-invariant fluxes on the hexagon and parallelogram loops defined in Fig.~\ref{fig:kagome_lattice}(a).}
  \label{tab:flux}
  \renewcommand{\arraystretch}{1.15}
  \setlength{\tabcolsep}{20pt}

  \begin{tabular}{@{}lcc@{}}
    \toprule
     & $\Phi_{\text{Hex}}$ & $\Phi_{\text{Para}}$ \\
    \midrule
    Ansatz I   & $0$   & $0$   \\
    Ansatz II  & $\pi$ & $0$   \\
    Ansatz III & $0$   & $\pi$ \\
    Ansatz IV  & $\pi$ & $\pi$ \\
    \bottomrule
  \end{tabular}
\end{table}

We first describe the AFMFT mean-field ans\"atze considered in this work. The six ans\"atze discussed here are shown in Fig.~\ref{fig:mean_field_ansatz_AFMFT}. In all ans\"atze considered below, the Lagrange multipliers $a_i^{\gamma}$ associated with the local constraints are assumed to be uniform, independent of site.
A detailed classification of Abrikosov fermion mean-field ans\"atze on the kagome lattice has been developed in Ref.~\cite{Lu-2011}. In particular, a previous study~\cite{Dodds-2013} discussed parent $\mathrm{U}(1)$ spin-liquid states in the nearest-neighbor sector and related $\mathbb{Z}_{2}$ states obtained by enlarging the set of allowed mean-field channels to include pairing terms and, in some cases, further-neighbor amplitudes. Since the Hamiltonian defined in Eq.~\eqref{eq:kagome-Hamiltonian} contains only nearest-neighbor interactions, we first consider nearest-neighbor hopping-type ans\"atze. Within our notation, ans\"atze I, II, and IV can be connected to parent $\mathrm{U}(1)$ states discussed previously~\cite{Dodds-2013}, whereas ansatz III is treated here as an additional nearest-neighbor ansatz.
For completeness, we present in Appendix~\ref{app:Analysis of Ansatz VII, gauge inequivalent to Ansatz III, with the same flux pattern} an analysis of the previously studied $\mathrm{U}(1)$ parent state having the same flux pattern as Ansatz III~\footnote{For clarity, we summarize here the correspondence between our mean-field states and the $\mathrm{U}(1)$ parent states introduced in Ref.~\cite{Dodds-2013}. In our notation, Ansatz I, Ansatz II, and Ansatz IV correspond to the $\mathrm{U}(1)[0,0]$, $\mathrm{U}(1)[\pi,\pi]$, and $\mathrm{U}(1)[0,\pi]$ states introduced there, respectively. Ansatz VII, discussed in Appendix~\ref{app:Analysis of Ansatz VII, gauge inequivalent to Ansatz III, with the same flux pattern}, corresponds to the $\mathrm{U}(1)[\pi,0]$ state. By contrast, Ansatz III introduced in the present work shares the same flux pattern as that $\mathrm{U}(1)[\pi,0]$ state, but is gauge inequivalent to it and was not considered in that work.}.
To explore whether a $\mathbb{Z}_{2}$ mean-field state can also arise within the nearest-neighbor model, we further introduce pairing channels in addition to hopping channels and allow extra phase factors in the pairing amplitudes, thereby defining ans\"atze V and VI.
According to the structure of the mean-field channels, the six ans\"atze can be broadly divided into two groups: ans\"atze I--IV [see Fig.~\ref{fig:mean_field_ansatz_AFMFT}(a)--(d)] and ans\"atze V and VI [see Fig.~\ref{fig:mean_field_ansatz_AFMFT}(e) and (f)].
Ans\"atze I--IV are hopping-type ans\"atze, in which only the hopping channels $\langle \chi_{ij} \rangle$ and $\langle E_{ij}^{z} \rangle$ take nonzero values. 
By contrast, ans\"atze V and VI include both hopping and pairing channels. Ansatz V introduces an additional phase factor $\phi$ in the pairing amplitudes on kagome triangles, whereas Ansatz VI is a more general phase-decorated ansatz characterized by two independent phases, $\phi$ and $\theta$.
Although both ans\"atze V and VI include pairing, only ansatz V has IGG $\mathbb{Z}_{2}$, whereas ansatz VI retains $\mathrm{U}(1)$.

First, in order to distinguish ans\"atze I-IV, which belong to the same $\mathrm{U}(1)$ spin-liquid class, we list in Table~\ref{tab:flux} the values of the fluxes through the blue hexagon and the orange parallelogram defined in Fig.~\ref{fig:kagome_lattice}(a) for each ansatz.
Here, one may in principle define the flux using either the Wilson loop $P^{0}(C)$ constructed from the matrix $u_{ij}^{0}$ or the Wilson loop $P^{z}(C)$ constructed from $u_{ij}^{z}$. We note, however, that for the purpose of distinguishing these four phases, it is sufficient to use $P^{0}(C)$ alone.
Ans\"atze~I-IV are described solely by hopping channels at the mean-field level, and their mean-field parameters can be chosen to be real by exploiting the gauge freedom. 
Namely, since $u_{ij}^{0}\propto\tau^{z}$, the Wilson loop satisfies $U^{0}(C)\propto\tau^{0}$ for both the hexagonal loop and the parallelogram loop, because each loop contains an even number of bonds. 
Consequently, the corresponding loop variable can be written as $P^{0}(C)=\pm1$. 
Here, when a loop contains an odd number of bonds for which the mean-field matrix $u_{ij}^{0}$ has the opposite sign among the even number of bonds, one obtains $P^{0}(C)=-1$, which corresponds to a $\pi$-flux state on that loop. 
In other words, ans\"atze~I-IV correspond to the four possibilities of having either zero flux or $\pi$ flux on the hexagon and on the parallelogram.
As can be seen from Table~\ref{tab:flux}, these four mean-field ans\"atze are completely classified by the gauge-invariant fluxes. 
We note that such a classification of mean-field ans\"atze has already been discussed in the context of SBMFT~\cite{Messio-Cepas-2010}. 
We note that, to realize $\Phi_{\text{Para}}=\pi$, it is necessary to carry out the mean-field construction in an enlarged kagome unit cell, i.e., a unit cell doubled relative to the minimal one [see Fig.~\ref{fig:mean_field_ansatz_AFMFT}(c),(d)].

We next consider ans\"atze V and VI. These ans\"atze are motivated by the SBMFT construction of Ref.~\cite{Messio-Bieri-2017}, in which additional phase factors are attached to the pairing channels in order to realize a time-reversal-symmetry-breaking spin-liquid state. Our purpose here is to examine whether an analogous construction can be implemented within AFMFT.
First, Ansatz V is obtained by a direct translation of that SBMFT construction into the Abrikosov fermion framework. For uniform hoppings $t^{0}$ and $t^{z}$, we introduce on the triangles formed by dashed bonds an $\mathrm{SU}(2)$-invariant pairing channel $\Delta^{0}e^{i\phi}$ in addition to the pairing channel $\Delta^{0}$ on the solid-bond triangles, and determine the phase $\phi$ self-consistently together with the other mean-field parameters. In the Abrikosov fermion formulation, however, this additional phase $\phi$ can be removed by a suitable $\mathrm{SU}(2)$ gauge transformation. Therefore, although Ansatz V is motivated by the chiral SBMFT construction, the phase $\phi$ does not by itself imply an intrinsic breaking of time-reversal symmetry. Nevertheless, Ansatz V is classified as a $\mathbb{Z}_{2}$ spin-liquid state. A more detailed discussion of this point is provided in Appendix~\ref{app:Discussion of gauge-equivalent mean-field ansatz for Ansatz V in AFMFT}.
We then consider a more general phase-decorated pairing ansatz, denoted as Ansatz VI, in which the $\mathrm{SU}(2)$-invariant pairing channels on the three bonds forming a triangle take the values $\Delta^{0}$, $\Delta^{0}e^{i\phi}$, and $\Delta^{0}e^{i\theta}$, with two independent phases $\phi$ and $\theta$ determined self-consistently. Even in this case, however, a suitable $\mathrm{SU}(2)$ gauge transformation maps the mean-field state to a $\mathrm{U}(1)$ ansatz containing only hopping channels and having vanishing hexagon and parallelogram fluxes, $\Phi_{\mathrm{Hex}}=0$ and $\Phi_{\mathrm{Para}}=0$. Hence, Ansatz VI does not represent a genuinely time-reversal-symmetry-breaking spin-liquid state, and its IGG is $\mathrm{U}(1)$.
The gauge-transformed state, however, should not be identified with Ansatz I: although the two share the same flux pattern, the former is accompanied by sublattice-dependent onsite constraint fields $a_{i}^{\gamma}$ ($\gamma=x,y,z$), reflecting the nonuniform phases originally assigned to the bond pairing channels.
This means that Ansatz VI and Ansatz I are gauge inequivalent.
These observations show that the larger $\mathrm{SU}(2)$ gauge redundancy of the Abrikosov fermion representation, compared with the $\mathrm{U}(1)$ gauge redundancy of the Schwinger boson representation, makes it much more restrictive to realize intrinsically time-reversal-symmetry-breaking spin-liquid states simply by attaching phase factors to pairing channels. At least within the class of ans\"atze considered here, the apparently time-reversal-breaking phase factors can be gauged away or reduced to nonchiral states by suitable $\mathrm{SU}(2)$ gauge transformations.

\subsubsection{Spinon dispersion}
\label{sec:AF_Spinon dispersion}

\begin{figure*}[t]
  \centering
      \includegraphics[width=2\columnwidth,clip]{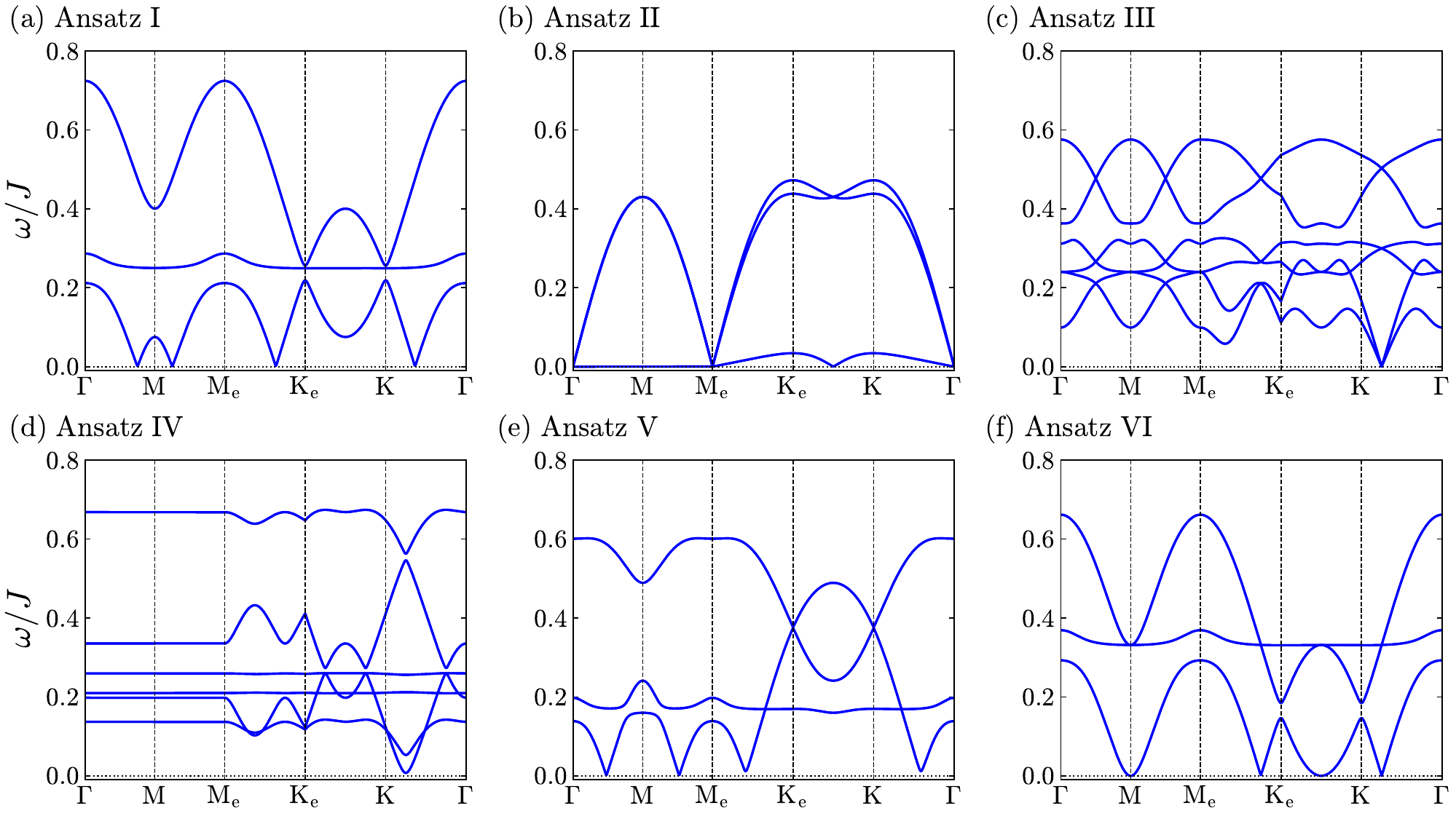}
      \caption{Spinon dispersion relations obtained from AFMFT for the ans\"atze shown in Fig.~\ref{fig:mean_field_ansatz_AFMFT}.
      We plot only the positive-energy branches $\epsilon_{\bm{k},n}\geq0$, which represent the quasiparticle excitations.
      The DM interaction is set to $d^{z}=0.10$.
      }
      \label{fig:AFMFT_dispersion}
\end{figure*}

We next discuss the spinon dispersion relations of the Abrikosov fermions obtained within AFMFT. The AFMFT results presented below were computed for a DM interaction strength $d^{z}=0.1$.
Figure~\ref{fig:AFMFT_dispersion} shows the zero-temperature Abrikosov fermion dispersions obtained from the mean-field ans\"atze I-VI.

We find that, except for Ansatz IV [Fig.~\ref{fig:AFMFT_dispersion}(d)], the Abrikosov fermion dispersion in each ansatz possesses at least one gapless point. This property is also reflected in the spin structure factors discussed in Sec.~\ref{sec:AF_Spin structure factor}.
Moreover, except for Ansatz III [Fig.~\ref{fig:AFMFT_dispersion}(c)], the spectrum still contains nearly flat bands with only weak dispersion over the entire momentum range, despite a finite DM interaction.
Since this feature appears not only in the $\mathrm{U}(1)$-IGG ans\"atze [Fig.\ref{fig:AFMFT_dispersion}(a)-(d) and (f)], but also in the $\mathbb{Z}_{2}$-IGG ansatz [Fig.~\ref{fig:AFMFT_dispersion}(e)], it is likely rooted in the geometry of the kagome lattice. In analogy to the flat band that arises in the tight-binding model of electrons on the kagome lattice, we thus expect such a weakly dispersing band structure to be a ubiquitous feature in AFMFT as well. Furthermore, these relatively flat bands leave clear signatures in the spin structure factors discussed in Sec.~\ref{sec:AF_Spin structure factor}.
The presence of such flat bands in the spinon spectrum implies the locality of the spinon excitations. This weakly dispersing band structure in momentum space stands in contrast to the Schwinger boson dispersion relations discussed in Sec.~\ref{sec:SB_Spinon dispersion}.

In the following, we describe the characteristic features of the spinon dispersion for each ansatz. 
For clarity, and to highlight how the spinon band dispersion is affected by the DM interaction, we show the AFMFT band dispersions in the absence of the DM interaction in Appendix~\ref{app:Spinon dispersion obtained by AFMFT without DM interaction} for comparison.

\paragraph*{Ansatz I}

The spinon dispersion obtained for this ansatz spans an energy range of $0\leq\omega\lesssim0.75$.
Focusing on $\omega=0$, one finds gapless points in each of the segments between the $\Gamma$ and $\mathrm{M}$ points, between the $\mathrm{M}$ and $\mathrm{M}_{\mathrm{e}}$ points, between the $\mathrm{M}_{\mathrm{e}}$ and $\mathrm{K}_{\mathrm{e}}$ points, and between the $\mathrm{K}$ and $\Gamma$ points.
Around these points, the dispersion is linear, resembling Dirac-like cones.
We also find the emergence of a nearly flat band around $\omega\simeq0.28$ over the entire momentum range.
This band becomes perfectly flat in the pure Heisenberg limit $d^{z}=0$ [see Fig.~\ref{fig:AFMFT_dispersion_without_DM}(a)].
This indicates that the very weak dispersion of this band originates from the DM interaction.
Focusing on the $\mathrm{K}$ and $\mathrm{K}_{\mathrm{e}}$ points, we further observe an energy gap separating the lowest doubly degenerate band from the next two doubly degenerate bands.
This gap is also induced by the DM interaction: for $d^{z}=0$, these three doubly degenerate bands become degenerate, resulting in an overall sixfold degeneracy [see Fig.~\ref{fig:AFMFT_dispersion_without_DM}(a)].

\paragraph*{Ansatz II}

The spinon dispersion obtained for this ansatz spans an energy range of $0\leq\omega\lesssim0.45$, and among the six ans\"atze considered it corresponds to the narrowest spinon bandwidth.
Focusing on $\omega=0$, Dirac-like dispersions are observed at the $\Gamma$ point, at the $\mathrm{M}_{\mathrm{e}}$ point, and along the segment between the $\mathrm{K}_{\mathrm{e}}$ and $\mathrm{K}$ points.
In addition, a flat band at $\omega=0$ appears along the path from $\Gamma$ to $\mathrm{M}_{\mathrm{e}}$ [see Fig.~\ref{fig:AFMFT_dispersion}(b)].
This flat band acquires a gentle dispersion along the $\mathrm{M}_{\mathrm{e}}$-$\Gamma$ segment, which is an effect of the DM interaction.
Indeed, in the pure Heisenberg limit $d^{z}=0$, a perfectly flat band is realized along this high-symmetry path [see Fig.~\ref{fig:AFMFT_dispersion_without_DM}(b)].

\paragraph*{Ansatz III}

The spinon dispersion obtained for this ansatz exhibits six doubly degenerate branches, because the number of sublattices in the unit cell is enlarged from three to six [see Fig.~\ref{fig:mean_field_ansatz_AFMFT}(c)].
The bands span an energy range of approximately $0\leq\omega\lesssim0.6$.
A Dirac-like dispersion is observed along the segment between the $\mathrm{K}$ and $\Gamma$ points.
The corresponding gapless point is fourfold degenerate, and we find that these gapless Dirac points remain robust even after introducing a finite DM interaction with $d^{z}=0.1$.
As mentioned above, this ansatz does not exhibit a nearly flat (weakly dispersive) band extending over the entire momentum range, which is present in other ans\"atze.
Indeed, such a band is absent already in the analysis at $d^{z}=0$.
In the pure Heisenberg limit, however, a flat band appears around $\omega\sim0.23$ along the $\Gamma$--$\mathrm{M}_{\mathrm{e}}$ segment, indicating that this flat structure is destroyed by the DM interaction.

\paragraph*{Ansatz IV}

The spinon dispersion obtained for this ansatz exhibits six doubly degenerate branches, as in Ansatz III, because the number of sublattices in the unit cell is enlarged from three to six [see Fig.~\ref{fig:mean_field_ansatz_AFMFT}(d)].
The bands span an energy range of approximately $0\lesssim\omega\lesssim0.7$.
Although the gap nearly closes at a point along the $\mathrm{K}$--$\Gamma$ line, it remains finite there, with only a tiny residual gap. Among the six ans\"atze considered, this is the only one whose spinon spectrum has no gapless points.
The slightly gapped dispersion observed along the $\mathrm{K}$--$\Gamma$ segment originates from a fourfold-degenerate Dirac-like structure in the pure Heisenberg limit, as in Ansatz III [see Fig.~\ref{fig:AFMFT_dispersion_without_DM}(d)].
With a finite DM interaction, this Dirac point becomes gapped, and the four gapped branches further split into two doubly degenerate dispersions.

\paragraph*{Ansatz V}

The spinon dispersion obtained for this ansatz spans an energy range of $0\leq\omega\lesssim0.6$.
Dirac-like gapless dispersions are found along the $\Gamma$--$\mathrm{M}$ and $\mathrm{M}$--$\mathrm{M}_{\mathrm{e}}$ segments.
We note that the gapless points that appear for $d^{z}=0$ along the $\mathrm{M}_{\mathrm{e}}$--$\mathrm{K}_{\mathrm{e}}$ and $\mathrm{K}$--$\Gamma$ segments acquire a small gap due to the DM interaction.
In addition, a nearly flat (weakly dispersive) band structure is visible around $\omega\sim0.2$, which is confirmed to be perfectly flat in the $d^{z}=0$ limit [see Fig.~\ref{fig:AFMFT_dispersion_without_DM}(e)].

\paragraph*{Ansatz VI}

The spinon dispersion obtained for this ansatz extends over approximately $0\leq\omega\lesssim0.65$ and exhibits gapless points at the $\mathrm{M}$ point, along the $\mathrm{M}_{\mathrm{e}}$--$\mathrm{K}_{\mathrm{e}}$ segment, along the $\mathrm{K}_{\mathrm{e}}$--$\mathrm{K}$ segment, and along the $\mathrm{K}$--$\Gamma$ segment.
Among them, the gapless points at the $\mathrm{M}$ point and along the $\mathrm{K}_{\mathrm{e}}$--$\mathrm{K}$ segment have a parabola-like shape, whereas those along the $\mathrm{M}_{\mathrm{e}}$--$\mathrm{K}_{\mathrm{e}}$ and $\mathrm{K}$--$\Gamma$ segments exhibit Dirac-like dispersions.
The overall spinon dispersion of Ansatz VI closely resembles that of Ansatz I, consistent with the fact that both realize $\mathrm{U}(1)$ spin-liquid ans\"atze with the same flux pattern.
As in Ansatz I, we find an almost flat band with very weak dispersion over the entire momentum range; however, it is located at $\omega\simeq0.38$, i.e., about $0.10$ higher in energy than in Ansatz I.
This nearly flat band also becomes perfectly flat in the pure Heisenberg limit $d^{z}=0$ [see Fig.~\ref{fig:AFMFT_dispersion_without_DM}(f)], indicating that its residual weak dispersion originates from the DM interaction.

\subsubsection{Spin structure factor}
\label{sec:AF_Spin structure factor}

\begin{figure*}[t]
  \centering
      \includegraphics[width=2\columnwidth,clip]{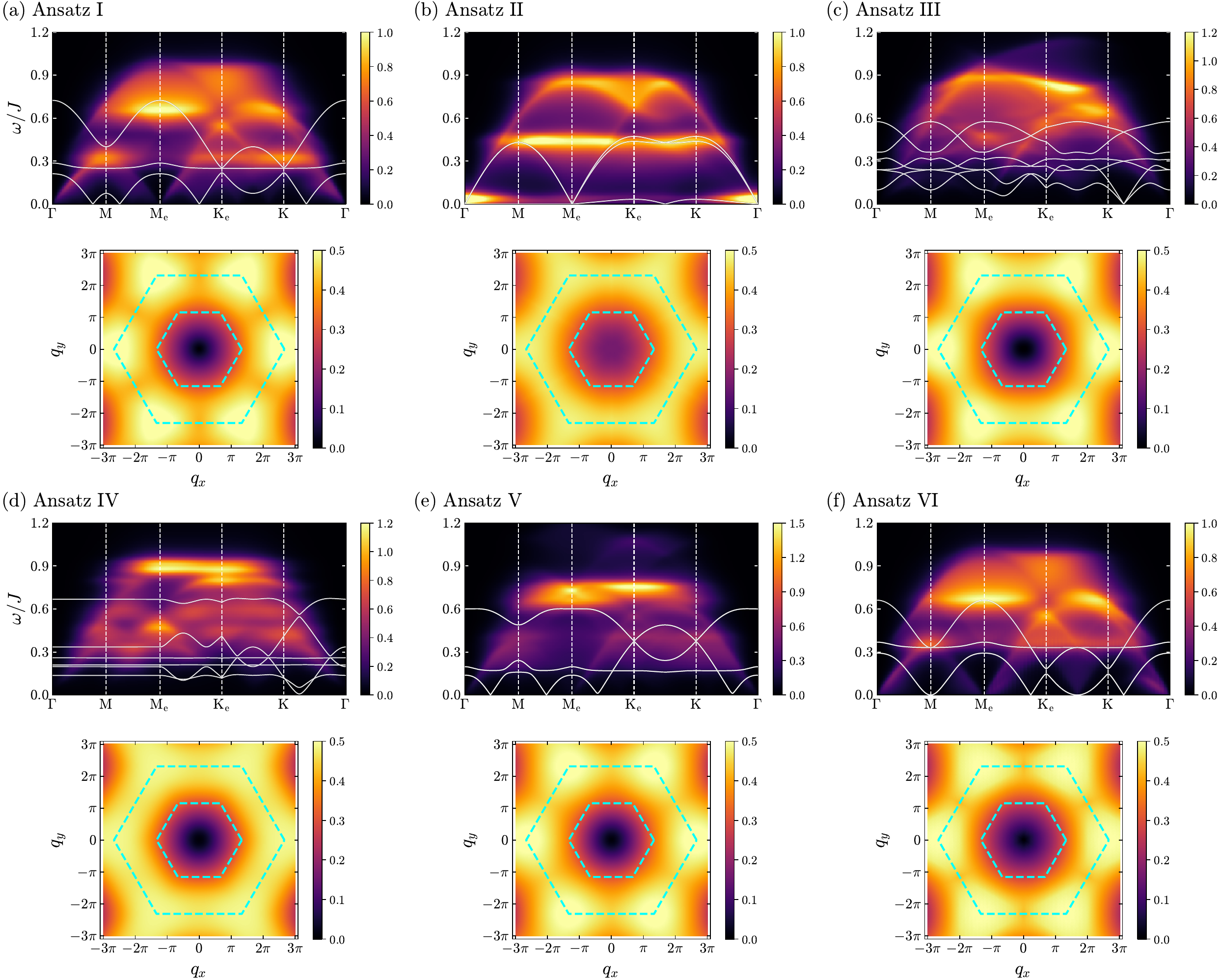}
      \caption{
      Dynamical (top) and static (bottom) spin structure factors obtained within AFMFT for each mean-field ansatz shown in Fig.~\ref{fig:mean_field_ansatz_AFMFT}.
      The dynamical spin structure factor $S(\bm{q},\omega)$ is evaluated along the high-symmetry path $\Gamma-\mathrm{M}-\mathrm{M}_{\mathrm{e}}-\mathrm{K}_{\mathrm{e}}-\mathrm{K}-\Gamma$. 
      The white curves overlaid on the dynamical panels show the corresponding single-spinon dispersions as guides to the eye. They should not be interpreted as the direct spectrum of $S(\bm{q},\omega)$ itself, because the spin structure factor is a spin-spin correlation function constructed from two-spinon excitations.
      In the plots of the static spin structure factor $S(\bm{q})$, the inner and outer cyan dashed hexagons indicate the first and extended Brillouin zones, respectively, corresponding to those shown in Fig.~\ref{fig:kagome_lattice}(b).
      }
      \label{fig:AFMFT_DSSF_and_SSF}
\end{figure*}

We next describe the static and dynamical spin structure factors calculated within AFMFT for each mean-field ansatz.
The results are shown in Fig.~\ref{fig:AFMFT_DSSF_and_SSF}.

Before discussing each ansatz in detail, we first focus on the characteristic features of the static spin structure factor $S(\bm{q})$.
Except for Ansatz II and Ansatz IV, the calculated static structure factor $S(\bm{q})$ exhibits a strong intensity at the $\mathrm{K}_{\mathrm{e}}$ point in momentum space.
This indicates an enhancement of spin correlations associated with the $\sqrt{3}\times\sqrt{3}$ ordering pattern.
We thus find that, in contrast to the static structure factor obtained within SBMFT discussed in Sec.~\ref{sec:SB_Spin structure factor}, AFMFT tends to enhance spin correlations with a different magnetic pattern.
We note, however, that gapless points in a fermionic spinon dispersion do not by themselves imply sharp Bragg peaks in the spin structure factor. Within AFMFT, the Abrikosov fermions do not condense, and the spin structure factor is evaluated as a two-spinon response rather than as a coherent ordered moment.
Below we discuss the characteristic features of the spin structure factors computed for each ansatz.

\paragraph*{Ansatz I}

The dynamical structure factor $S(\bm{q},\omega)$ obtained for this ansatz exhibits spectral weight distributed over the range $0\leq\omega\lesssim1$.
A characteristic feature is a dispersionless, flat high-intensity structure centered around the $\mathrm{M}_{\mathrm{e}}$ point at $\omega\sim0.65$ along the $\mathrm{M}$-$\mathrm{M}_{\mathrm{e}}$-$\mathrm{K}_{\mathrm{e}}$ path.
In the low-energy region the intensity is strongly suppressed; however, a faint weight remains at $\omega=0$ near the $\Gamma$ and $\mathrm{M}_{\mathrm{e}}$ points, reflecting the presence of gapless points in the spinon spectrum.
The static structure factor $S(\bm{q})$ shows enhanced intensity around the $\mathrm{K}_{\mathrm{e}}$ point, but it does not form a sharp Bragg peak.
Instead, the weight is broadly distributed outside the first BZ.

\paragraph*{Ansatz II}

The dynamical structure factor $S(\bm{q},\omega)$ obtained for this ansatz exhibits spectral weight over $0\leq\omega\lesssim0.9$.
A characteristic feature is a flat intensity structure extending across the $\mathrm{M}$-$\mathrm{K}$ segment around $\omega\sim0.45$.
Among the six ans\"atze, only this ansatz exhibits a non-negligible intensity at the $\Gamma$ point in the static structure factor $S(\bm{q})$, whereas the corresponding intensity is nearly zero for the others.
Since the static structure factor is obtained by integrating the dynamical structure factor over $\omega$, this feature can be attributed to the intensity distribution in the low-energy window $0\lesssim\omega\lesssim0.05$ along the $\Gamma$-$\mathrm{M}$ and $\mathrm{K}$-$\Gamma$ paths in the dynamical structure factor $S(\bm{q},\omega)$.
Indeed, for the dynamical structure factors computed from the other mean-field ans\"atze, one finds essentially no intensity at the $\Gamma$ point for any $\omega$.
This low-energy feature originates from an almost flat spinon dispersion lying at $\omega=0$ over a wide region of momentum space~[see Fig.~\ref{fig:AFMFT_dispersion}(b)].
It is useful to distinguish this relation between the flat spinon band and the low-energy spin response from the pinch-point singularities familiar from classical kagome or pyrochlore antiferromagnets~\cite{Henley-2010,Isakov-Moessner-Sondhi-2004,Yan-Pohle-Shannon-2018,Kiese-Ferrari-2023}. Classical pinch points originate from an emergent local constraint, often expressed as a divergence-free condition, and appear as singular structures in gauge-invariant static spin correlations~\cite{Henley-2010,Isakov-Moessner-Sondhi-2004}. In contrast, the flat band discussed here is a single-spinon mean-field band in a gauge-fixed Abrikosov-fermion description. It can enhance the low-energy two-spinon density of states and thereby affect $S(\bm{q},\omega)$ and $S(\bm{q})$, but it does not by itself imply a classical pinch-point singularity.
Furthermore, the static structure factor exhibits a rather featureless distribution of intensity outside the first Brillouin-zone boundary, and the intensities at the high-symmetry points $\mathrm{K}_{\mathrm{e}}$ and $\mathrm{M}_{\mathrm{e}}$ are comparable.

\paragraph*{Ansatz III}

The dynamical structure factor $S(\bm{q},\omega)$ obtained for this ansatz exhibits spectral weight distributed over the range $0\leq\omega\lesssim1$.
Unlike several other ans\"atze, this $S(\bm{q},\omega)$ does not display characteristic dispersionless spectral-weight features localized at specific momenta.
This can be naturally understood as a reflection of the fact that the spinon dispersion entering the calculation itself does not possess flat structures, as discussed in Sec.~\ref{sec:AF_Spinon dispersion}.
Although the spinon dispersion has Dirac points, the intensity at $\omega=0$ in the dynamical structure factor is almost completely suppressed, giving the appearance of a finite spin gap.
As for the static structure factor, the intensity is strong at the $\mathrm{K}_{\mathrm{e}}$ point and the overall appearance is very similar to that of Ansatz I.
Interestingly, however, the resulting dynamical structure factors are substantially different between the two.

\paragraph*{Ansatz IV}

The dynamical structure factor $S(\bm{q},\omega)$ obtained for this ansatz exhibits spectral weight distributed over the range $0\leq\omega\lesssim0.9$.
A characteristic feature is a dispersionless, flat high-intensity structure around $\omega\sim0.9$ extending across the region between the $\mathrm{M}_{\mathrm{e}}$ and $\mathrm{K}_{\mathrm{e}}$ points.
Another notable feature, among the six ans\"atze considered, is the broad continuum extending continuously from relatively low energies around $\omega\sim0.3$ up to $\omega\sim0.9$.
Moreover, consistent with the fact that the spinon dispersion has no gapless points, the dynamical structure factor $S(\bm{q},\omega)$ also exhibits a finite spin gap.
Furthermore, the static structure factor $S(\bm{q})$ shows a broadly spread intensity distribution outside the boundary of the first BZ, similar to Ansatz II, and the intensities at the high-symmetry points $\mathrm{K}_{\mathrm{e}}$ and $\mathrm{M}_{\mathrm{e}}$ are comparable.

\paragraph*{Ansatz V}

The dynamical structure factor $S(\bm{q},\omega)$ obtained for this ansatz exhibits spectral weight distributed over the range $0\leq\omega\lesssim0.8$.
We find a strong, dispersionless intensity feature around $\omega\sim0.8$ that extends along the path from the $\mathrm{M}$ point to the $\mathrm{K}$ point.
In addition, a small but finite intensity remains at $\omega=0$ near the $\Gamma$ and $\mathrm{M}_{\mathrm{e}}$ points, which is expected to reflect the gapless points in the spinon spectrum.
As for the static structure factor $S(\bm{q})$, the intensity is strong at the $\mathrm{K}_{\mathrm{e}}$ point, and the overall features of $S(\bm{q})$ are similar to those of Ansatz I and Ansatz VI.
In contrast, the dynamical structure factor $S(\bm{q},\omega)$ is found to be substantially different from those of Ansatz I and Ansatz VI.
This behavior distinguishes Ansatz V from the $\mathrm{U}(1)$ ans\"atze and suggests that the difference in IGG is reflected in the dynamical structure factor $S(\bm{q},\omega)$.
In particular, the dome-like structure seen in Ansatz I and Ansatz VI is absent in Ansatz V.
Instead, as noted above, a nearly flat high-energy spectral feature appears around $\omega\sim0.8$.
Such a flat high-energy structure is uncommon among the $\mathrm{U}(1)$ ans\"atze: except for Ansatz IV, the dominant high-intensity features of the other $\mathrm{U}(1)$ ans\"atze form dome-like structures rather than flat ones.
Furthermore, even in comparison with Ansatz IV, which also exhibits a flat high-energy feature, Ansatz V is distinguished by the absence of a continuum extending from the low-energy sector.
Thus, Ansatz V is characterized by a dynamical structure factor that exhibits a flat high-energy spectral feature without a continuous low-energy continuum.

\paragraph*{Ansatz VI}

The dynamical structure factor $S(\bm{q},\omega)$ obtained for this ansatz exhibits spectral weight distributed over the range $0\leq\omega\lesssim1$.
A characteristic feature is a dispersionless, flat high-intensity structure centered around the $\mathrm{M}_{\mathrm{e}}$ point at $\omega\sim0.65$ along the $\mathrm{M}$--$\mathrm{M}_{\mathrm{e}}$--$\mathrm{K}_{\mathrm{e}}$ path.
In the low-energy region the intensity is strongly suppressed; however, a faint weight remains at $\omega=0$ near the $\Gamma$ and $\mathrm{M}_{\mathrm{e}}$ points, reflecting the presence of gapless points in the spinon spectrum.
The static structure factor shows enhanced intensity around the $\mathrm{K}_{\mathrm{e}}$ point, but it does not form a sharp Bragg peak; instead, the weight is broadly distributed outside the first BZ.
These features closely resemble those of Ansatz I, which can be understood from the fact that both are associated with $\mathrm{U}(1)$ spin-liquid states having the same flux pattern.
In particular, the static structure factor $S(\bm{q})$ shows almost no visible difference between the two ans\"atze.
By contrast, fine differences remain visible in the dynamical structure factor $S(\bm{q},\omega)$.
For example, both ans\"atze exhibit a locally strong intensity at the $\mathrm{K}_{\mathrm{e}}$ point around $\omega\sim0.55$, but the intensity distribution is broader for Ansatz VI.
In addition, the spectral weight in the region extending from $\omega\sim0.65$ at the $\mathrm{K}_{\mathrm{e}}$ point to $\omega\sim0.60$ at the $\mathrm{K}$ point is more pronounced for Ansatz VI.
Thus, while the two ans\"atze are globally very similar, careful inspection of $S(\bm{q},\omega)$ reveals quantitative differences.

\subsection{Schwinger boson mean-field theory}
\label{sec:Results_Schwinger boson mean-field theory}

\subsubsection{Mean-field ansatz}
\label{sec:SB_Mean-field ansatz}
\begin{figure}[t]
  \centering
      \includegraphics[width=\columnwidth,clip]{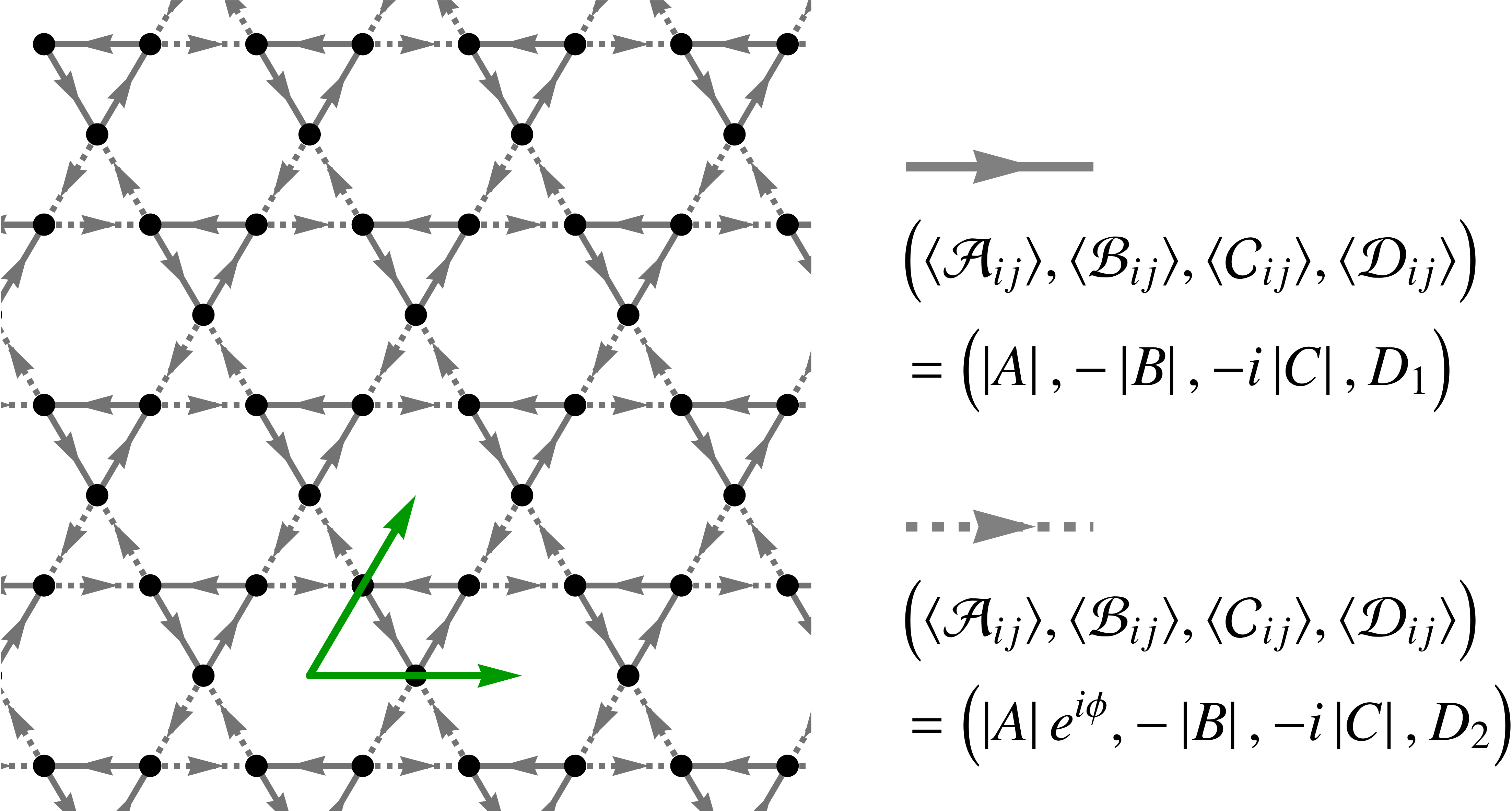}
      \caption{
      Mean-field ansatz used in the SBMFT. Solid and dashed bonds represent inequivalent links with different mean-field parameters $\left(\langle\mathcal{A}_{ij}\rangle,\langle\mathcal{B}_{ij}\rangle,\langle \mathcal{C}_{ij}\rangle,\langle\mathcal{D}_{ij}\rangle\right)$. In particular, the pairing amplitude $\langle\mathcal{A}_{ij}\rangle$ on dashed bonds carries an extra phase factor $e^{i\phi}$ relative to that on solid bonds, while $\langle\mathcal{D}_{ij}\rangle$ takes values $D_{1}$ and $D_{2}$.
      Such bond-dependent phases in the pairing channels break time-reversal symmetry.
      Arrows indicate the bond orientation from $i$ to $j$ used to define the mean fields.
      }
      \label{fig:mean_field_ansatz_SBMFT}
\end{figure}

We first describe the mean-field ansatz adopted in our SBMFT analysis.
In the study of kagome quantum spin liquids, the SBMFT has a long history, and a variety of mean-field ans\"atze have been proposed to date~\cite{Sachdev-1992,Li-Su-Shen-2007,Messio-Cepas-2010,Huh-Punk-Sachdev-2011,Fak-2012,Messio-Bernu-2012,Messio-Lhuillier-2013,Punk-2014,Halimeh-Punk-2016,Mondal-2017,Messio-Bieri-2017,Halimeh-Singh-2019,Mondal-2019,Mondal-2021,Lugan-Jaubert-2022,Rossi-Motruk-2023}.
Among these prior works, two ans\"atze stand out as being particularly successful in accounting for the neutron-scattering results~\cite{Han-2012}.
One is a theoretical study in which the coupling between Schwinger bosons and visons is incorporated perturbatively within the SBMFT framework~\cite{Punk-2014}.
The mean-field saddle point employed there is the prototypical $\pi$-flux ansatz, in which each hexagon of the kagome lattice carries a $\pi$ flux; its basic properties had already been examined in the early literature on kagome quantum spin liquids~\cite{Sachdev-1992}.
A notable limitation of this ansatz, however, is that at $\kappa=1.0$, which corresponds to an $S=1/2$ spin in the SBMFT parameterization, the Schwinger bosons are condensed, so that a quantum spin-liquid state cannot be realized at the mean-field level.
Indeed, Ref.~\cite{Punk-2014} does not determine the saddle point from a fully self-consistent mean-field solution.
The other is a theoretical proposal of a chiral spin-liquid state, where a finite chirality is induced by a bond-dependent DM interaction on the kagome lattice~\cite{Messio-Bieri-2017}.
This ansatz is characterized by (i) its ability to reproduce the dynamical structure factor on an energy scale consistent with the neutron-scattering experiment~\cite{Han-2012} and (ii) the existence of a mean-field solution at $\kappa=1.0$, which corresponds to an $S=1/2$ spin in the SBMFT parameterization, without Schwinger boson condensation.
Accordingly, our choice is motivated by previous work on the kagome Heisenberg antiferromagnet with a DM interaction, as suggested for herbertsmithite, where a time-reversal-symmetry-breaking $\mathbb{Z}_2$ spin-liquid saddle point was identified~\cite{Messio-Bieri-2017}.
They pointed out that the minimum-energy spinons can form a closed curve in the Brillouin zone, which makes the lower edge of the two-spinon continuum nearly flat over an extended region.
As a consequence, the low-energy dynamical structure factor $S(\bm{q},\omega)$ exhibits spectral weight broadly distributed in momentum space, providing a natural route to the weakly dispersing, momentum-broadened low-energy signal reported by inelastic neutron scattering on herbertsmithite.
Motivated by these observations, we adopt the corresponding mean-field ansatz as our starting point.

Figure~\ref{fig:mean_field_ansatz_SBMFT} summarizes the bond pattern of the ansatz used in this work.
Solid and dashed links denote two inequivalent sets of nearest-neighbor bonds carrying different mean-field parameters, and the arrows specify the bond orientation $i\!\to\! j$ employed in defining the directed mean fields.
Time-reversal symmetry is broken by allowing bond-dependent complex phases in the pairing channels; for example, the pairing amplitude on dashed bonds may acquire an additional phase factor $e^{i\phi}$ relative to that on solid bonds, leading to nontrivial gauge-invariant fluxes around lattice loops.
Finally, we comment on our treatment of the DM interaction.
Unlike Ref.~\cite{Messio-Bieri-2017}, where the DM interaction is incorporated via a link-dependent spin-rotation formulation that effectively introduces bond-dependent phases into the standard SBMFT bond operators, here we follow Sec.~\ref{sec:Method_Schwinger boson mean-field theory} and include the DM interaction explicitly through the SU(2)-breaking operators $\mathcal{C}^{z}_{ij}$ and $\mathcal{D}^{z}_{ij}$.
Since we consider a DM vector with only a finite out-of-plane component $d^{z}$, only the $z$ components of these operators appear; in the following, we drop the superscript and denote $\mathcal{C}^{z}_{ij}$ and $\mathcal{D}^{z}_{ij}$ simply as $\mathcal{C}_{ij}$ and $\mathcal{D}_{ij}$, respectively.

\subsubsection{Spinon dispersion}
\label{sec:SB_Spinon dispersion}

\begin{figure}[t]
  \centering
      \includegraphics[width=\columnwidth,clip]{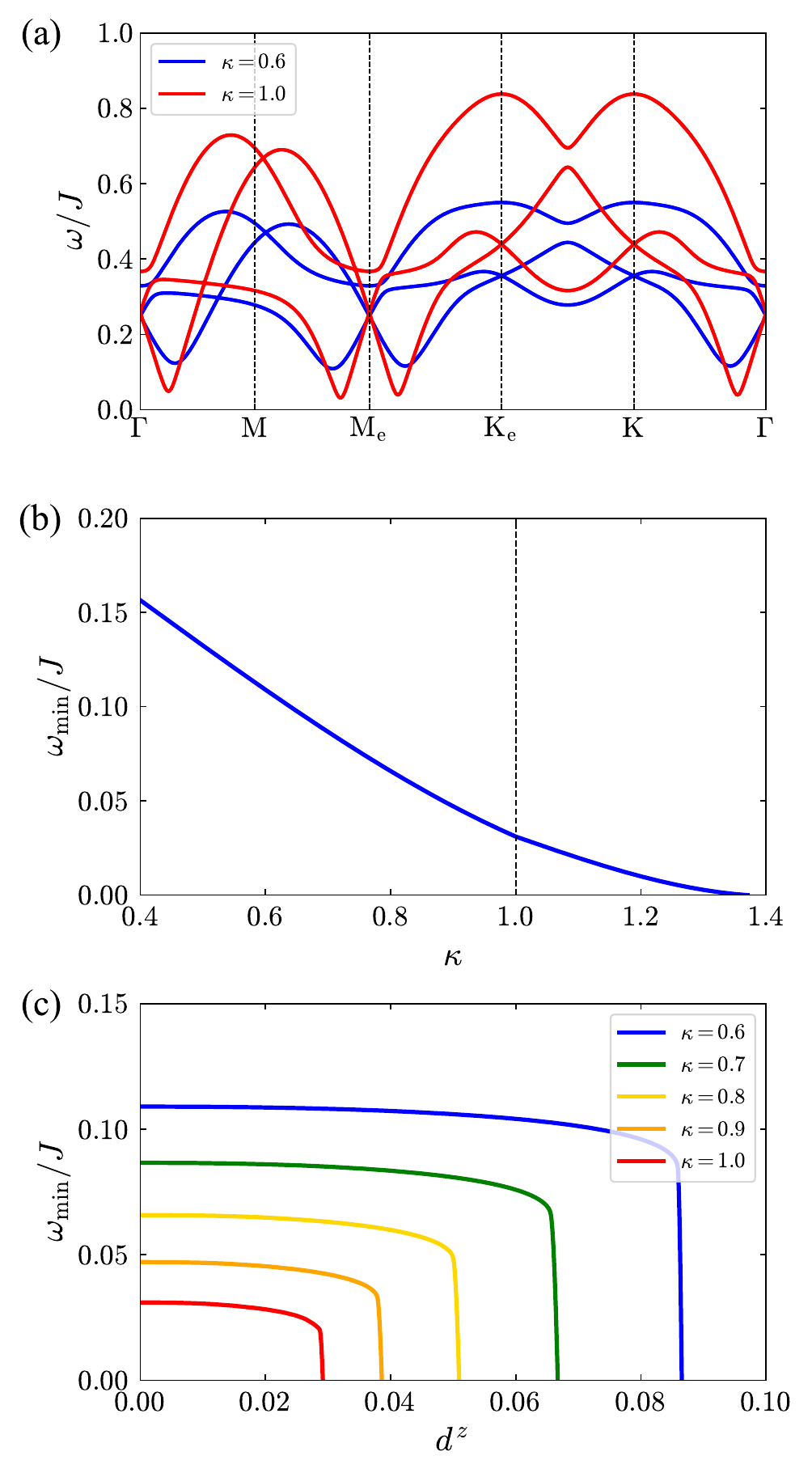}
      \caption{
      (a) Spinon dispersion obtained within the SBMFT, compared for $\kappa=0.6$ and $\kappa=1.0$. (b) $\kappa$ dependence of the spinon gap $\omega_{\text{min}}$. At $\kappa=1.0$, corresponding to an $S=1/2$ spin within the SBMFT parameterization, the spinons remain gapped, indicating that the system stays in a quantum spin-liquid regime. (c) Dependence of the spinon gap $\omega_{\text{min}}$ on the out-of-plane DM interaction $d^{z}$, shown for $\kappa=0.6-1.0$
      }
      \label{fig:SBMFT_dispersion}
\end{figure}

We next examine the spinon dispersion obtained within SBMFT.
As introduced in Sec.~\ref{sec:Method_Schwinger boson mean-field theory}, in SBMFT the parameter $\kappa = 2S$ controls the strength of quantum fluctuations, with smaller $\kappa$ corresponding to stronger quantum effects.
Equivalently, increasing $\kappa$ corresponds to moving toward the large-spin, more classical limit, in which quantum fluctuations are weaker and magnetic ordering is favored.
Thus, when we discuss $\kappa=1.0$, we mean that the average boson-number constraint gives the spin length $S=1/2$ within the SBMFT parameterization. Since the boson-number constraint is enforced only on average at the mean-field level, this should not be interpreted as an exact projection onto the spin Hilbert space or as an exact restoration of the spin sum rule.
Figure~\ref{fig:SBMFT_dispersion}(a) shows the spinon dispersion for the ansatz adopted in this work, comparing $\kappa=0.6$ with $\kappa=1.0$, which corresponds to an $S=1/2$ spin in this parameterization.
From this comparison, we find that increasing $\kappa$ broadens the bandwidth without qualitatively changing the overall shape of the dispersion.
Indeed, while the bands extend over $0.12\lesssim\omega\lesssim0.55$ for $\kappa=0.6$, they broaden to $0.03\lesssim\omega\lesssim0.85$ for $\kappa=1.0$.
This behavior can be understood from the fact that the mean-field parameters determining the bandwidth of the spinon dispersion increase with $\kappa$.

In SBMFT, the transition between a quantum spin-liquid phase and a magnetically ordered phase can be naturally described in terms of Bose condensation of spinons upon closure of the spinon gap.
This tendency can be understood as follows: as $\kappa$ increases and quantum fluctuations are reduced, the spinon bands broaden and eventually condense, driving the system into a magnetically ordered phase.
Consistent with this picture, Fig.~\ref{fig:SBMFT_dispersion}(a) shows that increasing $\kappa$ lowers several local minima of the spinon dispersion along the $q$ path toward zero energy.
The corresponding $\kappa$ dependence of the minimum gap $\omega_{\min}$ is shown in Fig.~\ref{fig:SBMFT_dispersion}(b).
From this, we find that the quantum spin-liquid phase can persist up to $\kappa\sim1.37$.
This result is consistent with the phase diagram obtained in a previous study~\cite{Messio-Bieri-2017}.
Furthermore, to examine the magnetic instability induced by the DM interaction for $\kappa=0.6,0.7,0.8,0.9,$ and $1.0$, we plot in Fig.~\ref{fig:SBMFT_dispersion}(c) the $d^{z}$ dependence of $\omega_{\min}$ for each $\kappa$.
From these results, we find that increasing the DM interaction destabilizes the spin-liquid phase and eventually drives the system into a magnetically ordered phase as the spinon gap closes.
The $d^{z}$ dependence of $\omega_{\min}$ is qualitatively similar for all values of $\kappa$: $\omega_{\min}$ decreases gradually over a wide range of $d^{z}$ and then drops rapidly as $d^{z}$ approaches its critical value, while remaining continuous throughout.
Since the gap at $d^{z}=0$ is smaller for larger $\kappa$, the system reaches the quantum critical point at a smaller value of $d^{z}$ as $\kappa$ increases.

\subsubsection{Spin structure factor}
\label{sec:SB_Spin structure factor}

\begin{figure*}[t]
  \centering
      \includegraphics[width=2\columnwidth,clip]{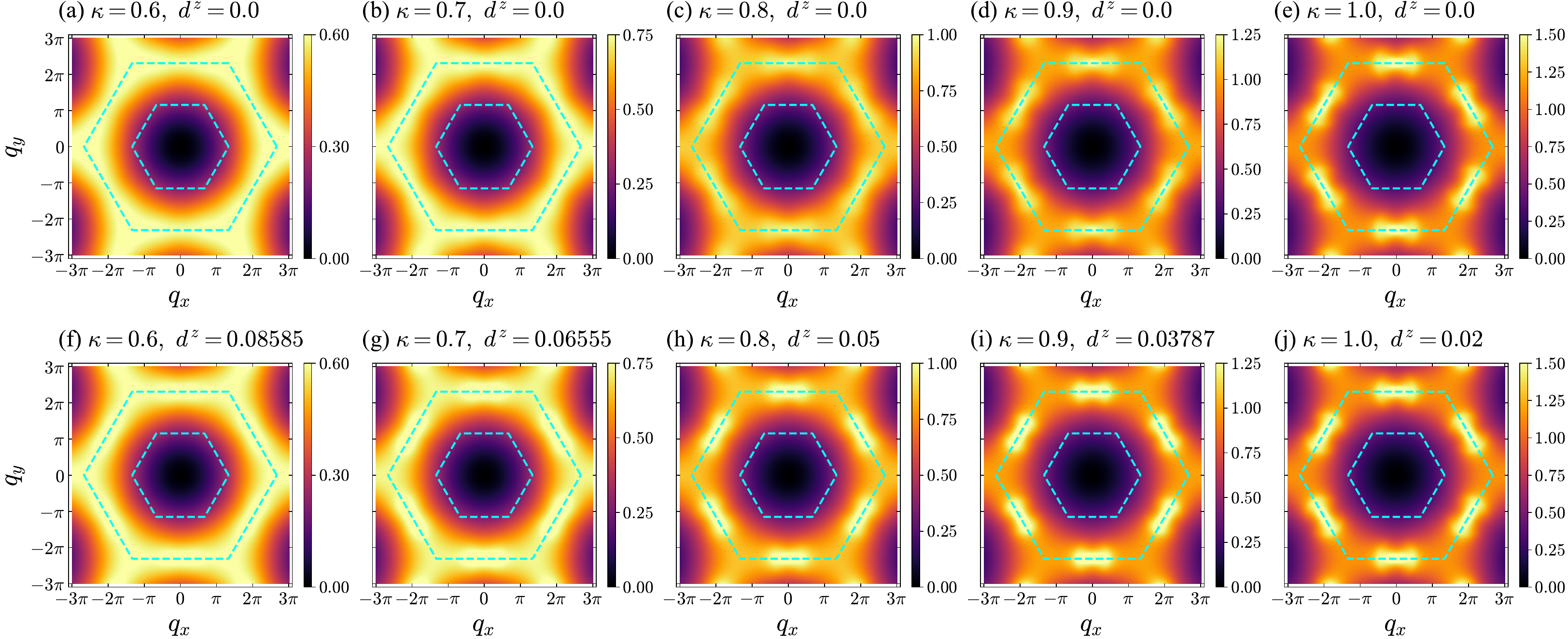}
      \caption{
        Ground-state static spin structure factor $S(\bm{q})$ in the kagome spin-liquid phase computed within SBMFT. Top row (a)-(e) : without DM interaction $d^{z}=0$; bottom row (f)–(j): with finite DM coupling $d^{z}$ (values indicated). From left to right, $\kappa$ increases ($\kappa=0.6, 0.7, 0.8, 0.9, 1.0$), corresponding to weaker quantum fluctuations for larger $\kappa$.
        In the plots of the static spin structure factor $S(\bm{q})$, the inner and outer cyan dashed hexagons indicate the first and extended Brillouin zones, respectively, corresponding to those shown in Fig.~\ref{fig:kagome_lattice}(b).
        }
      \label{fig:SBMFT_SSF}
\end{figure*}
\begin{figure*}[t]
  \centering
      \includegraphics[width=2\columnwidth,clip]{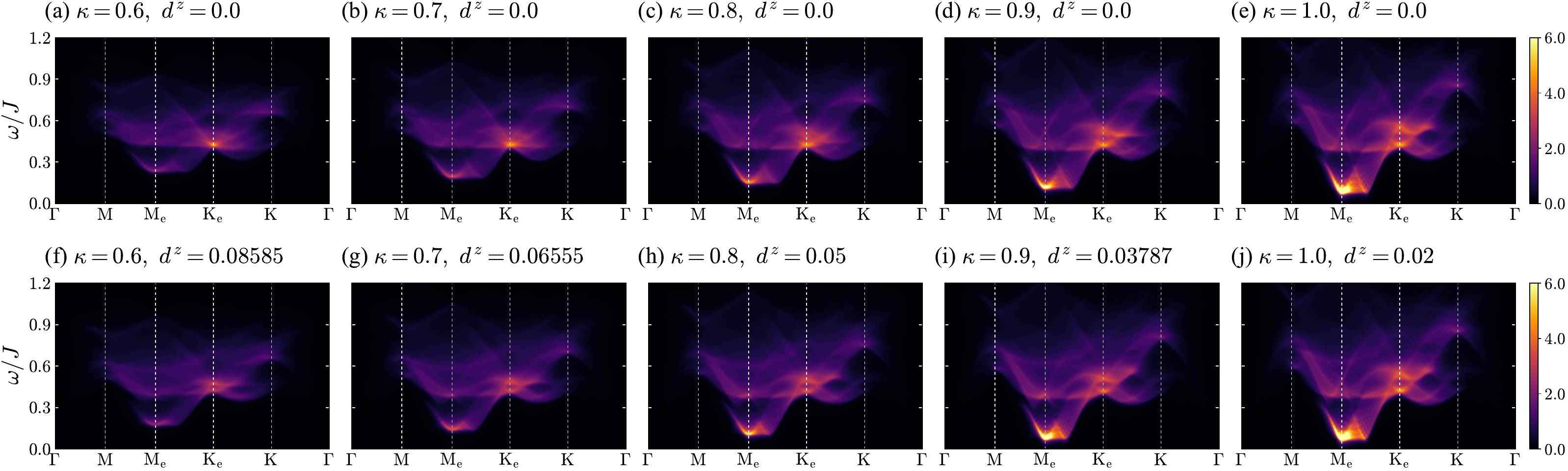}
      \caption{
      Dynamical spin structure factor $S(\bm{q},\omega)$ along the high-symmetry path $\Gamma-\mathrm{M}-\mathrm{M}_{\mathrm{e}}-\mathrm{K}_{\mathrm{e}}-\mathrm{K}-\Gamma$ obtained by SBMFT for the kagome spin-liquid ground state under the same conditions as in Fig.~\ref{fig:SBMFT_SSF}.
      }
      \label{fig:SBMFT_DSSF}
\end{figure*}

We next examine the static spin structure factor $S(\bm{q})$ and the dynamical structure factor $S(\bm{q},\omega)$ calculated within SBMFT.
The corresponding results are shown in Figs.~\ref{fig:SBMFT_SSF} and \ref{fig:SBMFT_DSSF}, respectively.
In both figures, the top row [(a)-(e)] corresponds to the pure Heisenberg model with $d^{z}=0$, while the bottom row [(f)-(j)] shows the results with a finite DM interaction.
From left to right, the panels correspond to $\kappa=0.6$, $0.7$, $0.8$, $0.9$, and $1.0$.
In panels (f)-(j), the value of $d^{z}$ is chosen for each $\kappa$ so as to place the system close to the magnetically ordered phase, and the corresponding values are indicated in the panels.

We first turn to the static spin structure factor $S(\bm{q})$ shown in Fig.~\ref{fig:SBMFT_SSF}, from which we can draw a particularly intriguing observation: the enhancement of spin correlations caused by weakening quantum effects and that caused by introducing the DM interaction are qualitatively consistent with each other.
Concretely, in the pure Heisenberg model with $d^{z}=0$, we find that, as $\kappa$ increases, the intensity at the $\mathrm{M}_{\mathrm{e}}$ point clearly grows and eventually exceeds that at the $\mathrm{K}_{\mathrm{e}}$ point.
Starting from $\kappa=0.6$, where the intensity distribution is broadly spread outside the boundary of the first BZ, increasing $\kappa$ leads to an evolution in which the weight becomes concentrated at the $\mathrm{M}_{\mathrm{e}}$ point.
Interestingly, for a fixed $\kappa$, increasing $d^{z}$ produces a similar trend, namely, it enhances the intensity at the $\mathrm{M}_{\mathrm{e}}$ point relative to that at the $\mathrm{K}_{\mathrm{e}}$ point.
This tendency is observed for all values of $\kappa$; a particularly transparent example is provided by the comparison between Fig.~\ref{fig:SBMFT_SSF}(c) and Fig.~\ref{fig:SBMFT_SSF}(h).
These observations suggest that the effects of the quantum parameter $\kappa$ and the DM interaction $d^{z}$ on the magnetic instability are qualitatively similar.
However, the color maps of the static structure factor shown in Fig.~\ref{fig:SBMFT_SSF} alone do not allow us to determine quantitatively how the correlations evolve, nor to disentangle whether the relative enhancement of the $\mathrm{M}_{\mathrm{e}}$ intensity over the $\mathrm{K}_{\mathrm{e}}$ intensity arises from an increase at $\mathrm{M}_{\mathrm{e}}$, a suppression at $\mathrm{K}_{\mathrm{e}}$, or a combination of both.
Nevertheless, understanding how these parameters influence spin correlations is an important issue for elucidating the instability of kagome quantum spin liquids.
We return to this point in Sec.~\ref{sec:Discussion}.

Next, we discuss the dynamical structure factor obtained within SBMFT, shown in Fig.~\ref{fig:SBMFT_DSSF}.
The influence of $\kappa$ and $d^{z}$ discussed above can also be identified in the dynamical structure factor $S(\bm{q},\omega)$.
For example, focusing on the pure Heisenberg model with $d^{z}=0$, one finds that, as $\kappa$ increases, the intensity at the $\mathrm{K}_{\mathrm{e}}$ point is reduced, while the intensity at the $\mathrm{M}_{\mathrm{e}}$ point is enhanced.
A similar trend is observed in the $d^{z}$ dependence: by comparing Fig.~\ref{fig:SBMFT_DSSF}(a) and Fig.~\ref{fig:SBMFT_DSSF}(f), one can clearly see that introducing $d^{z}$ suppresses the intensity at the $\mathrm{K}_{\mathrm{e}}$ point.
This tendency holds for all values of $\kappa$, indicating that the trend identified in the static structure factor $S(\bm{q})$ is also reflected in the dynamical structure factor $S(\bm{q},\omega)$.
As another interesting observation, one finds that the overall spectral spread of the dynamical structure factor increases as $\kappa$ becomes larger.
This reflects the fact that the underlying bosonic-spinon band dispersion broadens upon increasing $\kappa$, as seen in Fig.~\ref{fig:SBMFT_dispersion}(a).
At $\kappa=1.0$, corresponding to an $S=1/2$ spin, the spinon bandwidth is enlarged and $\omega_{\text{min}}$ is reduced to about $0.03$; correspondingly, the spin gap inferred from the dynamical structure factor becomes very small [see Fig.~\ref{fig:SBMFT_DSSF}(e) and (j)].

\subsection{Random phase approximation beyond Schwinger boson mean-field}
\label{sec:Random phase approximation beyond Schwinger boson mean-field}

\begin{figure}[t]
  \centering
      \includegraphics[width=\columnwidth,clip]{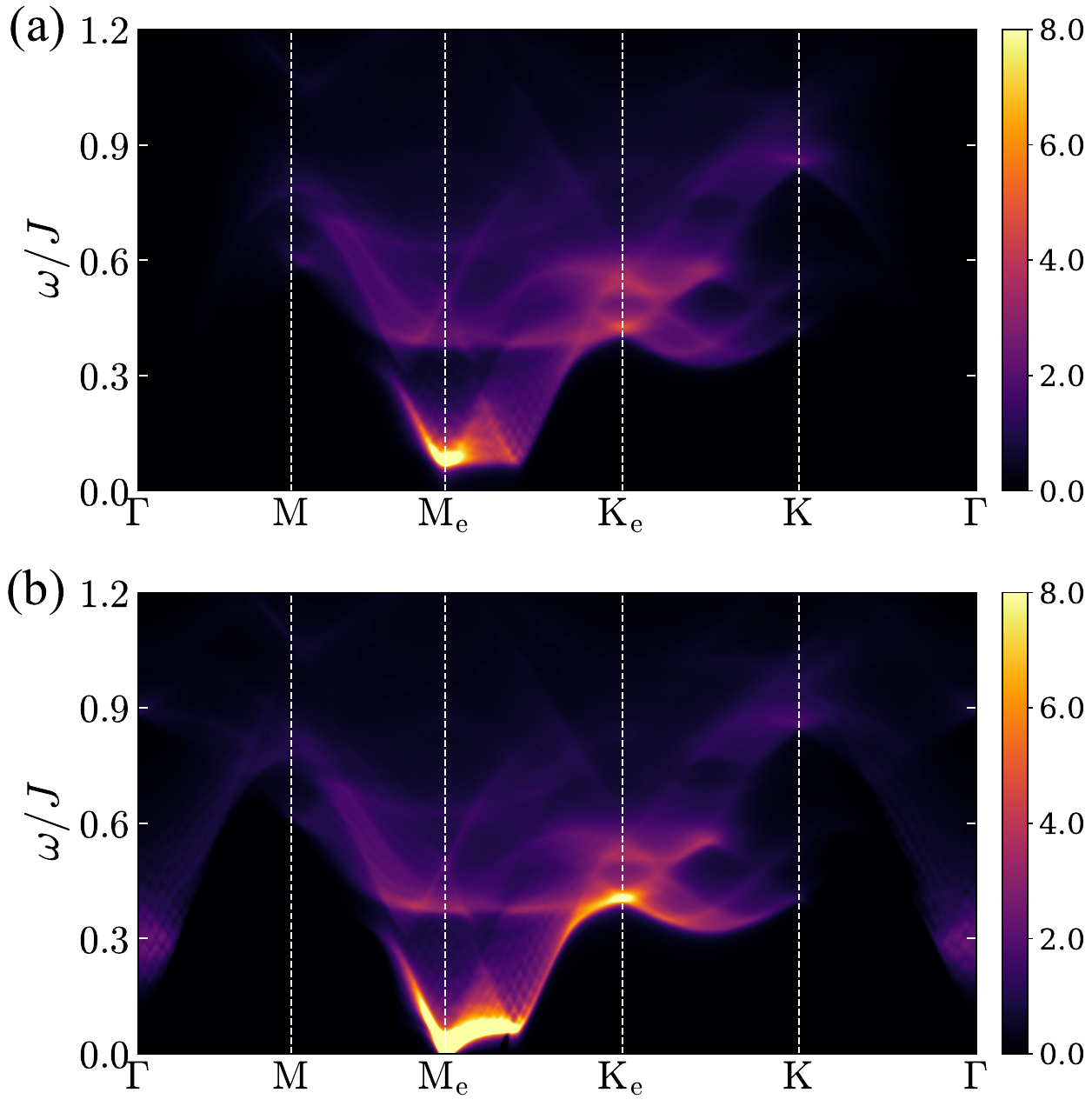}
      \caption{
      (a)
      Ground-state dynamical spin structure factor $S(\bm{q},\omega)$ obtained from SBMFT for $d^{z}=0.02$, reproduced from Fig.~\ref{fig:SBMFT_DSSF}(j), showing a finite spin gap.
      (b) RPA-corrected dynamical spin structure factor $S^{\text{RPA}}(\bm{q},\omega)$ obtained by performing an RPA calculation with respect to the Heisenberg term $J$ on top of the mean-field result in (a). The spin gap is strongly suppressed at the $\mathrm{M}_{\mathrm{e}}$ point.}
      \label{fig:RPA_DSSF}
\end{figure}

In this section, we present the results of our analysis of how the dynamical spin structure factor obtained within SBMFT is modified when many-body effects between spinons are incorporated within the RPA framework.
As emphasized in recent RPA studies based on the spinon picture~\cite{Ghioldi-Gonzalez-2018,Willsher-2025,Rao-Moessner-Knolle-2025}, a key prerequisite for such an RPA treatment is that the underlying mean-field ansatz already provides a reasonable description.
However, since the correct physics of kagome quantum spin liquids is not yet fully understood, selecting an appropriate ansatz is highly nontrivial.
Here, we adopt a pragmatic criterion based on experiment and choose an ansatz that reproduces the strong low-energy intensity at the $\mathrm{M}_{\mathrm{e}}$ point observed in inelastic neutron-scattering experiments~\cite{Han-2012}.
None of the six AFMFT ans\"atze captures this feature satisfactorily, whereas the SBMFT ansatz does.
We therefore choose the Schwinger boson mean-field ansatz, which is broadly consistent with the inelastic neutron-scattering results, as the starting point for the RPA analysis.

We show the RPA-corrected dynamical structure factor $S_{\text{RPA}}(\bm{q},\omega)$ in Fig.~\ref{fig:RPA_DSSF}(b).
As can be seen by comparison with Fig.~\ref{fig:RPA_DSSF}(a), the most significant RPA-induced modification occurs near the $\mathrm{M}_{\mathrm{e}}$ point in the low-energy regime.
In particular, the spin gap at the $\mathrm{M}_{\mathrm{e}}$ point is strongly suppressed by many-body effects.
By contrast, the high-energy part is only weakly affected; instead, the spectral weight is enhanced mainly in the range $0.05\lesssim\omega\lesssim0.4$ along the path from $\mathrm{M}_{\mathrm{e}}$ to $\mathrm{K}_{\mathrm{e}}$.
Importantly, the key features observed at the mean-field level---namely, the good agreement with neutron-scattering experiments in both the energy scale and the overall intensity distribution reported in a previous study~\cite{Messio-Bieri-2017}---are retained.
Thus, while preserving these salient characteristics, the inclusion of many-body effects successfully reduces the spin gap.

\section{Discussion}
\label{sec:Discussion}

\begin{figure}[t]
  \centering
      \includegraphics[width=\columnwidth,clip]{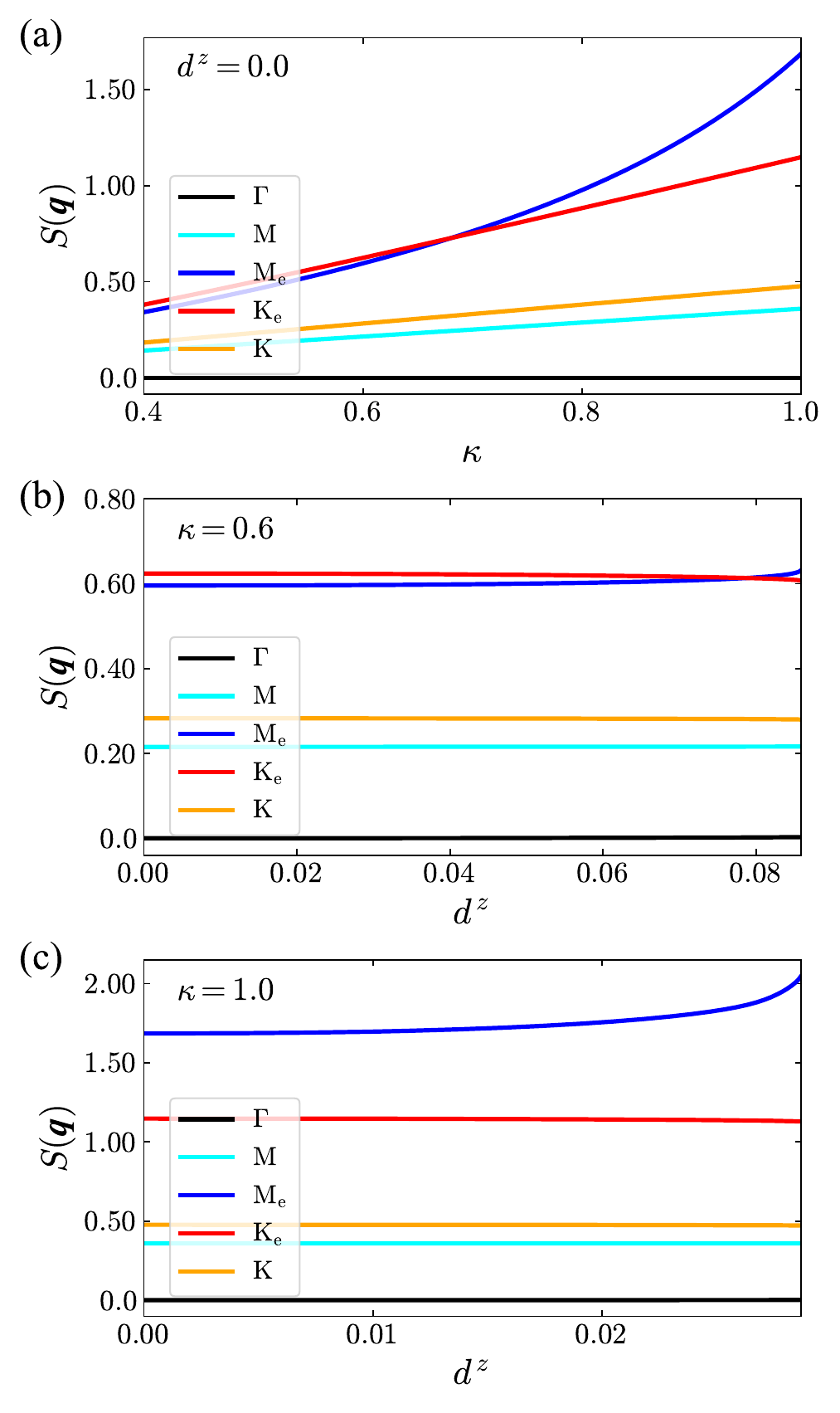}
      \caption{Parameter dependence of the static spin structure factor obtained within SBMFT at high-symmetry points.
      (a) $\kappa$ dependence of the static structure factor $S(\bm{q})$ for the pure Heisenberg model with $d^{z}=0$.
      (b) $d^{z}$ dependence of $S(\bm{q})$ in the quantum regime $\kappa=0.6$.
      (c) $d^{z}$ dependence of $S(\bm{q})$ at $\kappa=1.0$, corresponding to an $S=1/2$ spin.
      }
      \label{fig:SBMFT_SSF_dependence}
\end{figure}

In this section, based on the results in Sec.~\ref{sec:Results}, we revisit the differences between the results obtained within AFMFT and SBMFT.
As can be seen in Fig.~\ref{fig:AFMFT_DSSF_and_SSF}, the dynamical spin structure factor $S(\bm{q},\omega)$ obtained within AFMFT exhibits an overall convex-up parabolic distribution of spectral weight along the calculated $q$ path.
Moreover, because the regions of strong intensity lie near the top of this parabolic structure, AFMFT yields pronounced intensity on the high-energy side, whereas the intensity is suppressed on the low-energy side. This feature is not consistent with the low-energy neutron-scattering phenomenology emphasized here, which shows strong intensity also in the low-energy regime.
In particular, we find that none of the six ans\"atze considered exhibits strong intensity at the $\mathrm{M}_{\mathrm{e}}$ point in the region around $\omega\simeq0.044$ reported experimentally~\cite{Han-2012}. 
By contrast, the dynamical structure factor obtained within SBMFT exhibits an overall concave-down parabolic distribution of spectral weight, so that the strongest intensity appears on the low-energy side rather than on the high-energy side. At $\kappa=1.0$, corresponding to an $S=1/2$ spin, the SBMFT result reproduces the low-energy spectral intensity on an experimentally relevant energy scale and exhibits strong intensity around the $\mathrm{M}_{\mathrm{e}}$ point near $\omega \sim 0.06$, in reasonable agreement with neutron-scattering results~\cite{Han-2012}. Although the bosonic nature of the excitations in SBMFT leads to a finite spin gap, the mean-field result remains broadly consistent with the low-energy spectral features. These observations indicate that the strong low-energy intensity observed in neutron-scattering experiments can be naturally understood within the Schwinger-boson-based description considered here.
This comparison also clarifies the status of the AFMFT ans\"atze considered in this work. Since none of the AFMFT spectra exhibits the pronounced low-energy magnetic-instability feature near the $\mathrm{M}_{\mathrm{e}}$ point, it is difficult to identify a unique AFMFT ansatz as the most faithful representation of the experimental response based only on the present mean-field spin structure factors. Other features, such as the broad spinon dispersion and the overall continuum shape, are useful diagnostics, but they are not sufficient by themselves to select a single optimal AFMFT ansatz. We nevertheless note that, among these six AFMFT ans\"atze, Ansatz III gives the lowest unprojected mean-field ground-state energy at $d^{z}=0$. At the same time, variational Monte Carlo studies have shown that projected fermionic wave functions constructed from the $\mathrm{U}(1)$ Dirac spin-liquid state, which corresponds to Ansatz IV in our notation, provide an energetically favorable description of the spin-$1/2$ kagome Heisenberg antiferromagnet, with nearby $\mathbb{Z}_{2}$ descendants also being competitive~\cite{Ran-Hermele-2007,Iqbal-Becca-2011}. This suggests a promising future direction: many-body corrections or RPA-type treatments built on such AFMFT backgrounds may generate low-energy spectral weight and magnetic-instability signatures that are absent at the bare mean-field level.

To the best of our knowledge, a systematic comparison of the dynamical spin structure factor $S(\bm{q},\omega)$ characteristics of the same model using two mean-field approaches with different statistics has not been carried out in previous studies.
As discussed above, our results suggest that the difference in the statistics of the elementary excitations leaves clear signatures in the dynamical structure factor $S(\bm{q},\omega)$.
In particular, the overall parabolic structure of $S(\bm{q},\omega)$ and the locations of the intensity maxima can be regarded as consequences of the different statistics.
Indeed, this trend is also consistent with previous calculations of the dynamical structure factor for spin models on the triangular lattice:
within AFMFT the dynamical structure factor exhibits a convex-up parabolic structure and the spectral weight is concentrated near its vertex~[see Fig.~5 in Ref.~\cite{Willsher-2025}], whereas within SBMFT it exhibits a concave-down parabolic structure with the weight again concentrated near its vertex~[see Fig.~4 in Ref.~\cite{Ghioldi-Gonzalez-2018}].
From these structures, one can understand that the strong intensity in AFMFT appears predominantly on the high-energy side, while the strong intensity in SBMFT appears predominantly on the low-energy side.
Such behavior is determined by the underlying spinon dispersion, and since the dispersion depends strongly on whether the spinons are fermionic or bosonic on the lattice, we expect these qualitative differences not to be specific to the kagome-lattice model, but to extend to a broader class of frustrated magnets.
Our calculations are consistent with this interpretation in the kagome-lattice model.
These results further suggest that the dynamical structure factor may provide useful information for experimentally identifying the nature of the elementary excitations.
Conventionally, the statistics of excitations have often been inferred from thermodynamic quantities such as the temperature dependence of the specific heat and magnetic susceptibility.
However, such arguments essentially rely on the presence or absence of a Fermi surface, and in the absence of a Fermi surface it becomes difficult to determine whether the excitations are fermionic or bosonic.
Moreover, given that the Abrikosov fermion approach can, in principle, describe gapped spin liquids, the absence of a Fermi surface does not necessarily provide a decisive criterion for the excitation statistics.
In this respect, the present results indicate that characteristic features in $S(\bm{q},\omega)$ can serve as an alternative indicator, and we expect the dynamical structure factor to provide a useful diagnostic, particularly when discussing the statistics of excitations in gapped spin liquids.

The results of our analyses based on the two mean-field approaches suggest that an ansatz with bosonic low-energy excitations is more consistent with the experimental observations.
We therefore investigate, by tuning two parameters $\kappa$ and $d^{z}$, the characteristics of magnetic instabilities for this mean-field ansatz, which have not been discussed in previous studies~\cite{Messio-Bieri-2017}.
As discussed in Sec.~\ref{sec:SB_Spin structure factor}, from the resulting static spin structure factor $S(\bm{q})$ and dynamical spin structure factor $S(\bm{q},\omega)$, we found that increasing the parameter controlling quantum effects (thereby reducing quantum effects) and introducing the DM interaction have qualitatively similar consequences.
To understand the instability of kagome quantum spin liquids, we therefore quantify these trends, and summarize the results in Fig.~\ref{fig:SBMFT_SSF_dependence}.
This quantitative analysis reveals an intriguing behavior.
First, increasing $\kappa$ not only enhances the intensity at the $\mathrm{M}_{\mathrm{e}}$ point, as shown in Fig.~\ref{fig:SBMFT_SSF_dependence}(a), but also enhances the intensity at the $\mathrm{K}_{\mathrm{e}}$ point, which appears to be strongly suppressed in the plots of the static structure factor $S(\bm{q})$ [see Fig.~\ref{fig:SBMFT_SSF}] and the dynamical structure factor $S(\bm{q},\omega)$ [see Fig.~\ref{fig:SBMFT_DSSF}].
Interestingly, the enhancement is much stronger at $\mathrm{M}_{\mathrm{e}}$: around $\kappa\sim0.70$, the intensity at $\mathrm{M}_{\mathrm{e}}$ overtakes that at $\mathrm{K}_{\mathrm{e}}$, and continues to increase substantially up to $\kappa=1.0$.
At $\kappa=1.0$, which corresponds to the $S=1/2$ spin value in the SBMFT parameterization, the difference between the two intensities in $S(\bm{{q}})$ is found to widen to about $0.50$.
Turning to the effect of the DM interaction [see Figs.~\ref{fig:SBMFT_SSF_dependence}(b) and (c)], we find that, irrespective of the value of $\kappa$, it plays a qualitatively similar role: it suppresses the intensity at the $\mathrm{K}_{\mathrm{e}}$ point while enhancing that at the $\mathrm{M}_{\mathrm{e}}$ point.
By contrast, we confirm that the DM interaction has essentially no influence on the intensities at the other high-symmetry points, namely, the $\Gamma$, $\mathrm{K}$, and $\mathrm{M}$ points.
This behavior differs from the $\kappa$ dependence [see Fig.~\ref{fig:SBMFT_SSF_dependence}(a)], where $\kappa$ enhances the intensities at all points except the $\Gamma$ point.
Accordingly, the intensities at the $\mathrm{M}$ and $\mathrm{K}$ points also increase with $\kappa$, although the enhancement is more moderate than that at the $\mathrm{K}_{\mathrm{e}}$ and $\mathrm{M}_{\mathrm{e}}$ points.
Furthermore, the impact of the DM interaction exhibits a trend similar to the $d^{z}$ dependence of $\omega_{\text{min}}$ observed in Fig.~\ref{fig:SBMFT_dispersion}(c): as the system approaches the magnetic transition, the intensities change rapidly, with $d^{z}$ lowering the intensity at $\mathrm{K}_{\mathrm{e}}$ and raising that at $\mathrm{M}_{\mathrm{e}}$.
The pronounced DM-interaction effects visible in our static structure factor $S(\bm{q})$ [see Fig.~\ref{fig:SBMFT_SSF}] and dynamical structure factor $S(\bm{q},\omega)$ [see Fig.~\ref{fig:SBMFT_DSSF}] can thus be attributed to the fact that we evaluated these quantities at $d^{z}$ values close to the magnetic transition, as is also evident from the $d^{z}$ dependence of $S(\bm{q})$.
These results help clarify the mechanism by which the SBMFT ansatz adopted here reproduces the main features of the inelastic-neutron-scattering data.
Increasing $\kappa$ broadens the SBMFT bandwidth and thereby produces spectral weight over a broad energy range, while the DM interaction works in concert with this effect to enhance the response at the $\mathrm{M}_{\mathrm{e}}$ point. The combined effect naturally accounts for the pronounced low-energy intensity observed near the $\mathrm{M}_{\mathrm{e}}$ point in neutron-scattering experiments.

We now discuss what physical implications may follow from the fact that the many-body effects between spinons considered in this work close the gap in the dynamical structure factor $S(\bm{q},\omega)$.
This gap closing should be interpreted as a magnetic instability of the non-condensed SBMFT spin-liquid state, rather than simply as the realization of a stable gapless spin-liquid spectrum. In the Schwinger-boson language, the magnetically ordered phase beyond such an instability is associated with Bose condensation of Schwinger bosons. In this sense, the RPA result indicates that interactions beyond the bare SBMFT saddle point drive the system toward an ordered state with an ordering tendency centered near the $\mathrm{M}_{\mathrm{e}}$ point.
In perturbative studies that incorporate fluctuations beyond SBMFT, it has been pointed out that such beyond-mean-field effects become important near a quantum critical point, and it was shown that the existence of quasi-magnons, which naturally capture the instability of the spin-liquid phase already at the level of the mean-field description, can be discussed within a perturbative treatment of SBMFT.
Since our perturbative approach is based on a different scheme, we cannot make a fully parallel comparison; nevertheless, we expect that a similar line of reasoning applies in the sense that many-body effects encode the instability of the spin-liquid phase in kagome antiferromagnets.

\section{Summary}
\label{sec:Summary}
In summary, we have investigated the spin dynamics of kagome quantum spin liquids within two mean-field frameworks based on distinct elementary excitations: Abrikosov fermions and Schwinger bosons. In the AFMFT, we computed the dynamical spin structure factors for representative nearest-neighbor $\mathrm{U}(1)$ ans\"atze together with a nearest-neighbor $\mathbb{Z}_{2}$ spin-liquid state. In the SBMFT, we focused on a DM-interaction-induced time-reversal-symmetry-breaking ansatz relevant to the $S=1/2$ quantum spin-liquid regime and to neutron-scattering experiments.
The resulting dynamical structure factors exhibit qualitatively distinct features in the two approaches. While the AFMFT results are dominated by an overall convex-up parabolic structure, the SBMFT results instead show a pronounced concave-down parabolic structure. We therefore conclude that, within the set of ans\"atze considered here, the SBMFT results better capture the low-energy neutron-scattering phenomenology, in particular because they reproduce the strong low-energy intensity at the $\mathrm{M}_{\mathrm{e}}$ point reported in neutron-scattering experiments.
At the same time, the bosonic nature of the elementary excitations in SBMFT tends to produce a finite spin gap, as found in our mean-field calculation.
By incorporating many-body effects on top of SBMFT, we find that this gap is substantially reduced without destroying the overall spectral characteristics, suggesting a mechanism by which the Schwinger-boson approach can capture the gapless, or very small-gap, structure discussed in both theory and experiment.
These results and their physical interpretation are consistent with recent studies showing that perturbative calculations built on spinon mean-field theories can capture instabilities of quantum spin liquids~\cite{Ghioldi-Gonzalez-2018,Willsher-2025,Rao-Moessner-Knolle-2025}.

Moreover, our analyses of the kagome antiferromagnet with the DM interaction within the two mean-field approaches suggest that the dynamical spin structure factor contains useful information on the underlying spinon statistics.
This indicates that the dynamical spin structure factor $S(\bm{q},\omega)$ may serve as a useful diagnostic complementary to thermodynamic probes.
In particular, for gapped spin liquids---for which the temperature dependence of the specific heat and susceptibility often does not provide a decisive criterion---our results suggest that $S(\bm{q},\omega)$ can provide a promising alternative diagnostic.

Our analysis selects an ansatz that is suitable for describing neutron-scattering experiments, and the fact that this choice leads to bosonic excitations described by the Schwinger boson approach does not exclude the possibility that fermionic excitations may be identified from other experiments or theoretical considerations.
Therefore, a careful examination and judgement will remain necessary in identifying the elementary excitations of kagome quantum spin liquids.
Nevertheless, we emphasize that our calculations provide a particularly natural description of a kagome-lattice quantum spin-liquid state in the vicinity of a magnetic transition.
Another aspect that makes kagome quantum spin liquids an intriguing platform is the effect of an external magnetic field~\cite{Jeong-2011,Gomilsek-2017,Nishimoto-2013,Fu-2015,Khuntia-2020,Barthelemy-2022,Jeon-2023,He-Yu-2024,Kato-2024,Zheng-Zhang-2025,He-2025,Kato-2025}, which we did not address in this study.
It remains unclear whether the excitation structure under a magnetic field is also governed by the bosonic excitations that we concluded to be appropriate here.
Indeed, a previous study has reported results suggesting fermionic elementary excitations in kagome quantum spin liquids under a magnetic field~\cite{Zheng-Zhang-2025}.
How such results can be reconciled with our findings is an important open question, and one cannot rule out the possibility that the ground-state phase diagram in a magnetic field contains a transition at which the nature of the excitations changes from bosonic to fermionic.
If such a scenario is realized, the distinct differences between the dynamical structure factors obtained in AFMFT and SBMFT revealed in this work may appear in a pronounced manner, and further progress in the study of kagome quantum spin liquids is expected along this direction.

\begin{acknowledgments}
  The authors thank A.~Ono and R.~Iwazaki for fruitful discussions. D.S. is grateful to C.~D.~Batista for useful advice.
  D.S. also thanks J.~Knolle and R.~Eto for sharing early-stage results from their forthcoming analytical work on kagome magnets, which provided useful input for the present study. D.S. is also grateful to M.~O.~Takahashi for raising an essential question about the RPA methodology and for subsequent discussions.
  Parts of the numerical calculations were performed in the supercomputing systems in ISSP, the University of Tokyo.
  This work was supported by Grant-in-Aid for Scientific Research from
  JSPS, KAKENHI Grant Nos.~JP23H01129, JP23H04865, JP24K00563.
  D.S. acknowledges support from GP-Spin at Tohoku University.
\end{acknowledgments}

\appendix

\section{Analysis of Ansatz VII, gauge inequivalent to Ansatz III, with the same flux pattern}
\label{app:Analysis of Ansatz VII, gauge inequivalent to Ansatz III, with the same flux pattern}

\begin{figure}[t]
  \centering
      \includegraphics[width=\columnwidth,clip]{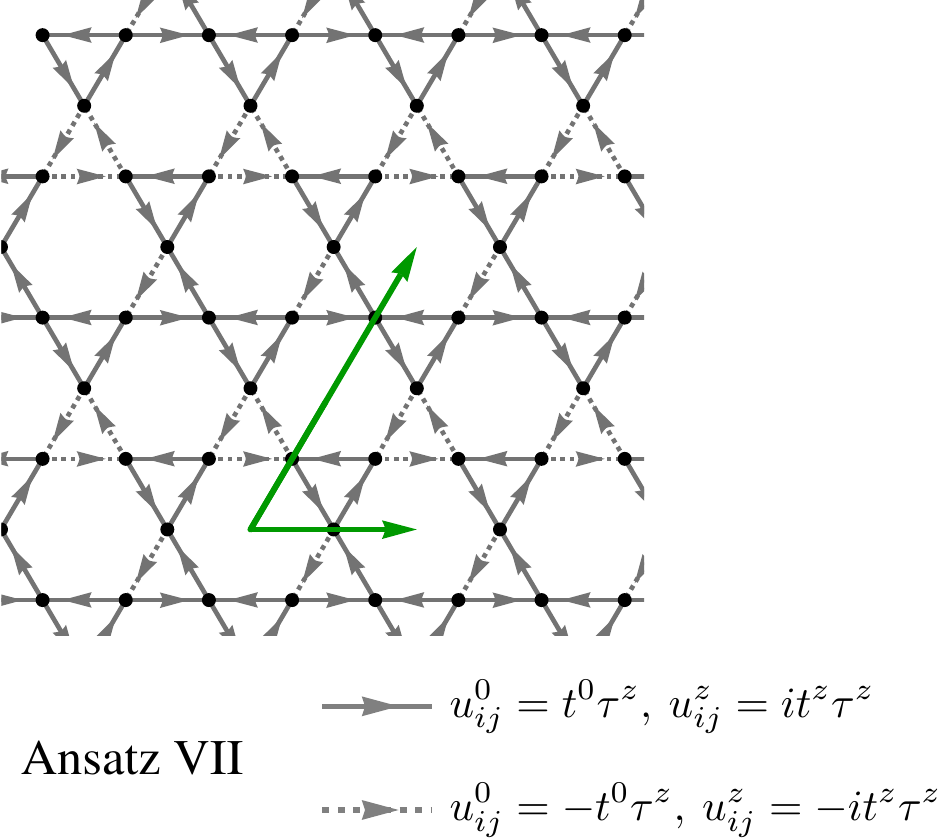}
      \caption{
    Mean-field Ansatz VII in the AFMFT. This ansatz shares the same flux pattern as Ansatz III introduced in Sec.~\ref{sec:AF_Mean-field ansatz}~[see Fig.~\ref{fig:mean_field_ansatz_AFMFT}(c)], but is gauge inequivalent to it, i.e., there exists no gauge transformation $W_i$ that connects Ansatz III and Ansatz VII. This ansatz has also been discussed in Ref.~\cite{Dodds-2013} as a $\mathrm{U}(1)$ spin-liquid state.
    }
    \label{fig:AnsatzVII}
\end{figure}

\begin{figure*}[t]
  \centering
      \includegraphics[width=2\columnwidth,clip]{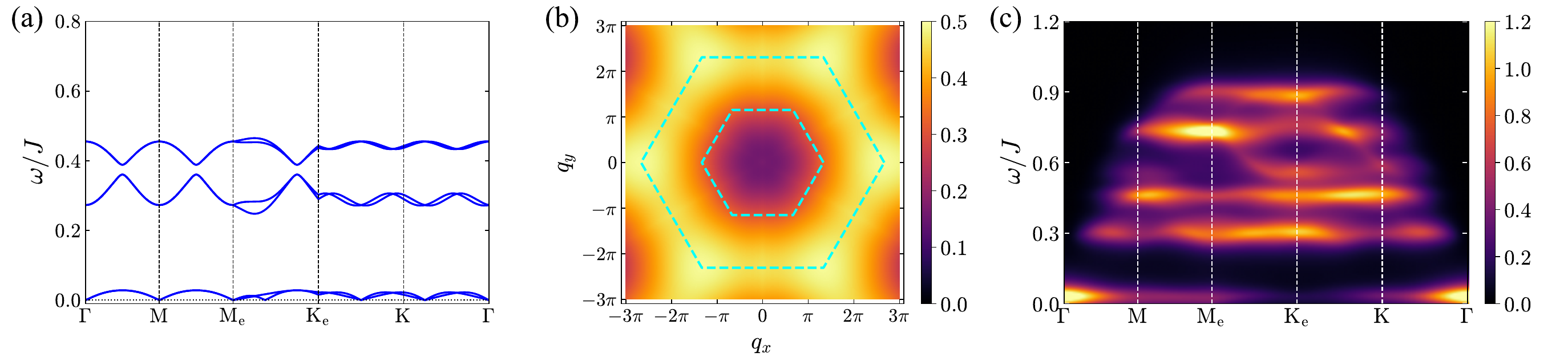}
      \caption{
      AFMFT results for the kagome Heisenberg antiferromagnet with a $z$-directed DM interaction $d^{z}=0.1$, obtained for Ansatz VII shown in Fig.~\ref{fig:AnsatzVII}. (a) Spinon dispersion relations. (b) Static spin structure factor $S(\bm{q})$. (c) Dynamical spin structure factor $S(\bm{q},\omega)$.
      }
      \label{fig:AnsatzVII_result}
\end{figure*}

In this Appendix, we present the spinon dispersion and spin structure factors calculated from the mean-field Ansatz VII shown in Fig.~\ref{fig:AnsatzVII}, which is not discussed in the main text. As in Ans\"atze I--IV considered in the main text, this ansatz assumes only hopping channels to be finite and corresponds to a $\mathrm{U}(1)$ mean-field ansatz. Applying to the present ansatz the same gauge-invariant flux analysis as that employed in the main text, we find that $\Phi_{\text{Hex}}=0$ and $\Phi_{\text{Para}}=\pi$. Therefore, at least in terms of the fluxes associated with the hexagon and parallelogram loops defined in Fig.~\ref{fig:kagome_lattice}(a), Ansatz VII has the same flux pattern as Ansatz III~[see Table~\ref{tab:flux}]. However, since one can verify that there exists no gauge transformation $W_i$ connecting these two ans\"atze, Ansatz VII should be regarded as a distinct mean-field ansatz. Here, using this Ansatz VII, we show in Fig.~\ref{fig:AnsatzVII_result} the calculated results for the antiferromagnetic Heisenberg model with a $z$-directed DM interaction $d^{z}=0.1$.
For completeness, we briefly summarize the characteristic features of the results obtained for Ansatz VII. 
The spinon dispersion extends over the range $0\leq\omega\lesssim0.45$. This overall energy range is comparable to that of Ansatz II discussed in the main text and is relatively narrow. Moreover, the spinon spectrum of Ansatz VII is composed of a set of weakly dispersive bands, each having a small bandwidth.
Compared with the ans\"atze discussed in the main text, the spinon bands are therefore less dispersive, suggesting a higher degree of spinon localization~[see Fig.~\ref{fig:AnsatzVII_result}(a)].
In addition, the static structure factor $S(\bm{q})$ exhibits a non-negligible intensity at the $\Gamma$ point, a feature shared with Ansatz II discussed in Sec.~\ref{sec:AF_Spin structure factor}.
As explained in the main text for Ansatz II, this feature originates from the low-energy spectral weight in the window $0\lesssim\omega\lesssim0.05$ along the $\Gamma$-$\mathrm{M}$ and $\mathrm{K}$-$\Gamma$ paths in the dynamical structure factor $S(\bm{q},\omega)$, which contributes to the static structure factor $S(\bm{q})$ upon integration over $\omega$.
This low-energy spectral weight can in turn be traced to the presence of an almost dispersionless spinon band lying very close to $\omega=0$ over a wide region of momentum space [see Fig.~\ref{fig:AnsatzVII_result}(a)].
Thus, Ansatz VII shares a similar low-energy excitation structure with Ansatz II. In contrast, its high-energy behavior is qualitatively different: whereas Ansatz II exhibits a dome-shaped and strongly dispersive structure at high energies, Ansatz VII shows relatively weakly dispersive spectral weight extending over approximately $0.3\lesssim\omega\lesssim1.0$.
This weak dispersion of the high-energy spectrum can be attributed to the small dispersion of the underlying spinon bands.

\section{Discussion of gauge-equivalent mean-field ansatz for Ansatz V in AFMFT}
\label{app:Discussion of gauge-equivalent mean-field ansatz for Ansatz V in AFMFT}

\begin{figure}[t]
  \centering
      \includegraphics[width=\columnwidth,clip]{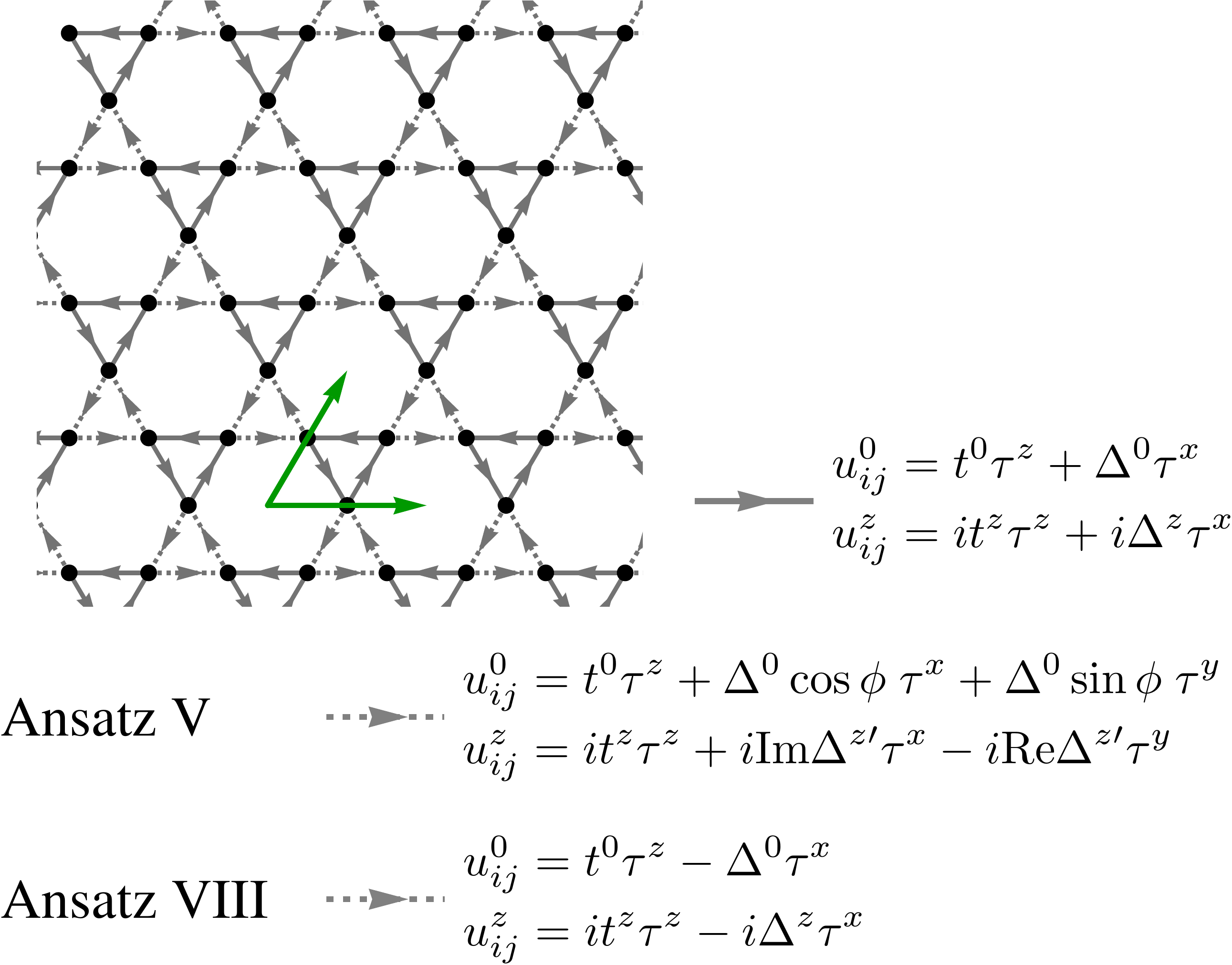}
      \caption{
      Two mean-field ans\"atze in the AFMFT. Ansatz V is the one analyzed in this work [see Fig.~\ref{fig:mean_field_ansatz_AFMFT}(e)], and as an alternative parametrization we introduce Ansatz VIII.
    The mean-field parameters on the solid bonds are identical for the two ans\"atze, whereas those on the dashed bonds differ. In Appendix~\ref{app:Discussion of gauge-equivalent mean-field ansatz for Ansatz V in AFMFT}, we show that Ansatz V and Ansatz VIII are related by a gauge transformation and are therefore physically equivalent.
      }
      \label{fig:equivalence_ansatzV}
\end{figure}

As discussed in the main text, the AFMFT possesses a local $\mathrm{SU}(2)$ gauge redundancy.
Ansatz V is introduced such that the singlet-pairing channel on the triangles formed by solid bonds and that on the triangles formed by dashed bonds differ by a relative complex phase $e^{i\phi}$.
At first sight, this phase seems to imply a time-reversal-symmetry-breaking mean-field state.
In this Appendix we show that this is only an apparent breaking: the phase degree of freedom can be removed by a local $\mathrm{SU}(2)$ gauge transformation.
Concretely, we demonstrate that the mean-field Ansatz V in Fig.~\ref{fig:equivalence_ansatzV} [see also Fig.~\ref{fig:mean_field_ansatz_AFMFT}(e)] is gauge equivalent to the manifestly time-reversal symmetry preserving Ansatz VIII in Fig.~\ref{fig:equivalence_ansatzV}.

Under a local $\mathrm{SU}(2)$ gauge transformation
\begin{equation}
\label{eq:gauge_transform_spinor}
\psi_i \to \psi_{i}W_{i},
\qquad
W_{i} \in \mathrm{SU}(2),
\end{equation}
the link matrices and the on-site term transform as
\begin{equation}
\label{eq:gauge_transform_links}
u_{ij}^{0,z} \to \widetilde{u}_{ij}^{0,z}=W_{i}u_{ij}^{0,z}W_{j}^{\dagger},
\qquad
a_{i}^{\gamma} \to \widetilde{a}_{i}^{\gamma} = W_{i}a_{i}^{\gamma}W_{i}^{\dagger}.
\end{equation}
Two mean-field ans\"atze are gauge equivalent if there exists a set $\{W_{i}\}$ satisfying Eq.~\eqref{eq:gauge_transform_links} on every bond.
Gauge-equivalent ans\"atze describe the same physical state; any gauge-invariant observable must be identical.

As discussed in Sec.~\ref{sec:Method_Abrikosov fermion mean-field theory}, a convenient gauge-invariant characterization of an ansatz is provided by Wilson-loop matrices.
For a directed closed loop $C:i_{1}\to i_{2} \to \cdots \to i_{n} \to i_{1}$, we define
\begin{equation}
\label{eq:Wilson_loop_def}
P^{\gamma}(C)
=
u_{i_{1} i_{2}}^{\gamma}
u_{i_{2} i_{3}}^{\gamma}\cdots
u_{i_{n} i_{1}}^{\gamma},
\qquad
\gamma=0,z.
\end{equation}
Under Eq.~\eqref{eq:gauge_transform_links}, the loop matrix transforms by conjugation at the base point,
\begin{equation}
\label{eq:Wilson_loop_conjugation}
P^{\gamma}(C) \to \widetilde{P}^{\gamma}(C)=W_{i_{1}}P^{\gamma}(C)W_{i_{1}}^{\dagger},
\end{equation}
so that $\mathrm{Tr}P^{\gamma}(C)$ and $\det P^{\gamma}(C)$ are gauge invariant.

In the present nearest-neighbor kagome ansatz, it is sufficient to evaluate the loop invariants on a generating set of elementary even-length loops.
We choose (i) the elementary hexagonal loop $C_{\text{Hex}}$ (six bonds) and (ii) the elementary parallelogram loop $C_{\text{Para}}$ (eight bonds) shown in Fig.~\ref{fig:kagome_lattice}.
By translational invariance, one representative of each loop type is sufficient.
Using the self-consistent mean-field solutions, we evaluated the loop matrices in Eq.~\eqref{eq:Wilson_loop_def} along the bond orientations of Fig.~\ref{fig:equivalence_ansatzV} and obtained the following gauge-invariant values:
\begin{align}
\label{eq:loopvals_hex}
\mathrm{Tr}\,P^{0,z}(C_{\text{Hex}}) &= 0.0,
&
\det P^{0,z}(C_{\text{Hex}}) &= 0.0,
\\
\label{eq:loopvals_para}
\mathrm{Tr}\,P^{0,z}(C_{\text{Para}}) &= 0.0,
&
\det P^{0,z}(C_{\text{Para}}) &= 0.0.
\end{align}
In addition, in the Ansatz V we verified that the dashed-bond parameter $\Delta^{z'}$ is not independent but satisfies
\begin{equation}
\label{eq:Deltazprime_relation}
\Delta^{z'} = i\Delta^{z}\,e^{i\phi}
\end{equation}
which ensures that $u^{0}$ and $u^{z}$ on dashed bonds are governed by the same angle $\phi$.
Equations~\eqref{eq:loopvals_hex}-\eqref{eq:loopvals_para} and \eqref{eq:Deltazprime_relation} show that Ansatz V and Ansatz VIII share the same gauge-invariant data on the elementary even loops and are compatible between the $\gamma=0$ and $\gamma=z$ sectors. These relations therefore provide the necessary consistency conditions for the existence of a local $\mathrm{SU}(2)$ gauge transformation connecting the two ans\"atze. While such agreement of loop invariants is not, in general, sufficient to establish gauge equivalence, in the present case we have explicitly verified that there exists a set of local $\mathrm{SU}(2)$ matrices $\{W_i\}$ satisfying
\begin{equation}
\label{eq:gauge_equiv_statement_appB}
u_{ij}^{0,z}\Big|_{\text{Ansatz VIII}}
=
W_{i}u_{ij}^{0,z}\Big|_{\text{Ansatz V}}W_{j}^{\dagger},
\qquad
\forall \langle ij\rangle,
\end{equation}
together with the corresponding relation for the on-site term. Hence Ansatz V and Ansatz VIII are gauge equivalent and represent the same physical mean-field state.

Finally, we comment on time-reversal symmetry.
Time reversal acts antiunitarily (complex conjugation accompanied by a spin rotation in the physical spin space).
In AFMFT, time-reversal symmetry is preserved if the mean-field ansatz is invariant under time reversal up to an $\mathrm{SU}(2)$ gauge transformation (projective implementation).
Ansatz VIII is manifestly time-reversal symmetry preserving in the gauge shown, since all mean-field amplitudes can be chosen real in $u^{0}$ and purely imaginary in $u^{z}$, which is consistent with the transformation of the $\sigma^{z}$ sector under time reversal.
Since Ansatz V is gauge equivalent to Ansatz VIII [Eq.~\eqref{eq:gauge_equiv_statement_appB}], Ansatz V in Fig.~\ref{fig:equivalence_ansatzV} also preserves time-reversal symmetry.
Thus, the complex phase $e^{i\phi}$ in Fig.~\ref{fig:equivalence_ansatzV} does not represent a physical chiral order parameter; it is a gauge artifact that can be removed by an appropriate gauge choice.

\section{Spinon dispersion obtained by AFMFT without DM interaction}
\label{app:Spinon dispersion obtained by AFMFT without DM interaction}
\begin{figure*}[t]
  \centering
      \includegraphics[width=2\columnwidth,clip]{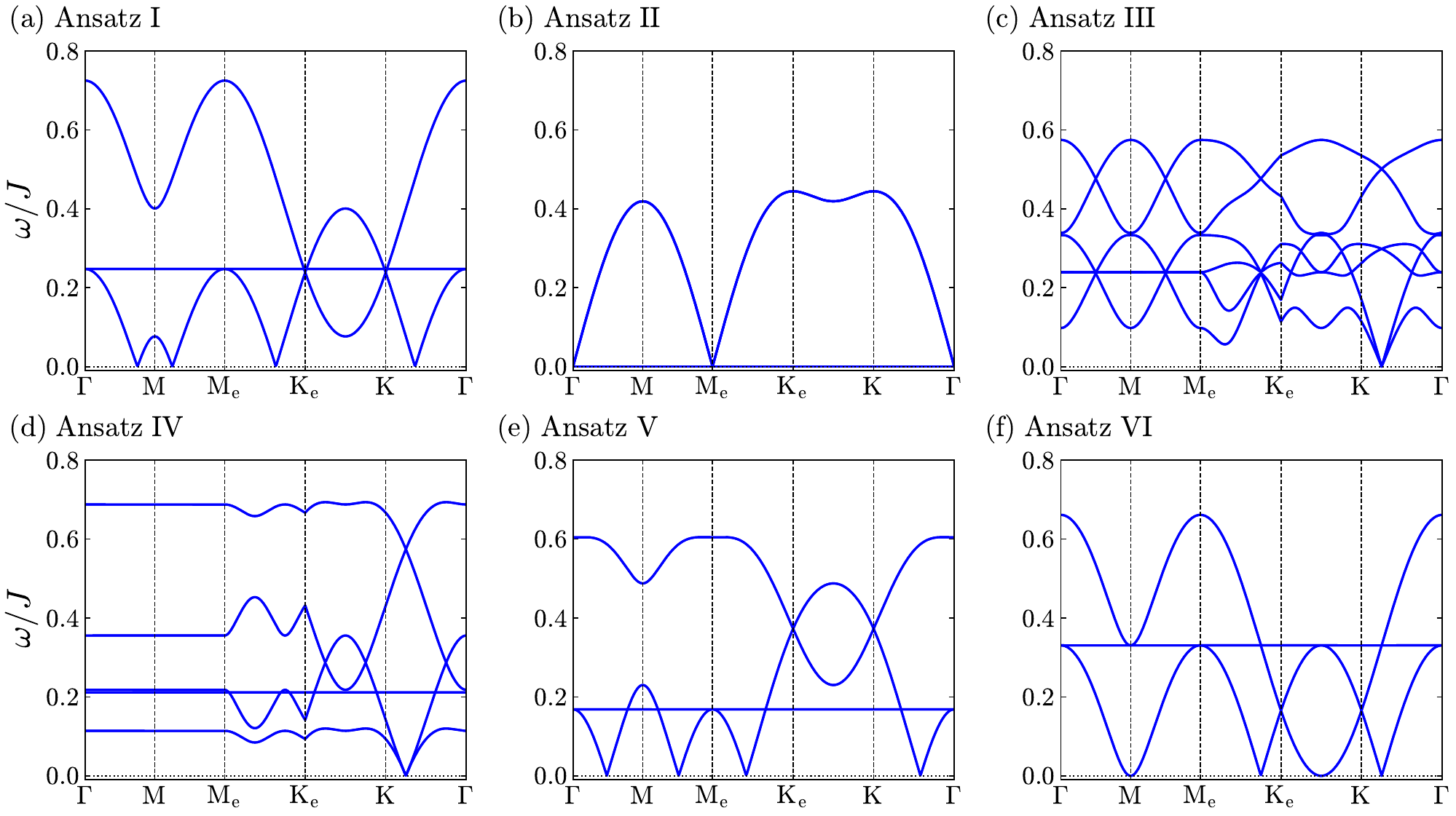}
      \caption{Spinon dispersion relations obtained from AFMFT for the ans\"atze shown in Fig.~\ref{fig:mean_field_ansatz_AFMFT} in the pure Heisenberg limit ($d^{z}=0$).
      We plot only the positive-energy branches $\epsilon_{\bm{k},n}\geq0$, which represent the quasiparticle excitations.
      }
      \label{fig:AFMFT_dispersion_without_DM}
\end{figure*}

Here we show in Fig.~\ref{fig:AFMFT_dispersion_without_DM} the spinon dispersions obtained within AFMFT at $d^{z}=0$, which were not presented in the main text, for each of the considered Ansatz I-VI.
By comparing them with the spinon dispersions at $d^{z}=0.1$ shown in the main text (see Fig.~\ref{fig:AFMFT_dispersion}), we can examine how the flat bands and the degeneracy points are modified by the DM interaction.

\section{Detailed calculation of the random-phase approximation}
\label{app:Detailed calculation of the random-phase approximation}

\begin{figure}[t]
  \centering
      \includegraphics[width=\columnwidth,clip]{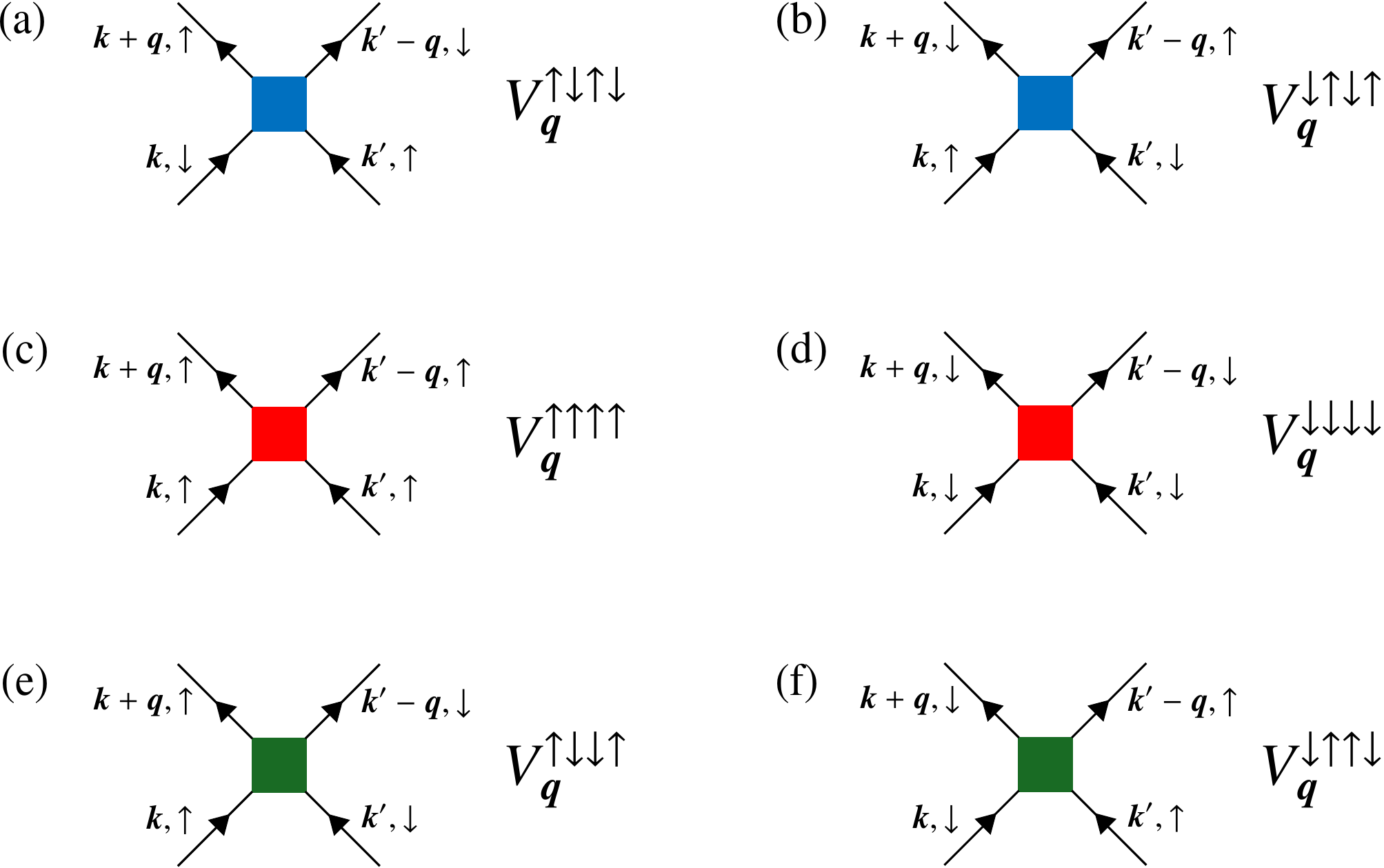}
       \caption{Six four-point vertices that appear in the perturbative calculation in this work. The vertex matrices used in the calculation are defined in Eqs.~\eqref{eq:vertex_blue}, \eqref{eq:vertex_red}, and \eqref{eq:vertex_green}; vertices shown in the same color share the same matrix representation.}
      \label{fig:vertex}
\end{figure}

In this appendix, we present the detailed implementation of the RPA calculation. In principle, the RPA vertex can be constructed from the original Hamiltonian. In the numerical calculation, however, we neglect the DM contribution to the vertex for simplicity because of its smallness and retain only the nearest-neighbor Heisenberg interaction,
\begin{align}
    H_{\mathrm{int}}
    =
    J \sum_{\langle i,j\rangle} \bm{S}_{i}\cdot\bm{S}_{j}
    =
    \frac{J}{2}\sum_{(i,j)} \bm{S}_{i}\cdot\bm{S}_{j},
\end{align}
where $\langle i,j\rangle$ runs over unordered nearest-neighbor bonds, whereas $(i,j)$ runs over all ordered nearest-neighbor pairs of sites, so that each bond is counted twice in the latter expression. This approximation is adopted because, in the present setup, $d^{z}/J=0.02$, so that the DM interaction is much smaller than the Heisenberg exchange.
We also note the approximation involved in this choice of perturbation Hamiltonian. Following previous spinon-based RPA calculations~\cite{Rao-Moessner-Knolle-2025,Willsher-2025}, we use the Heisenberg interaction as the perturbing interaction. A more microscopic treatment would isolate only the residual four-spinon interaction not already absorbed into the mean-field Hamiltonian, and the present implementation may therefore overestimate the strength of the magnetic instability quantitatively.
Expressing this Hamiltonian in the Schwinger boson representation, we obtain
\begin{align}
    \mathcal{H}
    &=\frac{J}{2}\sum_{\left(i,j\right)}\left[\frac{1}{2}\left(b_{i\uparrow}^{\dagger}b_{i\downarrow}b_{j\downarrow}^{\dagger}b_{j\uparrow}+b_{i\downarrow}^{\dagger}b_{i\uparrow}b_{j\uparrow}^{\dagger}b_{j\downarrow}\right)\right.\notag\\
    &\left.\qquad+\frac{1}{4}\bigl(n_{i\uparrow}-n_{i\downarrow}\bigr)\bigl(n_{j\uparrow}-n_{j\downarrow}\bigr)\right],
\end{align}
where $n_{i\mu}=b_{i\mu}^{\dagger}b_{i\mu}$.
The vertex function $V^{\mu\rho\lambda\nu}_{\bm{q}}$ is defined as Eq.~\eqref{eq:definition-vertex-function}.
For later convenience, we attach indices $(k,l)$ that indicate to which sublattices the sites $(i,j)$ belong, and reduce the vertex functions to a matrix form.
For example, for the $(k,l)=(\text{A},\text{B})$ component of $V_{\bm{q}}^{\uparrow\downarrow\uparrow\downarrow}$, we obtain
\begin{align}
    \left(V_{\bm{q}}^{\uparrow\downarrow\uparrow\downarrow}\right)_{\text{A}\text{B}}
    =\frac{M}{N}\sum_{i\in\text{A},j\in\text{B}}\frac{1}{2}J_{ij}e^{-i\bm{q}\cdot\left(\bm{r}_{i}-\bm{r}_{j}\right)}
    =J\cos\left(\bm{q}\cdot\bm{\delta}_{1}\right) \, .
\end{align}
Here, $\bm{\delta}_{1}$ is the displacement vector between the A and B sublattices, defined as
$\bm{\delta}_{1}=\bm{r}_{\text{B}}-\bm{r}_{\text{A}}$.
Similarly, we define $\bm{\delta}_{2}=\bm{r}_{\text{C}}-\bm{r}_{\text{B}}$ and $\bm{\delta}_{3}=\bm{r}_{\text{A}}-\bm{r}_{\text{C}}$~[see Fig.~\ref{fig:kagome_lattice}(a)].
We can then represent the six four-point vertex functions in matrix form as follows:
\begin{align}
    V_{\bm{q}}^{\uparrow\downarrow\uparrow\downarrow}
    &=V_{\bm{q}}^{\downarrow\uparrow\downarrow\uparrow}
    =J\begin{pmatrix}
        0&\cos{\left(\bm{q}\cdot\bm{\delta}_{1}\right)}&\cos{\left(\bm{q}\cdot\bm{\delta}_{3}\right)}\\
        \cos{\left(\bm{q}\cdot\bm{\delta}_{1}\right)}&0&\cos{\left(\bm{q}\cdot\bm{\delta}_{2}\right)}\\
        \cos{\left(\bm{q}\cdot\bm{\delta}_{3}\right)}&\cos{\left(\bm{q}\cdot\bm{\delta}_{2}\right)}&0
    \end{pmatrix},\label{eq:vertex_blue}\\
    V_{\bm{q}}^{\uparrow\uparrow\uparrow\uparrow}
    &=V_{\bm{q}}^{\downarrow\downarrow\downarrow\downarrow}
    =\frac{J}{2}\begin{pmatrix}
        0&\cos{\left(\bm{q}\cdot\bm{\delta}_{1}\right)}&\cos{\left(\bm{q}\cdot\bm{\delta}_{3}\right)}\\
        \cos{\left(\bm{q}\cdot\bm{\delta}_{1}\right)}&0&\cos{\left(\bm{q}\cdot\bm{\delta}_{2}\right)}\\
        \cos{\left(\bm{q}\cdot\bm{\delta}_{3}\right)}&\cos{\left(\bm{q}\cdot\bm{\delta}_{2}\right)}&0
    \end{pmatrix},\label{eq:vertex_red}\\
    V_{\bm{q}}^{\uparrow\downarrow\downarrow\uparrow}
    &=V_{\bm{q}}^{\downarrow\uparrow\uparrow\downarrow}
    =-\frac{J}{2}\begin{pmatrix}
        0&\cos{\left(\bm{q}\cdot\bm{\delta}_{1}\right)}&\cos{\left(\bm{q}\cdot\bm{\delta}_{3}\right)}\\
        \cos{\left(\bm{q}\cdot\bm{\delta}_{1}\right)}&0&\cos{\left(\bm{q}\cdot\bm{\delta}_{2}\right)}\\
        \cos{\left(\bm{q}\cdot\bm{\delta}_{3}\right)}&\cos{\left(\bm{q}\cdot\bm{\delta}_{2}\right)}&0
    \end{pmatrix}\label{eq:vertex_green}.
\end{align}

Here, we define the bare Matsubara susceptibility as a matrix in sublattice space with indices $(k,l)$.
The local bare spin correlation is
\begin{align}
    \label{eq:local-bare-spin-correlation}
    \chi_{ij}^{\alpha\alpha^{\prime}(0)}(\tau)
    &=\left\langle T_{\tau}S_{i}^{\alpha}(\tau)S_{j}^{\alpha^{\prime}}(0)\right\rangle_{0}\\
    &\simeq\sum_{p,q}A_{pq}^{\alpha\alpha^{\prime}}
    \overline{\mathcal{F}}_{ij}^{p}(\tau)\mathcal{F}_{ij}^{q}(\tau),
\end{align}
where the subscript 0 indicates the expectation value with respect to the SBMFT Hamiltonian, and the contractions $\overline{\mathcal{F}}_{ij}^{p}(\tau)$ and $\mathcal{F}_{ij}^{q}(\tau)$ are evaluated from the two-point Matsubara functions described in Sec.~\ref{sec:Calculation of the spin structure factor}. The corresponding sublattice-resolved susceptibility is
\begin{align}
    \label{eq:bare-matsubara-susceptibility-sublattice}
    \chi_{k,l}^{\alpha\alpha^{\prime}(0)}(\bm{q},i\omega_{n})
    &=\frac{M}{N}\sum_{i\in k,j\in l}e^{-i\bm{q}\cdot\left(\bm{r}_{i}-\bm{r}_{j}\right)}\notag\\
    &\qquad\times\int_{0}^{\beta}d\tau\,e^{i\omega_{n}\tau}
    \chi_{ij}^{\alpha\alpha^{\prime}(0)}(\tau).
\end{align}
The retarded bare susceptibility is obtained after the analytic continuation $i\omega_{n}\to\omega+i\delta$.
In the following, we focus on the eight components
$\chi^{++(0)}(\bm{q},i\omega_{n})$,
$\chi^{+-(0)}(\bm{q},i\omega_{n})$,
$\chi^{-+(0)}(\bm{q},i\omega_{n})$,
$\chi^{--(0)}(\bm{q},i\omega_{n})$,
$\chi^{\uparrow\uparrow(0)}(\bm{q},i\omega_{n})$,
$\chi^{\uparrow\downarrow(0)}(\bm{q},i\omega_{n})$,
$\chi^{\downarrow\uparrow(0)}(\bm{q},i\omega_{n})$,
and $\chi^{\downarrow\downarrow(0)}(\bm{q},i\omega_{n})$, and carry out the calculation.
Here, we defined $S_{i}^{\uparrow}=b_{i\uparrow}^{\dagger}b_{i\uparrow}$ and $S_{i}^{\downarrow}=b_{i\downarrow}^{\dagger}b_{i\downarrow}$.
In the two matrix equations below, all susceptibility components carry the common argument $(\bm{q},i\omega_{n})$, which is omitted to avoid clutter.
By setting up Dyson-type equations relating the bubble diagrams to the connected four-point vertices, we obtain the following matrix equations:
\begin{widetext}
\begin{align}
    \label{eq:Dyson_eq_pm}
    &\begin{pmatrix}
        \chi^{++}_{\text{RPA}}&\chi^{+-}_{\text{RPA}}\\
        \chi^{-+}_{\text{RPA}}&\chi^{--}_{\text{RPA}}
    \end{pmatrix}=
    \begin{pmatrix}
    \chi^{++(0)}&\chi^{+-(0)}\\
    \chi^{-+(0)}&\chi^{--(0)}
    \end{pmatrix}
    -\begin{pmatrix}
    \chi^{++(0)}&\chi^{+-(0)}\\
    \chi^{-+(0)}&\chi^{--(0)}
    \end{pmatrix}
    \begin{pmatrix}
    0&V_{\bm{q}}^{\uparrow\downarrow\uparrow\downarrow}\\
    V_{\bm{q}}^{\downarrow\uparrow\downarrow\uparrow}&0
    \end{pmatrix}
    \begin{pmatrix}
        \chi^{++}_{\text{RPA}}&\chi^{+-}_{\text{RPA}}\\
        \chi^{-+}_{\text{RPA}}&\chi^{--}_{\text{RPA}}
    \end{pmatrix},
\end{align}
\begin{align}
    \label{eq:Dyson_eq_ud}
    &\begin{pmatrix}
        \chi^{\uparrow\uparrow}_{\text{RPA}}&\chi^{\uparrow\downarrow}_{\text{RPA}}\\
        \chi^{\downarrow\uparrow}_{\text{RPA}}&\chi^{\downarrow\downarrow}_{\text{RPA}}
    \end{pmatrix}=
    \begin{pmatrix}
    \chi^{\uparrow\uparrow(0)}&\chi^{\uparrow\downarrow(0)}\\
    \chi^{\downarrow\uparrow(0)}&\chi^{\downarrow\downarrow(0)}
    \end{pmatrix}
    -\begin{pmatrix}
    \chi^{\uparrow\uparrow(0)}&\chi^{\uparrow\downarrow(0)}\\
    \chi^{\downarrow\uparrow(0)}&\chi^{\downarrow\downarrow(0)}
    \end{pmatrix}
    \begin{pmatrix}
    V_{\bm{q}}^{\uparrow\uparrow\uparrow\uparrow}&V_{\bm{q}}^{\uparrow\downarrow\downarrow\uparrow}\\
    V_{\bm{q}}^{\downarrow\uparrow\uparrow\downarrow}&V_{\bm{q}}^{\downarrow\downarrow\downarrow\downarrow}
    \end{pmatrix}
    \begin{pmatrix}
        \chi^{\uparrow\uparrow}_{\text{RPA}}&\chi^{\uparrow\downarrow}_{\text{RPA}}\\
        \chi^{\downarrow\uparrow}_{\text{RPA}}&\chi^{\downarrow\downarrow}_{\text{RPA}}
    \end{pmatrix}.
\end{align}
\end{widetext}
Note that both the Matsubara susceptibility $\chi^{\alpha\alpha^{\prime}}(\bm{q},i\omega_{n})$ and the four-point vertex $V_{\bm{q}}^{\mu\nu\rho\lambda}$ are $3\times 3$ matrices in sublattice space with indices $(k,l)$. Therefore, the matrices appearing above are $6\times 6$ matrices.

If we formally rewrite Eqs.~\eqref{eq:Dyson_eq_pm} and \eqref{eq:Dyson_eq_ud} as
\begin{align}
    \bm{\chi}_{\text{RPA}}(\bm{q},i\omega_{n})=\bm{\chi}^{(0)}(\bm{q},i\omega_{n})-\bm{\chi}^{(0)}(\bm{q},i\omega_{n})\bm{V}_{\bm{q}}\bm{\chi}_{\text{RPA}}(\bm{q},i\omega_{n}),
\end{align}
we can solve these equations as
\begin{align}
    \bm{\chi}_{\text{RPA}}(\bm{q},i\omega_{n})=\left(\bm{I}+\bm{\chi}^{(0)}(\bm{q},i\omega_{n})\bm{V}_{\bm{q}}\right)^{-1}\bm{\chi}^{(0)}(\bm{q},i\omega_{n}).
\end{align}
The RPA-corrected retarded function is obtained by analytic continuation, $\bm{\chi}_{\text{RPA}}(\bm{q},\omega+i\delta)=\bm{\chi}_{\text{RPA}}(\bm{q},i\omega_{n}\to\omega+i\delta)$. The component $\chi_{\text{RPA}}^{\alpha\alpha^{\prime}}(\bm{q},\omega+i\delta)$ is then obtained as
\begin{align}
    \chi^{\alpha\alpha^{\prime}}_{\text{RPA}}(\bm{q},\omega+i\delta)=\sum_{k,l}\left[\chi_{\text{RPA}}^{\alpha\alpha^{\prime}}(\bm{q},\omega+i\delta)\right]_{k,l} \, .
\end{align}
Using the fluctuation--dissipation theorem, the RPA-corrected dynamical spin structure factor can be computed from
\begin{align}
    S^{\alpha\alpha^{\prime}}_{\text{RPA}}(\bm{q},\omega)=\frac{1}{\pi}\frac{1}{1-e^{-\beta\omega}}\operatorname{Im}\chi^{\alpha\alpha^{\prime}}_{\text{RPA}}(\bm{q},\omega+i\delta).
\end{align}
Furthermore, the RPA-corrected dynamical spin structure factor $S_{\mathrm{RPA}}(\bm{q},\omega)$ plotted in Sec.~\ref{sec:Random phase approximation beyond Schwinger boson mean-field} is obtained as
\begin{align}
    S_{\mathrm{RPA}}(\bm{q},\omega)=\frac{1}{M}\sum_{\alpha=x,y,z}S^{\alpha\alpha}_{\mathrm{RPA}}(\bm{q},\omega)\, .
\end{align}

\bibliography{./refs}

\end{document}